\documentclass[
	final,
	aps,
	floatfix,
	twocolumn,
	]{revtex4-1}
\let\Twocolumn

\newif\ifTwocolumn
\Twocolumntrue

\usepackage{graphicx}
\usepackage{amsmath,amssymb,amsfonts}
\usepackage{gensymb}
\usepackage{bm}
\usepackage{upgreek}
\usepackage{wasysym}
\usepackage{bibentry}
\usepackage[T1]{fontenc}
\usepackage{textcomp}

\usepackage[scaled=0.92]{helvet}
\usepackage{courier}
\usepackage{color}
\usepackage{scalerel}
\usepackage{hyperref}
\usepackage{multirow}
\usepackage{cases}
\newcommand{\upd}{{\ensuremath{\textrm{d}}}}

\renewcommand{\vec}[1]{\mathbf{#1}}

\newcommand{\coscos}{(\cos\vartheta_1)\,(\cos\vartheta_2)}
\definecolor{gray}{gray}{.5}
\DeclareRobustCommand{\numcirc}[1]{\unitlength1ex\begin{picture}(2.5,2.5)\put(1.25,0.75){\circle{2.5}}\put(1.25,0.75){\makebox(0,0){#1}}\end{picture}}
\DeclareRobustCommand{\symbCS}{\includegraphics[height=0.7em,angle=90]{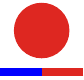}}
\DeclareRobustCommand{\symbJcH}{\includegraphics[height=0.7em,angle=90]{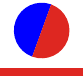}}
\DeclareRobustCommand{\symbJcS}{\includegraphics[height=0.7em,angle=90]{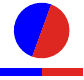}}
\DeclareRobustCommand{\symbJcSm}{\includegraphics[height=0.7em,angle=90]{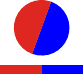}}
\DeclareRobustCommand{\symbJcJc}
{\includegraphics[height=1.2em, angle=90]{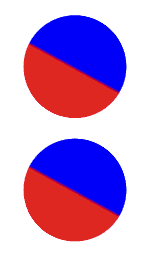}}

\ifTwocolumn
\newenvironment {multeqline} {\multline} {\endmultline}
\else
\newenvironment {multeqline} {\equation} {\endequation}
\fi

\begin{document}
\title{Critical Casimir interactions between Janus particles}
\author{M.~Labb{\'e}-Laurent}
\email{laurent@is.mpg.de}
\author{S.~Dietrich}
\email{dietrich@is.mpg.de}
\affiliation{
	Max-Planck-Institut f\"ur Intelligente Systeme,  
	Heisenbergstr.\ 3, D-70569 Stuttgart, Germany}
\affiliation{
	IV. Institut f\"ur Theoretische Physik, 
	Universit\"at Stuttgart, 
	Pfaffenwaldring 57, 
	D-70569 Stuttgart, Germany
}
\date{\today}
%
\begin{abstract}
  Recently there is strong experimental and theoretical interest in studying the self-assembly and the phase behavior of patchy and of Janus particles, which form colloidal suspensions. Although in this quest a variety of effective interactions have been proposed and used in order to achieve directed assembly, the critical Casimir effect stands out as being particularly suitable in this respect because it provides both attractive and repulsive interactions as well as the potential of a sensitive temperature control of their strength.
  Specifically, we have calculated the critical Casimir force between a single Janus particle and a laterally homogeneous substrate as well as a substrate with a chemical step. We have used the Derjaguin approximation and compared it with results from full mean field theory. A modification of the Derjaguin approximation turns out to be generally reliable. Based on this approach we have derived the effective force and the effective potential between two Janus cylinders as well as between two Janus spheres.
\end{abstract}
\maketitle
\section{Introduction}
The critical Casimir effect has been predicted \cite{Fisher:1978} as a classical analogue of the celebrated Casimir effect in quantum electrodynamics \cite{Casimir:1948}. The former is induced by the confinement of order parameter  fluctuations in a system close to its critical point $T_c$, whereas the latter is due to the confinement of vacuum fluctuations.
Upon approaching $T_c$ the bulk correlation length $\xi$, characterizing the exponential decay of the two-point order parameter correlation function, increases algebraically as $\xi\left(t=(T-T_c)/T_c\to 0^\pm\right)=\xi_0^\pm |t|^{-\nu}$, with a bulk critical exponent $\nu$ and non-universal amplitudes $\xi_0^\pm$. If $\xi$ becomes comparable with the size of the system, the so-called critical Casimir force arises which acts as an effective force on the confining surfaces of the system. The energy scale of the critical Casimir effect is set by $k_B T$ and its strength can be sensitively tuned by minute temperature changes. This effective force can be attractive as well as repulsive depending on the boundary conditions for the order parameter at the surfaces. Contrary to the quantum mechanical Casimir effect, the sign of the critical Casimir force can be chosen by modifying exclusively the surface chemistry of the confinement.

The first experimental evidence for critical Casimir forces was provided only indirectly by studying the thickness of thin wetting films in classical binary liquid mixtures \cite{Fukuto:2005,Rafai:2007} near demixing, as well as in mixtures of $^3\text{He}$ / $^4\text{He}$ \cite{Garcia:2002,Ueno:2003} and liquid $^4\text{He}$ close to their normal-superfluid transition \cite{Garcia:1999,Ganshin:2006}. Corresponding Monte Carlo simulations for the film geometry \cite{Hucht:2007,Vasilyev:2007,Vasilyev:2009,Hasenbusch:2009b,Hasenbusch:2010,Hasenbusch:2010a} are in very good quantitative agreement with the experiments.
The first direct measurement of the critical Casimir effect \cite{Hertlein:2008} was performed by monitoring optically the thermal motion of a single spherical colloid, immersed in a binary liquid mixture of water and 2,6-lutidine close to demixing and near a chemically homogeneous substrate. The experimental results are in excellent agreement with corresponding theoretical predictions \cite{Hertlein:2008,Gambassi:2009,Pousaneh:2012}, which make use of the Derjaguin approximation (DA) \cite{Derjaguin:1934} with Monte Carlo (MC) simulation results for the film geometry as input. 
A full MC simulation for the sphere-wall geometry has been performed only recently \cite{Hasenbusch:2013}.
Other theoretical studies rely on field-theoretical methods \cite{Burkhardt:1995, Eisenriegler:1995, Hanke:1998, Schlesener:2003, Eisenriegler:2004}.

Independently, at the same time strong experimental and theoretical interests emerged in patchy colloidal particles with chemically heterogeneous surface properties and in Janus particles with ``two faces'' --- a topic which has been popularized by the nobel prize lecture of de Gennes \cite{deGennes:1992}. These particles have the potential to be building blocks for directed self-assembly of new materials, such as the kagome open-lattice structure \cite{Chen:2011, Romano:2011, Fejer:2015}.
Topical reviews concerning both experimental and theoretical aspects of patchy particles are provided in Refs.~\cite{Pawar:2010} and \cite{Bianchi:2011}. From an experimental point of view, the fabrication of such particles poses a research challenge in itself \cite{Prasad:2009, Walther:2009, Yi:2013}, followed by the experimental observation of their (self-)assembly behavior \cite{Zhang:2015, Yu:2014, Chen:2011}.

In principle, any anisotropic surface structure gives rise to an orientation dependent behavior caused by surface mediated interactions, e.g., due to surface charges \cite{Hong:2006, BianchiIPC:2011} or critical fluctuations.
In this sense, the critical Casimir effect is a viable candidate to achieve controlled self-assembly, as demonstrated experimentally by the trapping of homogeneous colloids adjacent to chemically patterned substrates \cite{Soyka:2008,Troendle:2011}, in very good agreement with corresponding theoretical predictions \cite{Troendle:2009,Troendle:2011}.

The study of Janus particles exposed to the critical Casimir effect represents a rather new research issue, encompassing a few promising experimental investigations \cite{Iwashita:2013,Iwashita:2014}. The critical Casimir effect provides a controllable effective interaction which can be directed by both attraction and repulsion between the patches of the particles, depending on the design and surface treatment of the particles. The surfaces can also be modified in order to change boundary conditions for the order parameter of the underlying continuous phase transition of the solvent, e.g., by producing a surface with only weak adsorption preference for one of the two species forming the binary liquid solvent \cite{Mohry:2010}, though here we shall consider only the strong adsorption limit.

Concerning the modeling of effective interactions between patchy particles, the main body of theoretical research is, however, dominated by the simple Kern-Frenkel model \cite{Kern:2003}. This model assumes short-ranged on-off 
``bond-like'' interactions and is employed in simulations of self-assembly behavior \cite{Zhang:2004}, in numerical studies of the phase diagram of patchy particles \cite{Sciortino:2009, Sciortino:2010, Preisler:2014, Vissers:2013}, in Monte Carlo simulations \cite{Fantoni:2011}, or is embedded into other theoretical frameworks, ranging from Wertheim's association theory over integral-equation theory to self-consistent phonon theory \cite{Marshall:2012, Giacometti:2014, Shin:2014}.
In contrast, at the critical point $T=T_c$ of the solvent the critical Casimir forces are long-ranged, can be both attractive and repulsive, but the strengths of attraction and repulsion differ. %
In order to understand the behavior of patchy particles in a critical solvent it is therefore necessary to work out the distinguishing features of the critical Casimir interaction compared to those of the simple Kern-Frenkel model, which assumes a square-well potential and the interaction between two patches only.

Our theoretical analysis of the critical Casimir interaction between Janus particles is structured as follows: In Sec. \ref{sec:background} we present the theoretical background of our method, starting with a brief introduction to finite size scaling for the present system. Within a two-pronged approach, we outline both the full numerical mean field calculations valid in spatial dimension $d=4$, as well as the Derjaguin approximation used for $d=3$ and $d=4$.

Since previously a significant theoretical effort was put into the investigation of the interaction with patterned substrates \cite{Parisen:2010, ParisenToldin:2013, ParisenToldin:2015, Troendle:2010, Troendle:2011}, in Sec. \ref{sec:janus_vs_steps} we first consider a cylindrical Janus particle close to a homogeneous substrate. The Derjaguin approximation implies an intriguing link in the description between the presence of chemical steps on a striped surface and of the chemical step on a Janus particle. We investigate this link which is confirmed by the order parameter distribution to occur in modified form also within mean field theory (MFT). This result is then employed for a Janus particle floating above a substrate with a chemical step.

In Sec. \ref{sec:janus_cyls} we draw on this knowledge in order to establish within the Derjaguin approximation the force and the effective interaction potential between two Janus cylinders close to each other, but with a constraint on their orientation. In Sec. \ref{sec:janus_spheres} we present the force and the interaction potential between two Janus spheres for arbitrary orientations of the two particles. The derivation of the corresponding scaling function can be found in Appendices A, B, and C.
Finally, in Sec. \ref{sec:conclusions} we conclude and provide an outlook.
\section{Theoretical background}
\label{sec:background}
\subsection{Finite size scaling}
Close to a critical point of a fluid, thermal fluctuation become correlated over macroscopic distances and are, to a large extent, independent of microscopic details. Upon approaching the critical demixing point $T_c$ of a binary liquid mixture at its critical concentration, the bulk 
correlation length diverges as $\xi_\pm\left(t=(T-T_c)/T_c\to0^\pm\right)=\xi_0^\pm|t|^{-\nu}$, with the critical exponent $\nu\simeq0.63$ in $d=3$ and $\nu=1/2$ 
in $d=4$ \cite{Pelissetto:2002}.
The sign of $t$ is chosen such that $t>0$ corresponds to the homogeneous, mixed state, whereas $t<0$ corresponds to the two phase region.
Many experiments are performed advantageously in binary liquid mixtures with a lower critical point \cite{Hertlein:2008, Soyka:2008, Gambassi:2009, Troendle:2011, Iwashita:2013, Iwashita:2014}; in this case one has $t=(T_c-T)/T_c$.

According to finite size scaling, in the vicinity of its bulk critical point a (partially) finite system is described by universal scaling functions, which depend only on the shape of the sample and on coarse features of the system, summarized by universality classes. Here, we focus on the case of binary liquid mixtures, which belong to the Ising universality class, for which the scalar order parameter $\phi$ is defined as the deviation of the number density of one species from its value at criticality.

Accordingly, the critical Casimir force is described by an universal scaling function uniquely determined by the bulk universality class \cite{Pelissetto:2002} (here: Ising), the surface universality class \cite{Binder:1983, Diehl:1986} (here: normal transition with symmetry-breaking boundary conditions $(+)$ and $(-)$), the spatial dimensional (here: $d=3$ and $d=4$ in mean field theory), and the geometry of the confinement \cite{Krech:book, Brankov:2000, Gambassi:2009Review} (here: cylinders, spheres, and planar walls).

In the case of the \emph{film} geometry with two flat, parallel, homogeneous, and macroscopically large walls at distance $l$, renormalization group theory predicts the following form for the critical Casimir force $f_{(a,b)}$ per area of the wall \cite{Krech:1992all}:
\begin{equation} 
  \label{eq:planar-force}
  f_{(a,b)}(l,T)=k_BT\,\frac{1}{l^d}\,k_{(a,b)}(\Theta=\mathrm{sign}(t)\, l/\xi_\pm),
\end{equation} 
where the subscript $(a,b)$ indicates the pair of  boundary conditions (BC) $(a)$ and $(b)$ characterizing the two walls.
In the absence of a bulk ordering field and for infinitely strong surface fields, the scaling function $k_{(a,b)}$ depends only on a single scaling variable, which is given by
the sign of the reduced temperature $t$ and the film thickness $l$ in units of the bulk correlation length $\xi_\pm$
(with $\pm$ taken for $t\gtrless0$).
We emphasize that Eq.~\eqref{eq:planar-force} describes the behavior of the \textit{singular} contribution to the effective force acting on the confining walls, in addition to any background forces, e.g., van der Waals forces.

At the critical point $T=T_c$, $\xi_\pm$ diverges and the scaling function of the force $k_{(a,b)}$ between two walls reduces to an universal number referred to as the critical Casimir amplitude (see Ref.~\cite{Krech:book}; the notation differs slightly)
\begin{equation}
k_{(a,b)}(l/\xi_\pm=0)=\Delta_{(a,b)},
\end{equation}
which leads to an algebraic decay $\sim l^{-d}$ of the critical Casimir force as a function of the film thickness.
In contrast, off criticality the critical Casimir force decays exponentially as 
a function of $l/\xi_\pm$. 
For the symmetry-breaking BCs $(-,-)$ or $(+,-)$ valid for binary liquid mixtures and for $t>0$, the critical Casimir force is expected to decay as (see Refs.~\cite{Krech:1997, Gambassi:2009, Troendle:2009})
\begin{equation} 
  \label{eq:exponential-decay}
  k_{(+,\pm)}(l/\xi_+\gg1)=\mathcal{A}_\pm \left(\frac{l}{\xi_+}\right)^d \exp(-l/\xi_+),
\end{equation}
where $\mathcal{A}_\pm$ are universal amplitudes \cite{Gambassi:2009}.

\subsection{Mean field theory}
Within MFT, the bulk and surface critical phenomena belonging to the Ising universality class are described by the standard Landau-Ginzburg-Wilson fixed point Hamiltonian \cite{Binder:1983, Diehl:1986}
\begin{multline}
\mathcal{H}[\phi(\vec{r})]=\ifTwocolumn \\ \fi
\int_V\upd^d\vec{r}\,\left(\frac{1}{2}\left(\nabla\phi(\vec{r})\right)^2+\frac{\tau}{2}\left(\phi(\vec{r})\right)^2 + \frac{u}{4!}\left(\phi(\vec{r})\right)^4\right)\\
+ \int_{\partial V}\upd^{(d-1)}\vec{s}\,\left(\frac{c}{2}\left(\phi(\vec{s})\right)^2-h_1\phi(\vec{s})\right),
\label{eq:Hamiltonian}
\end{multline}
which is a functional of the order parameter profile $\phi(\vec{r})$ of the fluid such as the difference between the local concentration of one of the two species and its critical value in a binary liquid mixture. The Hamiltonian consists of a bulk term representing a $d$-dimensional liquid-filled volume $V$ and a term describing the confining surface $\partial V$ of this volume, e.g., provided by the surfaces of colloids immersed in the binary mixture, with $\left.\phi(\vec{r})\right|_{\partial V} = \phi(\vec{s})$ evaluated at the boundary $\partial V$. Within MFT, the parameter $\tau$ is proportional to the reduced temperature $t$ as $\tau=t/(\xi_0 ^+)^2$ \cite{Krech:book}, while the coupling constant $u>0$ ensures the stability of $\mathcal{H}[\phi(\vec{r})]$ for $t<0$ in the demixed phase; $u$ is dimensionless in $d=4$. In order to treat off-critical concentrations, the expression in Eq.~\eqref{eq:Hamiltonian} can be extended to contain a term proportional to a bulk field $h$. The surface enhancement $c$ and the symmetry breaking surface field $h_1$ determine the BC. We focus on the so-called normal surface universality class, which is generic for liquids, with $c=0$ and the two fixed point values $h_1=\pm\infty$. This leads to a divergence of $\phi\to\pm\infty$ at the surface of the colloids corresponding to what is denoted as the $(+)$ and $(-)$ BC \cite{Diehl:1986}. Concerning the numerical implementation, the divergence is realized by a short distance expansion close to the surface \cite{Hanke:1999a,Kondrat:2007}.
Within MFT, only the order parameter configuration with the largest statistical weight $\exp\left(-\mathcal{H}[\phi(\vec{r})]\right)$ is considered and fluctuations of the order parameter are neglected. Within this approximation the free energy follows from $\delta\mathcal{H}[\phi]/\delta\phi|_{\phi=\langle\phi\rangle}=0$. The MFT order parameter profile defined as $m=\langle\phi\rangle/\phi_t$ minimizes the Hamiltonian $\mathcal{H}$, where $\phi_t=\sqrt{6/u}/\xi_0^+$ is the non-universal amplitude of the bulk order parameter $\phi_b=\phi_t\,|t|^\beta$, $\beta=1/2$ in $d=4$ and $\beta\simeq 0.33$ in $d=3$. MFT captures correctly the critical behavior above the upper critical dimension $d_c=4$, with logarithmic corrections in $d=4$. In the context of renormalization group theory, the MFT results represent also the leading order contribution within an expansion in terms of $\epsilon=4-d$. There are only two independent non-universal bulk amplitudes \cite{Binder:1983,Diehl:1986}, such as $\phi_t$ and $\xi_0^+$.


For a film confined by two planar walls, the MFT scaling functions of the critical Casimir force have been determined analytically \cite{Krech:1997} and, inter alia, the critical Casimir amplitudes for symmetry breaking BC have been found as $\Delta_{(+,+)}=\Delta_{(-,-)}=-\Delta_{(+,-)}/4=48[K(1/\sqrt{2})]^4/u$ where $K$ is the complete elliptic integral of the first kind.

For the geometries studied here within MFT, the Hamiltonian $\mathcal{H}[\phi]$ has been minimized numerically using a three-dimensional finite element method in order to obtain the order parameter profiles. The system is assumed to be translationally invariant along an extra dimension in $d=4$. The critical Casimir forces are determined directly from the order parameter profile using the stress-tensor method \cite{Schlesener:2003,Kondrat:2009,Krech:1997}.

The scaling function $k_{(a,b)}(\Theta)$ in Eq.~\eqref{eq:planar-force} covers the full range $\Theta\in\mathbb{R}$ with $\Theta<0$ for $t<0$ and $\Theta>0$ for $t>0$, respectively. We note that the scaling variable $\Theta=\mathrm{sign}(t)\, l/\xi_\pm$ contains distinct denominators $\xi_0^\pm$ for $t\gtrless 0$ in accordance with the universal ratio $R_\xi=\xi_0^+/\xi_0^-=1.96$ in $d=3$ \cite{Pelissetto:2002} and $R_\xi=\sqrt{2}$ in $d=4$ \cite{Tarko:all}. Here, we focus on $t\geq0$, for which the solvent is in the homogeneous, mixed phase. This relates to the common experimental situation in which the critical behavior near the lower critical point of a binary liquid mixture is studied upon approaching $T_c$ along a thermodynamic path from below (e.g., Refs.~\cite{Hertlein:2008, Soyka:2008, Troendle:2009, Bonn:2009, Iwashita:2013, Iwashita:2014}).

\subsection{Derjaguin approximation}
The Derjaguin approximation (DA) is a common technique to extend theoretically results for planar geometry, which can be derived directly, to curved objects, which are more common in practice. This approximation builds on the additivity of forces. Accordingly, a curved surface is sliced into infinitesimally small surface elements and the total force is calculated by summing up the individual planar wall-wall contributions $k_{(a,b)}$ from the surface elements vis-\`{a}-vis, with $(a)$ and $(b)$ indicating the BC at the two surfaces.
In the case of a spherical object, its surface is divided into thin rings of radius $\rho $ \cite{Hanke:1998, Gambassi:2009}, whereas the surface of cylindrical objects is decomposed into parallel pairs of infinitesimally narrow stripes at lateral positions $\pm\rho$ \cite{Troendle:2010, Labbe:2014}. For both types of objects, the distance of each element from a planar wall is given by $D(\rho) = D + R(1-\sqrt{1-\rho^2/R^2})$, where $D$ is the closest distance between the particle surface with radius $R$ and the planar wall. Since the DA holds only in the limit of large particle radii $R$, i.e., $\Delta = D/R\to 0$, it is often \cite{Hanke:1998, Gambassi:2009, Troendle:2010} used in conjunction with the further ``parabolic distance approximation'' $D(\rho) \approx D(1+\rho^2/(2RD))$.
For comparison, the surface-to-surface distance $D(\rho) = D + 2 R(1-\sqrt{1-\rho^2/R^2})$ either between two spheres or between two cylinders increases twice as fast with $\rho$; correspondingly, within the ``parabolic distance approximation'' one has in these two cases $D(\rho) \approx D(1+\rho^2/(RD))$.

For Janus particles, the basic DA approach remains the same. However, for them the force contribution switches spatially between $k_{(+,+)}=k_{(-,-)}$ and $k_{(+,-)}=k_{(+,-)}$ due to the variation of the BC across the surface(s). Assuming again additivity and neglecting edge effects, the summation over these force contributions can be performed after appropriately subdividing the surface and grouping the surface elements according to the various pairs of BC. For two Janus spheres this is presented in detail in Appendix \ref{sec:app_janus_spheres}.

The DA for these geometries is based on the scaling function of the force for the film geometry.  For $d=4$ this is adopted directly from our independent MFT calculations for two parallel walls (see below). In $d=3$ the scaling function of the force for the film geometry has been obtained from MC simulations \cite{Vasilyev:2007,Vasilyev:2009,
Dantchev:2004,Hasenbusch:2010a,Hasenbusch:2012}. Here, we rely on the numerical estimate referred to as ``approximation (i)'' in Figs.~9 and 10 of Ref.~\cite{Vasilyev:2009}.
The systematic uncertainty of the overall amplitude of these scaling functions can, in the worst case, reach up to 10\%--20\% 
\cite{Vasilyev:2009}, which also affects our predictions.
However, the impact on the scaling functions \emph{normalized} by the critical amplitude $\Delta_{(+,+)}$ is greatly reduced and only on the relative level of at most $5\%$ \cite{Troendle:2010}.

It has been shown that the DA is most reliable for $t\geq0$ \cite{Labbe:2014, Hasenbusch:2013}, whereas for $(+,-)$ BC and $t<0$ clear deviations from the DA occur, which can be explained in terms of the formation of an interface surrounding the particles \cite{Labbe:2014}.
\section{General aspects concerning Janus particles}
\label{sec:janus_vs_steps}
\subsection{Implications of the DA for a cylindrical particle above a substrate}

\begin{figure*}[t!]
  \centering
  \includegraphics[width=10.5cm]{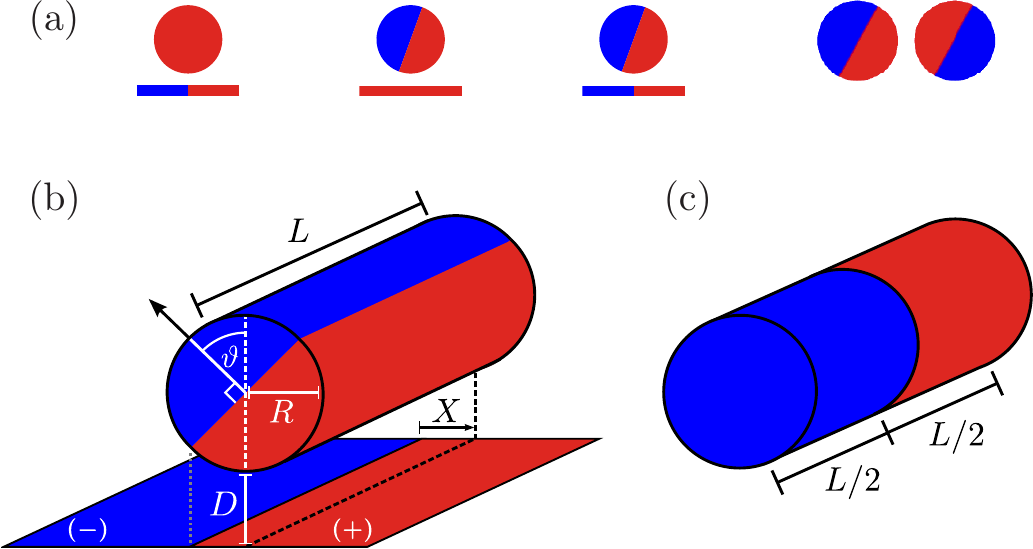}
  \caption{
(a) Sideview sketches of all types of configurations considered in the present study: chemically homogeneous cylinder vs. substrate with a chemical step --- Janus cylinder (as in (b)) vs. homogeneous substrate --- Janus cylinder vs. substrate with a chemical step --- two Janus cylinders or two Janus spheres without a substrate.
(b) Janus cylinder in $d=3$ with the chemical step along the cylinder axis, shown in proximity and parallel to a planar substrate. The orientation of the Janus cylinder is given by the angle $\vartheta$ between the normal of the equatorial plane of the Janus cylinder and the substrate normal. The substrate may also feature a chemical step parallel to the cylinder, at a lateral position $X$ which measures the distance between the projection of the cylinder axis (dotted line) and the chemical step at the substrate.
(c) Second variant of a Janus cylinder in $d=3$, with the chemical step perpendicular to the axis of the cylinder. For a discussion of this variant see Ref.~\cite{Labbe:2014}.
}
  \label{fig:sketch_cyls}
\end{figure*}

Before we address the subject of the effective interaction between \textit{two} Janus particles, we assess the quality of the DA for the configuration of a \textit{single} Janus cylinder above a substrate. The analysis in this section follows Fig.~\ref{fig:sketch_cyls}(a) by first restating the case of a homogeneous particle above a substrate with a chemical step, then by introducing a Janus particle above a homogeneous substrate before considering a Janus particle above a substrate step, which will connect to the case of two Janus particles.

First, we clarify the ambiguous definition of a (hyper-)cylinder in higher dimensions $(d\geq 4)$. In the present context, a cylinder in $d=4$ is a geometrical object with radius R and two lengths $L$ and $L_4$, defined by the volume $x^2 + y^2 \leq R^2,\ 0 \leq z \leq L$ and $0\leq w \leq L_4$, where $w$ is the coordinate in the extra dimension and $L_4$ is the length in that direction \footnote{Another definition of the volume of a hypercylinder would be $x^2 + y^2 + z^2 \leq R^2,\ 0\leq w \leq L_4$, which we dismiss for formal reasons: The projection of this object onto three dimensions renders a sphere instead of a cylinder. Thus this object does not fulfill the expectation for a basic extension of a cylinder from three to four dimensions.}. We will use the $(d-2)$ dimensional length $\mathcal L$ in order to denote $\mathcal L=L$ in $d=3$ and $\mathcal L = L\times L_4$ in $d=4$.

The Janus character due to the BC at the surface of a cylinder can be realized in two distinct ways in $d=3$ [see Figs.~\ref{fig:sketch_cyls}(b) and (c)] and in three ways in $d=4$. 
The two possibilities in $d=3$ are evident with the chemical step, separating two domains of BC, either running along the length of the cylinder, cutting it into two half-cylinders [Fig.~\ref{fig:sketch_cyls}(b)], or perpendicular to the length of the cylinder, cutting it in two cylinders of half the length [Fig.~\ref{fig:sketch_cyls}(c)].
It has been demonstrated that the latter case can be constructed within DA by a straightforward combination of two cylinders (see Ref.~\cite{Labbe:2014}).
The former case, however, requires a new analysis, which is carried out in the present study.
The third case, occurring for a cylinder in $d=4$, has the step in the BC in the extra dimension, rendering two equal sized hypercylinders with different BC. This is of limited practical use regarding the comparison with results in $d=3$.
We therefore restrict our description to the ``natural'' choice of a Janus cylinder being composed of two half-cylinders, both in $d=3$ and in $d=4$.

In order to set the stage, we recall the case of a chemically homogeneous cylinder close to a substrate with a chemical step (see Fig.~\ref{fig:sketch_cyls}(a)). The lateral position of the cylinder axis relative to a chemical step in parallel on the substrate is denoted by $X$ (see Fig.~\ref{fig:sketch_cyls}(b)). Moreover, we always consider the cylinder to be parallel to the substrate (and to the step).
%
%
The critical Casimir force $F^{(cs)}_{\symbCS}(X, D, R, T)$ between a \textit{homogeneous} \textit{c}ylindrical particle of length $L$ and radius $R$, and a \textit{s}ubstrate with a step, at a lateral position $X$, has the scaling form (see Eq.~(D1) in Appendix D of Ref.~\cite{Troendle:2010})
\begin{equation}
 F^{(cs)}_{\symbCS}(X, D, R, T) = k_B T\,\frac{\mathcal L}{R^{d-1}}\,\frac{K^{(cs)}_{\symbCS}(\Xi, \Delta, \Theta)}{\Delta^{d-1/2}},
 \label{eq:scalform_F_cyl_step}
\end{equation}
with the dimensionless scaling variables $\Xi=X/\sqrt{R D}$, $\Delta = D/R$, and $\Theta = \pm D/\xi_\pm(T)$ (with $\mathrm{sign}(\Theta)=\mathrm{sign}(t)$) in $d$ dimensions. The scaling function $K^{(cs)}_{\symbCS}(\Xi, \Delta, \Theta)$ of the force $F^{(cs)}_{\symbCS}$ can be decomposed as \cite{Troendle:2010}
\begin{equation}
 K^{(cs)}_{\symbCS}(\Xi, \Delta, \Theta) = \begin{cases}
 K^{(cs)}_{(\textcolor{red}{+},\textcolor{red}{+})}(\Delta,\Theta)-\Delta K^{(cs)}_{\symbCS}(|\Xi|,\Delta,\Theta) \ifTwocolumn \\ \fi & \ifTwocolumn \llap{\text{for $\Xi>0,$}} \else \text{for $\Xi>0,$} \fi \\
 K^{(cs)}_{(\textcolor{red}{+},\textcolor{blue}{-})}(\Delta,\Theta)+\Delta K^{(cs)}_{\symbCS}(|\Xi|,\Delta,\Theta) \ifTwocolumn \\ \fi & \ifTwocolumn \llap{\text{for $\Xi\leq0,$}} \else \text{for $\Xi\leq0,$} \fi 
\end{cases}
\label{eq:DA_K_cyl_step}
\end{equation}
where (see Eq.~(D3) in Ref.~\cite{Troendle:2010})
\begin{equation}
 K^{(cs)}_{(+,\pm)}(\Delta\to 0,\Theta)=\sqrt{2}\int_1^{\infty}\upd\alpha\ \frac{k_{(+,\pm)}(\alpha\Theta)}{\alpha^d\sqrt{\alpha-1}}
\label{eq:DA_K_cyl_hom}
\end{equation}
is the scaling function of the force within DA for a homogeneous \textit{c}ylindrical particle ($+$ or $-$) close to a homogeneous \textit{s}ubstrate ($+$ or $-$), and thus does not depend on $\Xi$. The scaling function $k_{(+,\pm)}$ for the slab geometry serves as an input, which is obtained either from MFT calculations for the film geometry in $d=4$ or from an interpolation of MC data provided in Ref. \cite{Vasilyev:2009} for $d=3$. The choice of signs in Eq.~\eqref{eq:DA_K_cyl_step} reflects $\Xi\gtrless 0$, chosen such that the direction of positive $X$ points to the side of the step with the same BC as the colloid (see Fig. 1(a)) which is $(+)$ in the present notation.

The excess scaling function $\Delta K^{(cs)}_{\symbCS}$ involving the step position $X$ is given within DA by (see Eq.~(D6) in Ref.~\cite{Troendle:2010})
\begin{equation}
 \Delta K^{(cs)}_{\symbCS}(|\Xi|,\Delta\to 0,\Theta)=\frac{1}{\sqrt{2}}\int_{1+\Xi^2/2}^{\infty}\upd\alpha\ \frac{\Delta k(\alpha\Theta)}{\alpha^d\sqrt{\alpha-1}},
\label{eq:DA_deltaK_cyl_step}
\end{equation}
where $\Delta k = k_{(+,+)}-k_{(+,-)}<0$ is the difference between the slab scaling functions for distinct BC, which is negative for all temperatures $\Theta$. Note that $ \Delta K^{(cs)}_{\symbCS}$ depends only on the absolute value of the scaled distance $\Xi$, because the inverted position is equivalent to a switch of the BC of the step, which is covered by Eq.~\eqref{eq:DA_K_cyl_step}.

\begin{figure}[t!]
  \centering
  \includegraphics[width=0.47\textwidth]{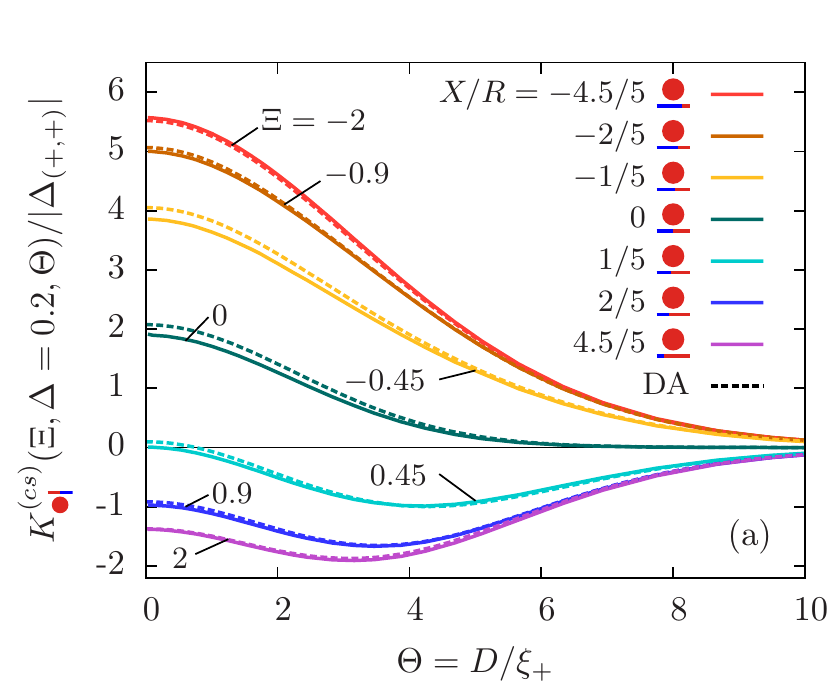}\hfill
  \includegraphics[width=0.47\textwidth]{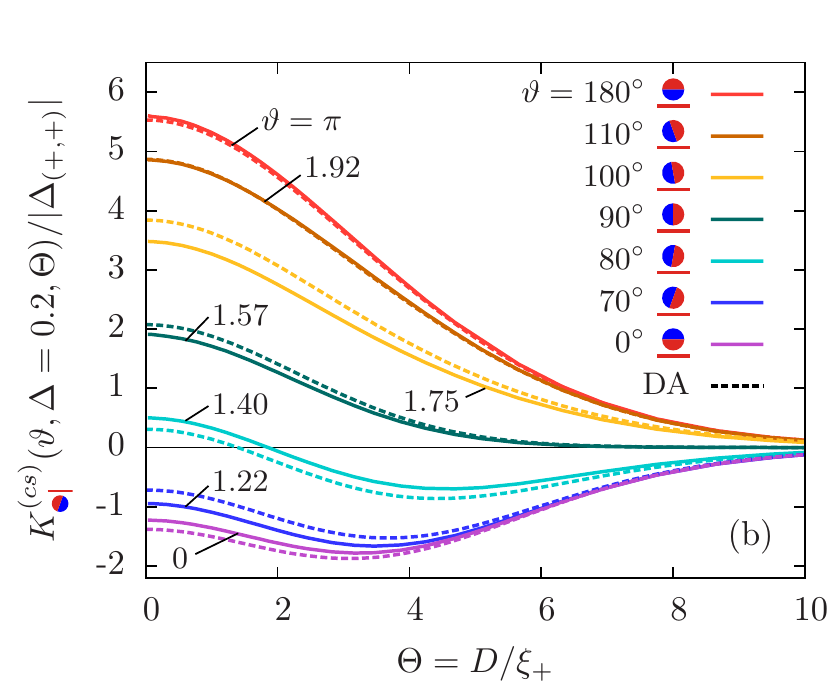}
  \caption{(a) Scaling function $K^{(cs)}_{\symbCS}$ of the force between a homogeneous cylindrical particle above a substrate with a chemical step at various scaled lateral positions $\Xi$. (b) Scaling function of the force $K^{(cs)}_{\symbJcH}$ between a Janus cylinder and a homogeneous substrate for various orientations $\vartheta$. The full MFT results are shown as solid lines, whereas the corresponding DA scaling functions are shown as dashed lines. The DA yields a qualitatively adequate approximation for the MFT scaling functions, with varying quantitative deviations in (a) and (b).}
  \label{fig:Kcyl_step_Theta}
\end{figure}

As a function of the scaled temperature $\Theta$, in Fig. \ref{fig:Kcyl_step_Theta}(a) we compare the scaling function of the force $K^{(cs)}_{\symbCS}$ obtained within DA for $d=4$ via Eqs.~\eqref{eq:DA_K_cyl_step}--\eqref{eq:DA_deltaK_cyl_step} (dashed curves) with the corresponding full MFT results (solid lines) determined by numerical minimization of the Hamiltonian.
As expected from Ref.~\cite{Troendle:2010}, in Fig.~\ref{fig:Kcyl_step_Theta}(a) the DA scaling function approximates the full MFT results well for the geometry of a \textit{homogeneous} \textit{c}ylinder above a \textit{s}ubstrate step, shown for various scaled step positions $\Xi$ on both sides of the step.

In accordance with the second sketch in Fig.~\ref{fig:sketch_cyls}(a), we now consider a Janus \textit{c}ylinder, but placed above a homogeneous \textit{s}ubstrate. The corresponding critical Casimir force $F^{(cs)}_{\symbJcH}(\vartheta, \Delta, \Theta)$ depends on the orientation angle $\vartheta$ (Fig.~\ref{fig:sketch_cyls}(b)) of the Janus cylinder. The scaling form remains the same as in the previous case, i.e.,
\begin{equation}
 F^{(cs)}_{\symbJcH}(\vartheta, D, R, T) = k_B T\,\frac{\mathcal L}{R^{d-1}}\,\frac{K^{(cs)}_{\symbJcH}(\vartheta, \Delta, \Theta)}{\Delta^{d-1/2}}.
 \label{eq:scalform_F_januscyl_hom}
\end{equation}

Comparing in Fig.~\ref{fig:sketch_cyls}(a) the sketch for the case of a homogeneous cylinder near a stepped substrate with the case of a Janus cylinder above a homogeneous substrate, one realizes that for a suitable orientation $\vartheta$ of the Janus cylinder the same pairings of BC between the substrate and the particle enter the DA. Projecting the equatorial plane of a Janus cylinder onto a homogeneous substrate yields a distance $X=X_J=R\,\cos\vartheta$ between the (left) edge of the projection and the projection of the cylinder axis (Fig.~\ref{fig:sketch_cyls}(b)). Conversely, the projection of the axis of a homogeneous cylinder onto a substrate with a chemical step renders a distance $X$ between them (Fig.~\ref{fig:sketch_cyls}(b)). Choosing $X=X_J=R\,\cos\vartheta$, within DA the sums of the surface elements vis-\`{a}-vis for these two configurations are the same and thus yield the same force. In terms of the present scaling function the relation $X=X_J$ translates into $\cos\vartheta=\Xi\sqrt{\Delta}$.
This implies that within DA the scaling function $K^{(cs)}_{\symbJcH}$ of the force between a Janus cylinder and a homogeneous substrate follows from Eqs.~(\ref{eq:DA_K_cyl_step}--\ref{eq:DA_deltaK_cyl_step}) upon substituting $X=R\cos\vartheta$ therein.
Figure~\ref{fig:Kcyl_step_Theta}(b) shows for a Janus cylinder next to a homogeneous wall as function of the scaled temperature $\Theta$ the full MFT results (solid lines) for various orientations $\vartheta$ (chosen independently from Fig.~\ref{fig:Kcyl_step_Theta}(a)). The corresponding DA scaling functions are shown as dashed lines. In Fig.~\ref{fig:Kcyl_step_Theta}(b), for the same distance $\Delta=1/5$, the DA scaling functions appear to deviate slightly more from the corresponding full MFT results than those in Fig.~\ref{fig:Kcyl_step_Theta}(a).

\begin{figure}[t!]
  \centering
  \includegraphics[width=0.47\textwidth]{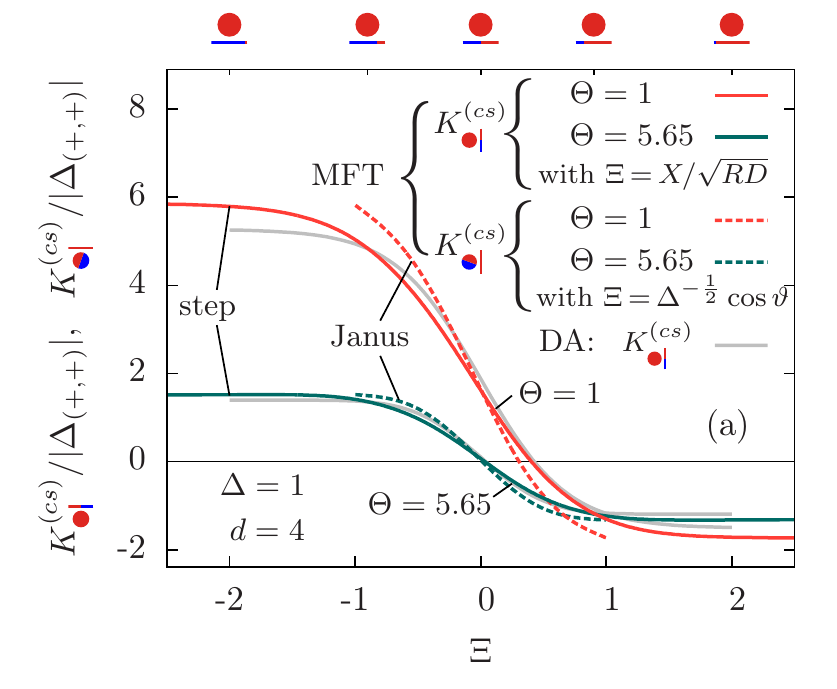}
  \hfill
  \includegraphics[width=0.47\textwidth]{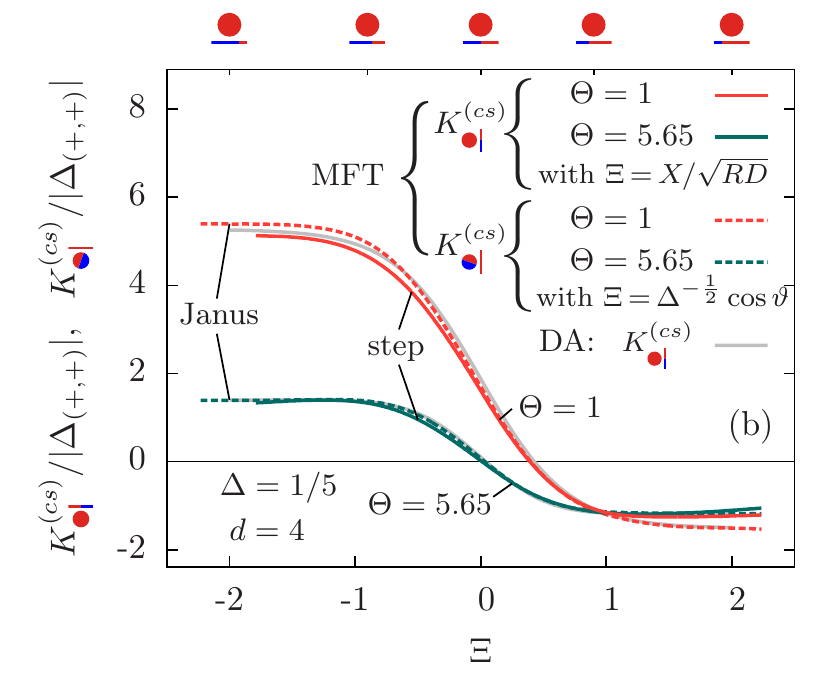}
  \caption{Comparison of the scaling functions of the force $K^{(cs)}_{\symbCS}$ between a homogeneous cylinder above a chemical step on the substrate (solid lines) and $K^{(cs)}_{\symbJcH}$ for a Janus cylinder above a homogeneous substrate (dashed lines). The DA (valid for $\Delta\ll 1$) implies the same scaling function in both cases (gray lines), provided the tilt angle $\vartheta$ of the Janus cylinder (see Fig.~\ref{fig:sketch_cyls}(b)) is related to the scaled step position on the substrate as $\Xi=\Delta^{-1/2}\cos\vartheta$.  The full mean field results for $K^{(cs)}_{\symbCS}$ (step) and $K^{(cs)}_{\symbJcH}$ (Janus) are shown for $\Delta=1$ in (a) and $\Delta=1/5$ in (b), each for the two scaled temperatures $\Theta=1$ (red) and $\Theta=5.65$ (green). From (a) it can be seen that within full MFT the correspondence between the case of a homogeneous cylinder above a chemical step on the substrate and a Janus particle above a homogeneous substrate does not hold in general. It holds roughly for $\Theta=5.65$ and further away from $T_c$, but not close to $T_c$ (such as for $\Theta=1$). However, for $\Delta=1/5$ in (b), i.e., close to the DA limit of $\Delta\ll 1$, the correspondence of the two scaling functions within DA carries over to the MFT results. As a guide to the eye, visualizations of the geometry corresponding to certain values of $\Xi$ are provided at the top of the panels.}
  \label{fig:Kcyl_step_to_janus}
  \vspace{1em}
\end{figure}

In order to asses quantitatively the difference between DA and full MFT, it is more suitable to compare the corresponding scaling functions $K^{(cs)}_{\symbCS}$ and $K^{(cs)}_{\symbJcH}$ of the force for fixed scaled temperature $\Theta$ as function of the scaling variable $\Xi=X/\sqrt{R D}$, which either corresponds to the lateral position $X$ of the axis of a homogeneous cylinder relative to a chemical step on the substrate, or to the orientation $\cos\vartheta=\Xi\sqrt{\Delta}$ of a Janus cylinder above a homogeneous substrate. Accordingly, for the two scaled temperatures $\Theta=1$ and $\Theta=5.65$ in Fig.~\ref{fig:Kcyl_step_to_janus} we show the full MFT scaling function $K^{(cs)}_{\symbCS}(\Xi, \Delta, \Theta)$ of the force for the homogeneous cylinder-step geometry [Eq.~\eqref{eq:scalform_F_cyl_step}] as solid lines and the full MFT scaling function $K^{(cs)}_{\symbJcH}(\vartheta, \Delta, \Theta)$ of a Janus cylinder next to a homogeneous substrate [Eq.~\eqref{eq:scalform_F_januscyl_hom}] as dashed lines. In the spirit of the aforementioned equivalence within DA, the orientation angle $\vartheta$ of the Janus cylinder is related to the distance $X$ between the projected axis of a homogeneous cylinder and the chemical step at the wall via the DA relation $\Xi=\Delta^{-1/2}\cos\vartheta$. For $\Delta=1$ in Fig.~\ref{fig:Kcyl_step_to_janus}(a), there is a visible difference between the two scaling functions. However, for $\Delta=1/5$ in Fig.~\ref{fig:Kcyl_step_to_janus}(b), which is closer to the DA limit $\Delta\ll 1$, the difference is considerably smaller. For comparison, in gray the scaling function of the force within DA is shown, which approximates both MFT scaling functions for $\Delta\ll 1$.

Thus it appears that the MFT results of both geometries approach each other in the limit of $\Delta\to 0$. This raises the question whether the relation between the two configurations, as implied by DA, reflects a more general foundation beyond DA.

\subsection{Comparison of forces in terms of order parameter profiles}
Contrary to the DA, the MFT minimization technique renders equilibrium order parameter profiles for each scaled temperature $\Theta$. Nonetheless, the DA implies a certain structure of the order parameter profile, even though in general it is ignorant concerning the profile.

\ifTwocolumn
\begin{figure*}[t!]
  \centering
  \includegraphics{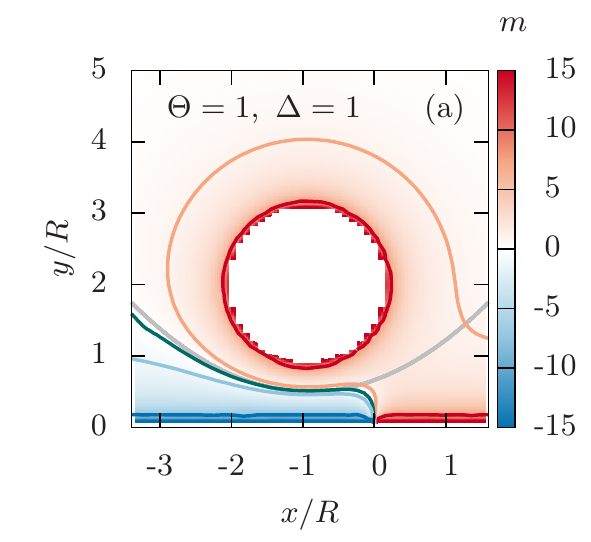}\hspace{1em}
  \includegraphics{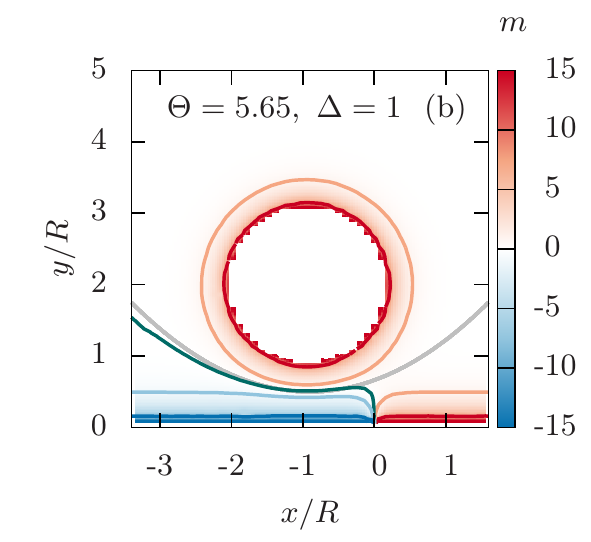}\\
  \includegraphics{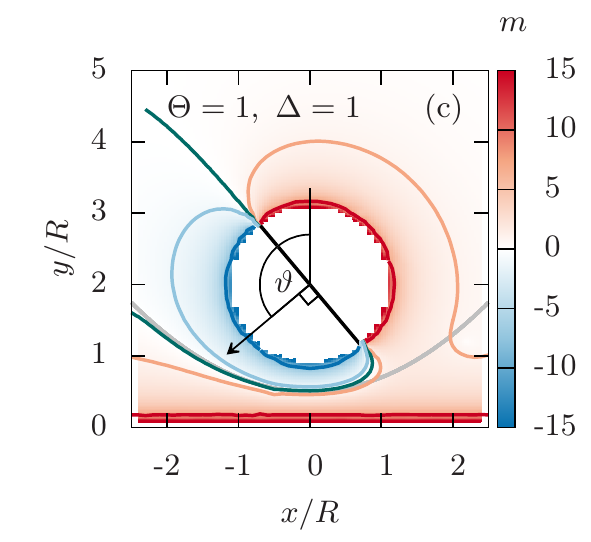}\hspace{1em}
  \includegraphics{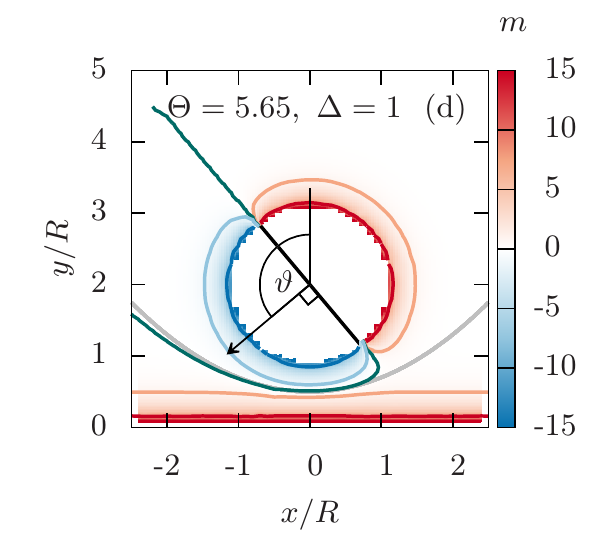}
  \caption{Reduced order parameter profiles $m$ as obtained from MFT in $d=4$ and in units of the amplitude $\phi_t$ of the bulk order parameter $\phi_b=\phi_t\,|t|^{1/2}$. The values of the order parameter are color coded, with red for positive values and blue for negative values, following the convention for the BC in Fig.~\ref{fig:sketch_cyls}. For $\Theta=1$ (a) depicts a homogeneous cylinder with $(+)$ BC at $X=-0.9\,R$ above a substrate with a chemical step between $(+)$ BC for $x>0$ and $(-)$ BC for $x < 0$. Panel (b) features the same geometry at $\Theta=5.65$, i.e., further away from $T_c$.
  For comparison, in (c) a Janus cylinder above a homogeneous substrate with $(+)$ BC is shown for $\Theta=1$ and in (d) for $\Theta=5.65$. The orientation of the Janus cylinder is taken as $\vartheta=130 ^\circ$, so that $\cos\vartheta=-0.64$.
  We have included certain isolines of the profile as a guide to the eye. The green line represents the zero crossing of the profiles, which has a special significance discussed in the main text. The gray curve indicates the zero crossing expected (at the same temperate) for the profile in the case that both the particle and the substrate are homogeneous, but with opposite BC.}
  \label{fig:profiles_janus_vs_step}
\end{figure*}
\fi

The reduced MFT order parameter profiles $m(\vec{r})$ for a homogeneous cylinder above a chemical step are depicted in Fig.~\ref{fig:profiles_janus_vs_step}(a) for $\Theta=1$ and in Fig.~\ref{fig:profiles_janus_vs_step}(b) for $\Theta=5.65$. In this example, the geometric parameters have been chosen such that $D=R$, i.e., $\Delta = 1$; the colloid with $(+)$ BC is positioned at $X=-0.9\,R$ on the left side of the step with opposite $(-)$ BC there, and the cylinder axis is normal to the cut plane of the order parameter profiles, which are invariant along the cylinder axis.
The profiles are taken for $\Theta > 0$ at the critical concentration, i.e., in the mixed phase, in which the order parameter differs from zero primarily only near the surfaces. Due to the opposing BC on the colloid and on the left half of the substrate surface, the profile must cross zero (green line), although this does not indicate the formation of an actual interface.
The gray line represents the zero crossing (at the same temperature) of the profile between a homogeneous particle and a homogeneous substrate, but with opposing BC. In the case of a chemical step on the substrate, the DA implicitly assumes that the order parameter profile follows that for a homogeneous substrate up to the lateral position $x=0$ of the step (Figs.~\ref{fig:profiles_janus_vs_step}(a) and (b)). Generally, Figs.~\ref{fig:profiles_janus_vs_step}(a) and (b) show that the actual zero crossing (green) follows closely the homogeneous case (gray), as assumed by the DA, up to a certain lateral position. However, the point of deviation between the green and the gray lines occurs at a lateral position which is to the left of the step position, because the actual zero crossing line (green) smoothly bends towards the step. The curvature of this bending depends on the temperature and broadens upon increasing the correlation length (i.e., decreasing $\Theta$).

\ifTwocolumn
\else
\begin{figure*}[t!]
  \centering
  \includegraphics{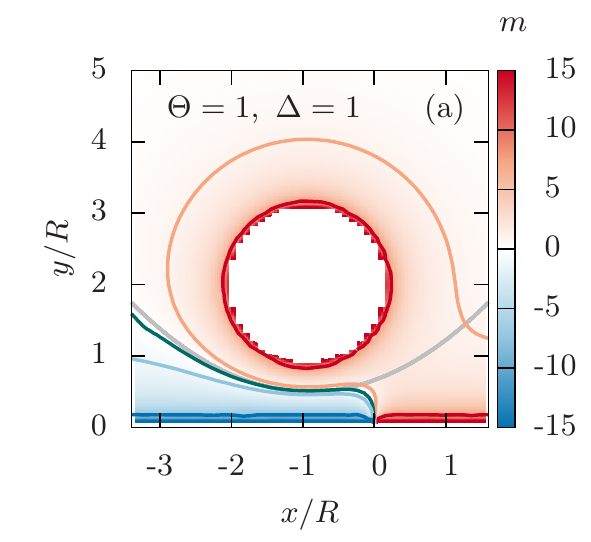}\hspace{1em}
  \includegraphics{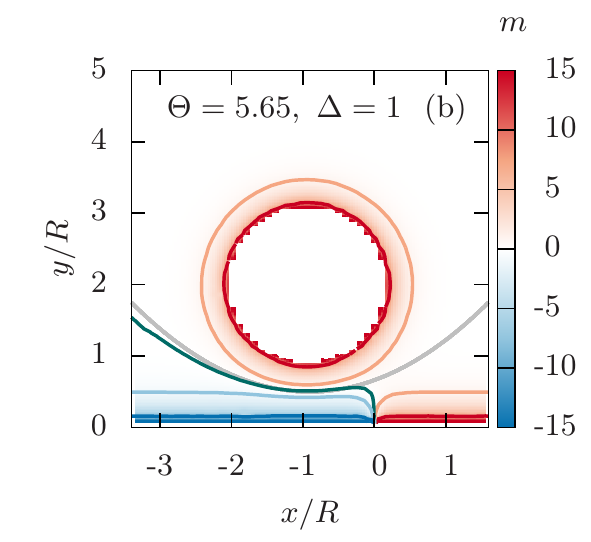}\\
  \includegraphics{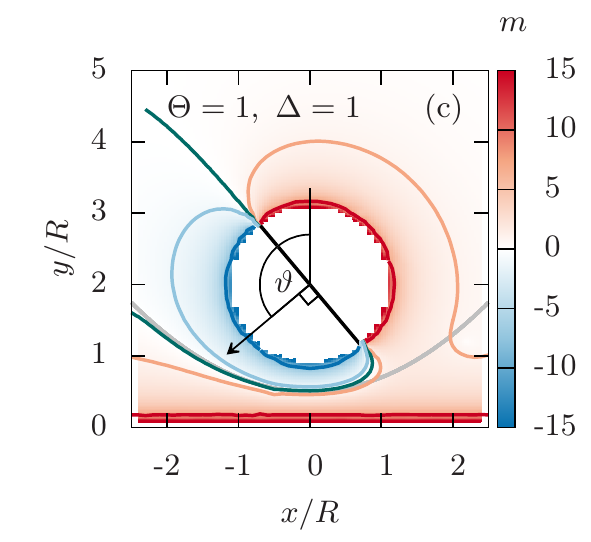}\hspace{1em}
  \includegraphics{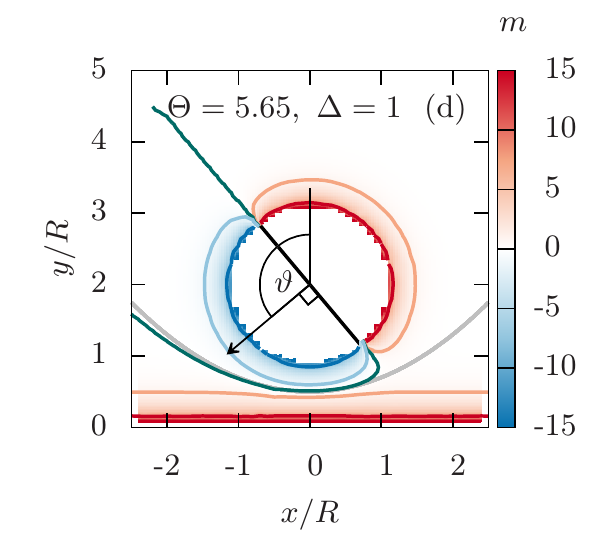}
  \caption{Reduced order parameter profiles $m$ as obtained from MFT in $d=4$ and in units of the amplitude $\phi_t$ of the bulk order parameter $\phi_b=\phi_t\,|t|^{1/2}$. The values of the order parameter are color coded, with red for positive values and blue for negative values, following the convention for the BC in Fig.~\ref{fig:sketch_cyls}. For $\Theta=1$ (a) depicts a homogeneous cylinder with $(+)$ BC at $X=-0.9\,R$ above a substrate with a chemical step between $(+)$ BC for $x>0$ and $(-)$ BC for $x < 0$. Panel (b) features the same geometry at $\Theta=5.65$, i.e., further away from $T_c$.
  For comparison, in (c) a Janus cylinder above a homogeneous substrate with $(+)$ BC is shown for $\Theta=1$ and in (d) for $\Theta=5.65$. The orientation of the Janus cylinder is taken as $\vartheta=130 ^\circ$, so that $\cos\vartheta=-0.64$.
  We have included certain isolines of the profile as a guide to the eye. The green line represents the zero crossing of the profiles, which has a special significance discussed in the main text. The gray curve indicates the zero crossing expected (at the same temperate) for the profile in the case that both the particle and the substrate are homogeneous, but with opposite BC.}
  \label{fig:profiles_janus_vs_step}
\end{figure*}
\fi

\begin{figure*}[ht!]
  \centering
  \includegraphics[width=0.56\textwidth]{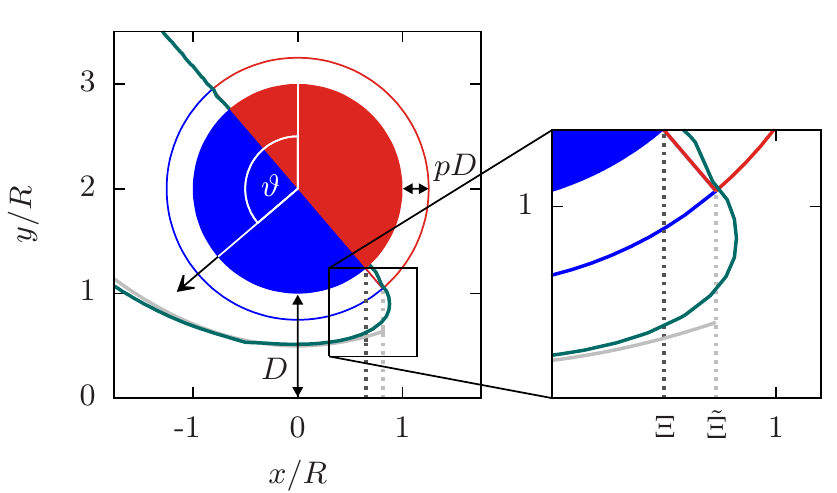}
  \caption{A generic sketch depicting the essential features of a Janus cylinder at distance $D$ above a homogeneous substrate, tilted by an angle $\vartheta$, akin to Figs.~\ref{fig:profiles_janus_vs_step}(c) and (d). An example for the actual zero crossing line of the order parameter profile, as found within full MFT, is shown in green. The zero crossing implied by the DA is shown in gray (solid light gray line, light and dark gray vertical dotted lines). In DA, the zero crossing is taken into account up to the scaled position $\Xi=\Delta^{-1/2}\cos\vartheta$ of the step in BC of the Janus particle, projected onto the substrate along the normal of the substrate (dark gray dotted line). The improved DA relation $\tilde\Xi(\vartheta)$ in Eq.~\eqref{eq:DAext_janus_step_rel} follows the same principle, but applied to a fictitious particle of increased radius $R + p D$, with the rescaling parameter $p$, resulting in the solid light gray zero crossing line and the light gray vertical dotted line. The inset provides a magnified view of the relevant features.}
  \label{fig:profile_sketch}
  \vspace{1em}
\end{figure*}

In Fig.~\ref{fig:profiles_janus_vs_step}(c) [(d)], the configuration of a Janus cylinder above a homogeneous substrate is shown in comparison to (a) [(b)], for the scaled temperature $\Theta=1$ [$\Theta=5.65$]. The orientation $\vartheta$ of the Janus cylinder has been chosen such that the configuration (a) [(b)] and the configuration (c) [(d)] yield forces within MFT which are approximately equal to each other. For both scaled temperatures, this was found to be the case for $\vartheta\approx 130^\circ$, which deviates significantly from the DA relation $\vartheta=\cos^{-1}(X/R) = 154^\circ$ for $X/R=-0.9$. Such a deviation is expected to occur away from the DA limit of $\Delta\ll 1$ [compare Figs.~\ref{fig:Kcyl_step_to_janus}(a) and (b)].
For the Janus particles, we find that the zero crossing of the profiles (green line) again follows the one for a homogeneous colloid (gray line), but now bending towards the Janus equator on the particle. A systematic analysis reveals that one always finds equal values of the force in MFT for the step on the surface and for the Janus particle whenever the bending and the extension of the zero crossing line are closely mirroring each other in the two geometries.
The reason for the equality of these forces within MFT goes right back to Eq.~\eqref{eq:Hamiltonian}. The Hamiltonian depends on the gradient of the order parameter profile, which relates to the bending of the zero line, but only via its square, which is independent of the direction of the bending. In Figs.~\ref{fig:profiles_janus_vs_step}(c) and (d) there is also an upper green zero crossing line, which is absent in (a) and (b). This line contributes only little to the force because it is relatively straight and because in that region the order parameter is small.

Based on the knowledge of the full MFT order parameter profiles, we construct a phenomenological relation beyond the DA relation of $\Xi=\Delta^{-1/2}\cos\vartheta$, which seeks to incorporate the bending of the zero crossing line. The base of this idea follows from Ref.~\cite{Labbe:2014}, where a similar principle was used successfully in order to reconcile DA with MFT results.

In Fig.~\ref{fig:profile_sketch}, we sketch the essential features of a Janus cylinder of radius $R$, close to a homogeneous wall at distance $D$; the actual zero crossing line of the order parameter profile is shown in green (which is taken from Fig.~\ref{fig:profiles_janus_vs_step}(c), but here serves to represent a generic case), and the zero crossing implied by DA is shown as a solid light gray line.
The dotted, vertical dark gray line indicates the original DA relation, which cuts off the solid gray zero crossing line (of the homogeneous system with opposing BC at the colloid and substrate surface) at the projected position of the Janus equator. The visual agreement of the zero crossing lines can be improved by considering the DA for a fictitious scaled colloid (the blue and red semi-rings), with an effective radius of $\tilde R = R + p D$ and an effective surface-to-surface distance $\tilde D = (1 - p) D$, so that the zero crossing line follows the solid light gray line. This yields an improved scaled position (dotted, vertical light gray line)
\begin{equation}
\tilde \Xi(\vartheta) = \tilde\Delta^{-1/2} \cos(\vartheta)= \sqrt{\frac{1}{1-p}}\cdot\sqrt{\frac{1}{\Delta}+p}\,\cos(\vartheta),
\label{eq:DAext_janus_step_rel}
\end{equation}
where $p$ is a free parameter which describes the rescaling of the particle size.

Independently, we have calculated the scaling functions of the force within full MFT as function of the position $X$ of a homogeneous cylinder relative to a stepped substrate and for the orientation $\vartheta$ for the Janus cylinder, at fixed scaled temperatures $\Theta$ and distances $\Delta$. Via linear interpolation within the two MFT scaling functions, we have extracted those values of $X$ and $\vartheta$ for which both scaling functions of the force render the same value, which in turn renders a relation between the numeric values of $\vartheta$ and $X$. The proposed model $\tilde\Xi(\vartheta)$ in Eq.~\eqref{eq:DAext_janus_step_rel} can be checked against this discrete set $\{\Xi, \vartheta\}$.
We note that the projected, scaled step position $\tilde\Xi$ is proportional to $\tilde\Delta^{-1/2}>\Delta^{-1/2}$ for $p > 0$, i.e., for the same orientation $\vartheta$, the scaled step position $\tilde\Xi$ is larger than $\Xi$. However, for values of $\Xi\gg 1$, the scaling function of the force saturates (see Fig.~\ref{fig:Kcyl_step_to_janus}) and relating $\Xi$ and $\vartheta$ numerically via the force within MFT becomes rather error-prone. This discredits fitting assumptions beyond linear order. However, the relation in Eq.~\eqref{eq:DAext_janus_step_rel}, linearized around $\vartheta\approx\frac{\pi}{2}$ by using $\cos(\vartheta)\approx\frac{\pi}{2}-\vartheta$, results in a reasonable fit for $p\approx 1/4$. Within fitting errors, the fit parameter $p$ does not depend noticeably on the scaled temperature $\Theta$ and the scaling variable $\Delta$. The value of the rescaling parameter $p=1/4$ is in line with the presentation in Fig.~\ref{fig:profile_sketch}, as it places the surface of the fictitious colloid halfway between the physical particle and the zero crossing line.

\begin{figure}[t!]
  \centering
  \includegraphics[width=0.47\textwidth]{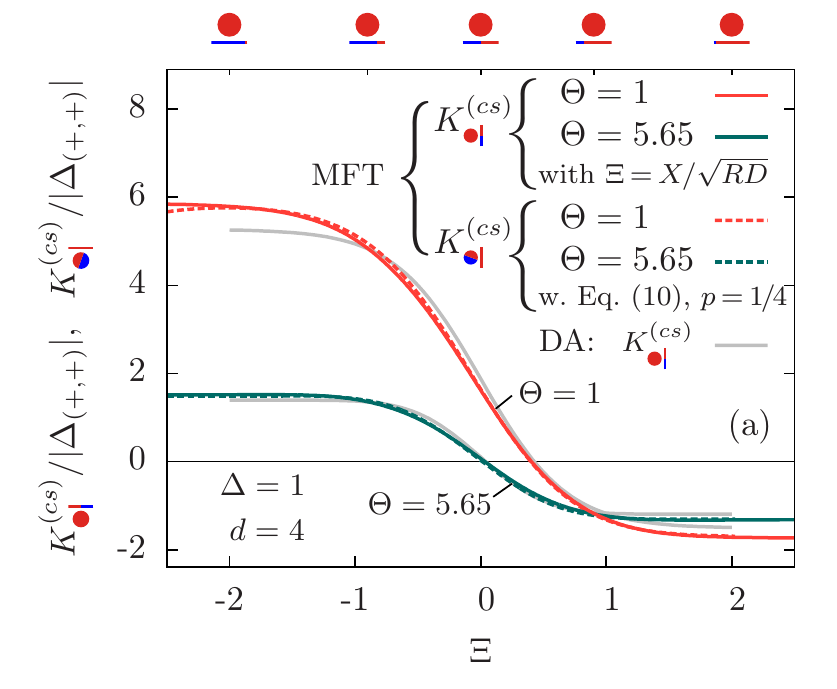}
  \hfill
  \includegraphics[width=0.47\textwidth]{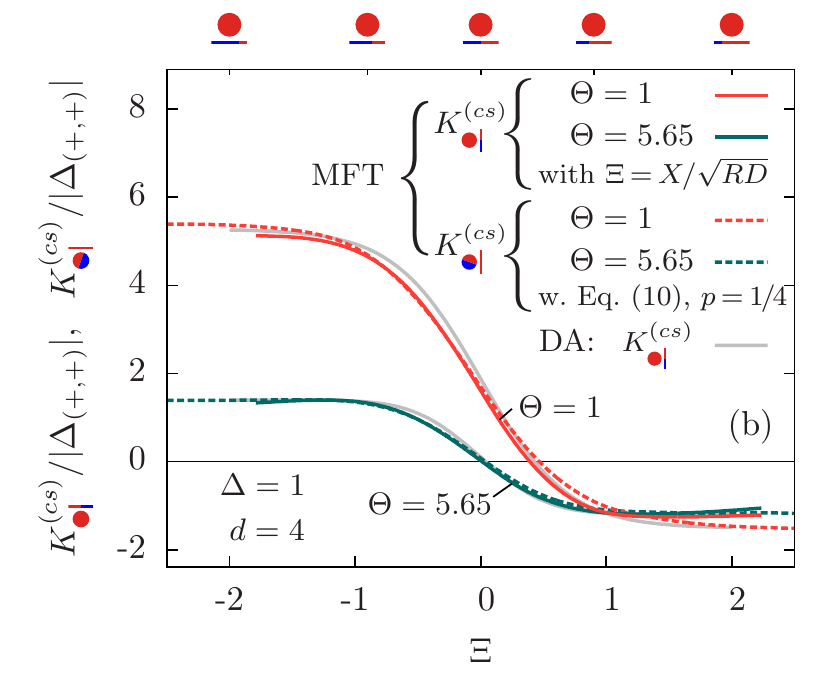}
  \caption{Same as Fig.~\ref{fig:Kcyl_step_to_janus} but replacing $\Xi$ by $\tilde{\Xi}=\tilde{\Delta}^{-1/2}\cos\vartheta$ (Eq.~\eqref{eq:DAext_janus_step_rel} with $p=1/4$) for $K^{(cs)}_{\symbJcH}$. In this case, the correspondence between the scaling functions of the two configurations \symbCS\enskip and \symbJcH\enskip holds within full MFT, for values of $\Delta$ outside the DA limit $\Delta\ll 1$.}
  \label{fig:Kcyl_step_to_janus_mapped}
  \ifTwocolumn\vspace{-2em}\fi
\end{figure}

For comparison, Fig.~\ref{fig:Kcyl_step_to_janus_mapped} demonstrates the improved performance of the phenomenological relation $\tilde{\Xi}=\tilde{\Delta}^{-1/2}\cos\vartheta$ in Eq.~\eqref{eq:DAext_janus_step_rel} with $p=1/4$ compared with that of the approach used in Fig.~\ref{fig:Kcyl_step_to_janus}, even for $\Delta=1$.


As a final remark, we emphasize that, in the above approach, within DA we counted the force to be normal to the substrate. An approach alternative to the DA considers the forces to be normal to the surface of the particle \cite{Dantchev:2012}, which, however, leads to the same formal expressions for the critical Casimir forces. The improved DA relation in Eq.~\eqref{eq:DAext_janus_step_rel} can be interpreted as a partial consideration of forces directed normal to the particle surface, with $p$ being a weighting factor for the two force directions (see Fig.~\ref{fig:profile_sketch}).

\subsection{Cylindrical Janus particle above a chemical step}
Here we analyze fully the case depicted in Fig.~\ref{fig:sketch_cyls}(b) of a single cylindrical Janus particle floating above a chemical step on the substrate. The cylindrical particle is taken to be oriented horizontally and all chemical steps are parallel to each other. 
Within DA, the configuration of a Janus particle above a step relates to the case of two walls each endowed with a chemical step, shifted with respect to each other \cite{Sprenger:2006}, but accounting for distinct distance relations between the surface elements appearing in DA. Since the presence of two chemical steps can have a profound effect on the order parameter profile, one has to check whether this spoils the usefulness of the relation introduced in Eq.~\eqref{eq:DAext_janus_step_rel}.

Within DA and for special configurations, the scaling function of the force $K^{(cs)}_{\symbJcS}(\vartheta, \Xi, \Delta\ll 1, \Theta)$ between a chemical step on the substrate and a Janus particle with orientation $\vartheta$ and its center shifted by $\Xi=X/\sqrt{R D}$ from the substrate step, attains certain limiting expressions. For an upright orientation $\vartheta = 0$ it has the same value as the scaling function $K^{(cs)}_{\symbCS}(\Xi, \Delta\ll 1, \Theta)$ of the force between a homogeneous cylinder and a stepped substrate. If the Janus cylinder is positioned far away from the step, i.e., $\Xi \gg 1$, $K^{(cs)}_{\symbJcS}$ reduces to the scaling function of a Janus cylinder above a homogeneous substrate, so that $K^{(cs)}_{\symbJcS}(\vartheta, \Xi\to\infty, \Delta\ll 1, \Theta) = K^{(cs)}_{\symbJcH}(\vartheta,\Delta\ll 1, \Theta) = K^{(cs)}_{\symbCS}(\Xi(\vartheta), \Delta\ll 1, \Theta)$ (where $\Xi(\vartheta)=\Delta^{-1/2}\cos\vartheta$ or is given by Eq.~\eqref{eq:DAext_janus_step_rel}; analogously for $\Xi\to -\infty$).

Thus, similar to $K^{(cs)}_{\symbCS}$ in Eqs.~\eqref{eq:DA_K_cyl_step} and \eqref{eq:DA_K_cyl_hom}, the scaling function $K^{(cs)}_{\symbJcS}$ can be decomposed as $K^{(cs)}_{\symbJcS}(\vartheta,\Xi, \Delta, \Theta) = K^{(cs)}_{(+,\pm)}\mp\Delta K^{(cs)}_{\symbJcS}(\vartheta,|\Xi|,\Delta,\Theta)$, where
$K^{(cs)}_{(+,\pm)}$ again refers to the scaling function of the force between a homogeneous cylinder and a homogeneous substrate (the rules when to use the upper and lower signs depend on $\vartheta$ and $\Xi$; see below):
\begin{equation}
 K^{(cs)}_{(+,\pm)}(\Delta\ll 1,\Theta)=\sqrt{2}\int_1^{1+\Delta^{-1}/2}\upd\alpha\ \frac{k_{(+,\pm)}(\alpha\Theta)}{\alpha^d\sqrt{\alpha-1}}.
\label{eq:DA_K_cyl_hom_finite}
\end{equation}
However, here the rhs of Eq.~\eqref{eq:DA_K_cyl_hom_finite} carries a finite upper limit of integration, i.e., without explicitly setting $\Delta\to 0$. But the expression is still valid only in the DA limit $\Delta\ll 1$. The dependence on nonzero values of $\Delta$ ensures consistency with the scaling function of the excess force $\Delta K^{(cs)}_{\symbJcS}(\vartheta, |\Xi|, \Delta\ll 1, \Theta)$. The latter depends on the position of the Janus cylinder relative to the substrate step (again only via the scaled absolute value $|\Xi|$ of the distance) and on the orientation $\vartheta\in[-\pi,\pi)$. The sign of the position $\Xi$ and the sign of the orientation $\vartheta$ can be chosen according to different conventions. Here, the coordinates are chosen such that $\vartheta>0$ rotates the normal of the equatorial plane of the Janus particle towards that side of the substrate which has the same BC, i.e., here, the rotation is counter-clockwise towards the side $\Xi<0$ (see Fig.~\ref{fig:sketch_cyls}(b)).
We note that the force is invariant under reflection at the plane normal to the substrate and containing the cylinder axis ($\vartheta\to -\vartheta$, $\Xi\to -\Xi$ and exchange of BC on the substrate), i.e., $K^{(cs)}_{\symbJcS}=K^{(cs)}_{\raisebox{0.5em}{\scalebox{1}[-1]{\symbJcS}}}$.
Utilizing this symmetry, the decomposition reads
\ifTwocolumn\begin{widetext}\fi
\begin{equation}
K^{(cs)}_{\symbJcS}(\vartheta,\Xi, \Delta, \Theta) =\begin{cases}
K^{(cs)}_{(\textcolor{red}{+},\textcolor{red}{+})}(\Delta,\Theta)-\Delta K^{(cs)}_{\symbJcS}(\vartheta,|\Xi|,\Delta,\Theta) & \text{for }\Xi(\vartheta)\,\Xi> 0,\\
K^{(cs)}_{(\textcolor{red}{+},\textcolor{blue}{-})}(\Delta,\Theta)+\Delta K^{(cs)}_{\symbJcS}(-\vartheta,|\Xi|,\Delta,\Theta) & \text{for }\Xi(\vartheta)\,\Xi\leq 0.
\end{cases}
\label{eq:DA_K_Jcyl_step}
\end{equation}
\ifTwocolumn\end{widetext}\fi
(Note that, as indicated, in Eq.~\eqref{eq:DA_K_Jcyl_step}, only in the first factor of the conditions, $\Xi$ is replaced by $\Xi(\vartheta)=\Delta^{-1/2}\cos\vartheta$ or, alternatively, by Eq.~\eqref{eq:DAext_janus_step_rel}.) The condition $\Xi(\vartheta)\,\Xi\gtrless 0$ considers in which direction the Janus cylinder is tilting (e.g., $\Xi(\vartheta)\propto \cos\vartheta > 0 \Rightarrow$ upwards) and over which side of the step it levitates (via $\Xi$). 
Additionally, the equivalences $k_{(+,+)}=k_{(-,-)}$ and $k_{(+,-)}=k_{(-,+)}$ of the interaction between homogeneous, planar, and parallel walls lead to an invariance of the scaling function $K^{(cs)}_{\symbJcS}$ upon inverting the normal of the particle, i.e., $\vartheta\to\vartheta\pm\pi$ (such that $\vartheta\in[-\pi,\pi)$) and exchanging the BC of the substrate step (but without changing the position $\Xi$), so that $K^{(cs)}_{\symbJcS}=K^{(cs)}_{\symbJcSm}$.

The excess scaling function $\Delta K^{(cs)}_{\symbJcS}$ is obtained from the careful DA summation of the corresponding surface elements:
\vspace{0.2em}
\begin{multline}
 \Delta K^{(cs)}_{\symbJcS}(\vartheta,\Xi,\Delta\ll1,\Theta)=
 \left\{\begin{array}{lr}
 \ifTwocolumn
  	+1, &\text{if } |\Xi(\vartheta)|<|\Xi|  \\ &  \text{or } \vartheta < 0, \\
 \else
 	+1, &\text{if } |\Xi(\vartheta)|<|\Xi| \text{ or } \vartheta < 0, \\
 \fi
 	-1, & \text{otherwise}\end{array}\right\}\\
 \times
 \Bigg(\frac{1}{\sqrt{2}}\int_{1+\Xi(\vartheta)^2/2}^{1+\Delta^{-1}/2}\upd\alpha\ \frac{\Delta k(\alpha\Theta)}{\alpha^d\sqrt{\alpha-1}}
 \ifTwocolumn \\ \fi
  -\frac{\mathrm{sign}(\vartheta)}{\sqrt{2}}\int_{1+\Xi^2/2}^{1+\Delta^{-1}/2}\upd\alpha\ \frac{\Delta k(\alpha\Theta)}{\alpha^d\sqrt{\alpha-1}}\Bigg),
\label{eq:DA_deltaK_Jcyl_step}
\end{multline}
which has the structure of the difference between two expressions, resembling the scaling function corresponding to the chemical step on the substrate as in Eq.~\eqref{eq:DA_deltaK_cyl_step}. The intricate prefactor effectively exchanges $\Xi(\vartheta)\leftrightarrow\Xi$ if $|\Xi(\vartheta)| \geq |\Xi|$, which affects the sign only if $\vartheta\geq 0$.
Note that $\Delta K^{(cs)}_{\symbJcS}$ depends on $\vartheta$ only via the sign and via $|\Xi(\vartheta)|\propto |\cos\vartheta|$. One can verify that both the symmetry operations of reflection ($\vartheta\to-\vartheta$) as well as inversion ($\vartheta\to\vartheta\pm\pi$ such that $\vartheta\in[-\pi,\pi)$) yield the same result for the excess scaling function, i.e., that $\Delta K^{(cs)}_{\symbJcS}(-\vartheta,\ldots)=\Delta K^{(cs)}_{\symbJcS}(\vartheta\pm\pi,\ldots)$. Note that neither reflecting the position $\Xi\to-\Xi$ nor exchanging the BC affects $\Delta K^{(cs)}_{\symbJcS}$, but only $K^{(cs)}_{\symbJcS}$.

\ifTwocolumn
\begin{figure}[b!]
  \centering\vspace{-1.5em}
  \includegraphics[width=0.47\textwidth]{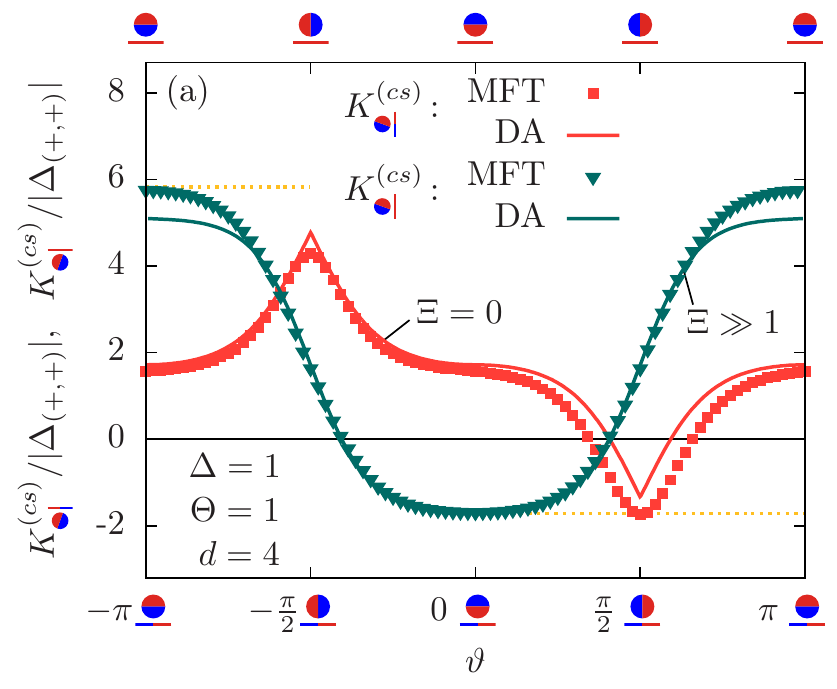}\hfill
  \includegraphics[width=0.47\textwidth]{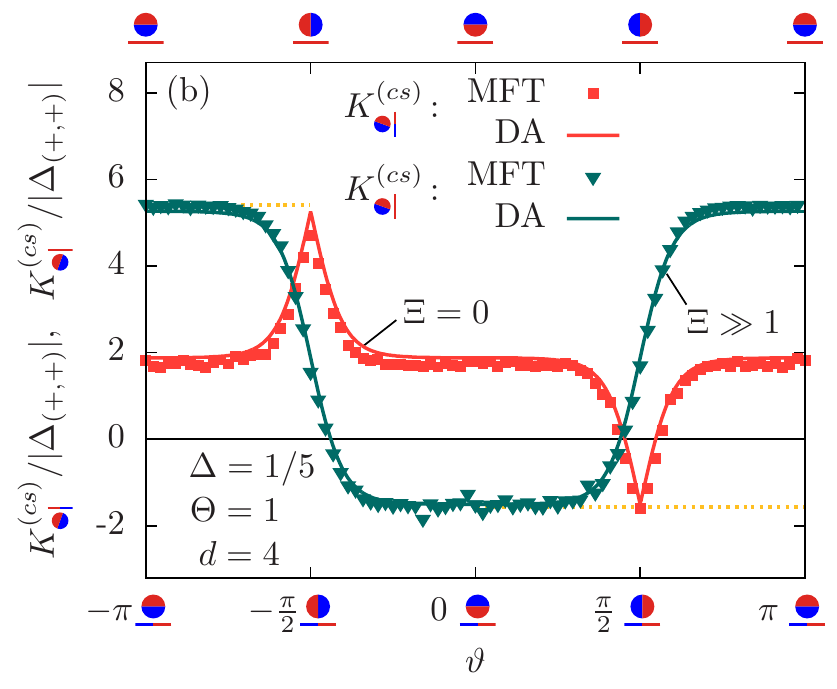}
  \caption{(a) Scaling function $K_{\symbJcS}^{(cs)}$ of the force between a Janus cylinder and a step on the substrate (in red), as a function of the particle orientation $\vartheta$ for $\Xi = 0$. The solid lines represent the results within DA, whereas the squares correspond to numerical MFT results for a separation $\Delta=D/R = 1$. The green lines and triangles represent the scaling function $K_{\symbJcH}^{(cs)}$ of the force which corresponds to the case of a homogeneous substrate, or equally, to the case of a step that is far away from the particle, i.e., $|\Xi|\gg 1$ [see Eq. \eqref{eq:DA_K_cyl_step}]. (b) The same, but for $\Delta=1/5$. Both in (a) and (b), the MFT values of the scaling functions $K^{(cs)}_{(+,+)}$ and $K^{(cs)}_{(+,-)}$ [Eq.~\eqref{eq:DA_K_cyl_hom_finite}] for the fully attractive ($<0$) and repulsive ($>0$) cases, respectively, of a homogeneous cylinder and homogeneous substrate are indicated by dotted golden lines. At the top of the panels, we indicate configurations with the Janus cylinder above a homogeneous substrate corresponding to certain points of the green curve for $\Xi\gg 1$. Similarly, at the bottom of the panels, configurations are shown with the Janus particle directly above the step corresponding to the red curve, i.e., $\Xi=0$.}
  \label{fig:cyl_da_mft_comparison}
\end{figure}
\fi

In Fig.~\ref{fig:cyl_da_mft_comparison}, we compare the DA with the full MFT results for the scaling function $K^{(cs)}_{\symbJcS}$ for two separations $\Delta=1$ in (a) and $\Delta=1/5$ in (b), with the step on the substrate fixed at $\Xi = 0$ (red sets of squares and lines). Within DA, this represents a peculiar configuration in that the orientations $\vartheta = \pm\pi/2$ of the Janus particle correspond to configurations in which both the step on the particle and the one on the substrate share a common vertical plane (see the sketches below the horizontal axis). At $\vartheta=-\pi/2$, due to opposing BC between all DA surface elements, the force (red lines) is repulsive ($>0$). For $\Delta=1$, around $\vartheta=-\pi/2$ the DA result slightly overestimates the MFT result. Similarly, the special orientation at $\vartheta = \pi/2$ leads to an attractive force ($<0$); here, however, and for $\Delta=1$, DA clearly underestimates the MFT results. The cusplike shape of the scaling function around the maximum and minimum is an artifact of the DA; MFT renders a smooth and broader curve. In general, the MFT results are slightly more attractive and less repulsive than predicted by DA. Nonetheless, for $\Delta\ll 1$ [Fig.~\ref{fig:cyl_da_mft_comparison}(b)] DA and MFT agree rather well, even at $\vartheta=\pm\pi/2$. This is reassuring because for these orientations the shortcomings of the DA are particularly pronounced. As implied by the DA and in view of its reliability, the overall shape of the scaling function $K^{(cs)}_{\symbJcS}(\vartheta, \Xi=0, \Delta\to 0, \Theta)$, within MFT and as a function of $\vartheta$, is consistent with the dependence of the scaling function of the force between two patterned, planar substrates on a lateral shift (see Ref.~\cite{Sprenger:2006}).

We point out that the DA curves shown in Fig.~\ref{fig:cyl_da_mft_comparison} are based on the improved relation given by Eq.~\eqref{eq:DAext_janus_step_rel}. For the original DA relation $\Xi(\vartheta) = \Delta^{-1/2}\cos\vartheta$, the agreement between DA and MFT turns out to be poorer in Fig.~\ref{fig:cyl_da_mft_comparison}(a), i.e., for $\Delta = 1$, but remains comparable to the good agreement evident in Fig.~\ref{fig:cyl_da_mft_comparison}(b), i.e., for $\Delta = 1/5$ (see also Fig.~\ref{fig:Kcyl_step_to_janus}).
We find that the explicit dependence on $\Delta$ introduced by Eqs.~\eqref{eq:DAext_janus_step_rel} and \eqref{eq:DA_K_cyl_hom_finite} does not improve the agreement between DA and MFT for the strongly attractive or repulsive configurations: in Fig.~\ref{fig:cyl_da_mft_comparison}(a) see the difference between the green line and the green symbols as well as the dotted golden lines which refer to MFT results for $K^{(cs)}_{(+,+)}<0$ and $K^{(cs)}_{(+,-)}>0$.
However, the dependence on $\Delta$ of the MFT scaling functions for the case of a homogeneous cylinder and substrate has a different cause \cite{Labbe:2014}. Within DA, a dependence on $\Delta$ has been introduced via the DA relation $\Xi(\vartheta)=\Delta^{-1/2}\cos\vartheta$ or via Eq.~\eqref{eq:DAext_janus_step_rel} along with the dependence on $\vartheta$. Thus, the good agreement between the slopes of the DA and MFT scaling functions shown in Fig.~\ref{fig:cyl_da_mft_comparison} as a function of $\vartheta$ for different $\Delta$ indicates the consistency of these relations beyond the DA limit.

\ifTwocolumn
\else
\begin{figure}[t!]
  \centering\vspace{-1.5em}
  \includegraphics[width=0.47\textwidth]{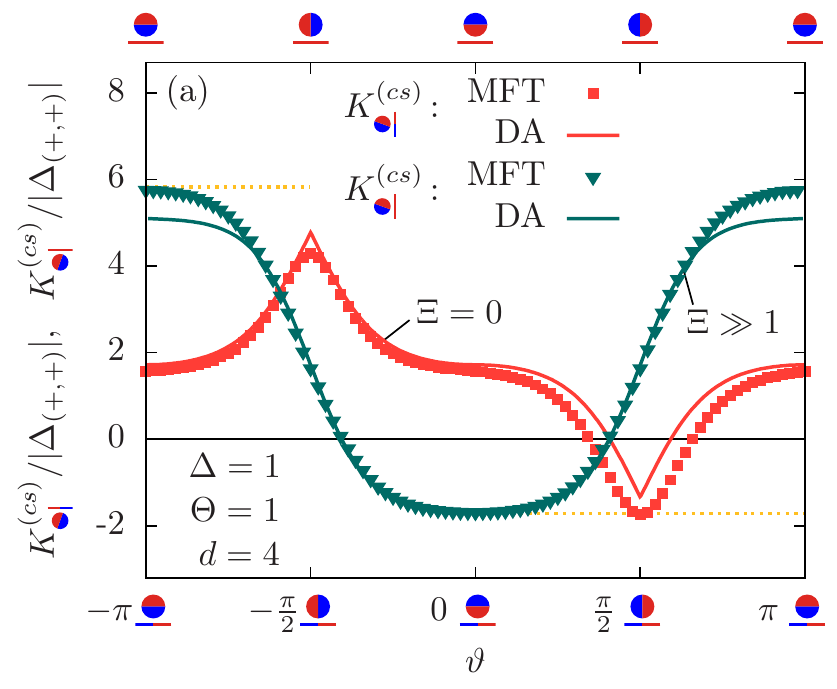}\hfill
  \includegraphics[width=0.47\textwidth]{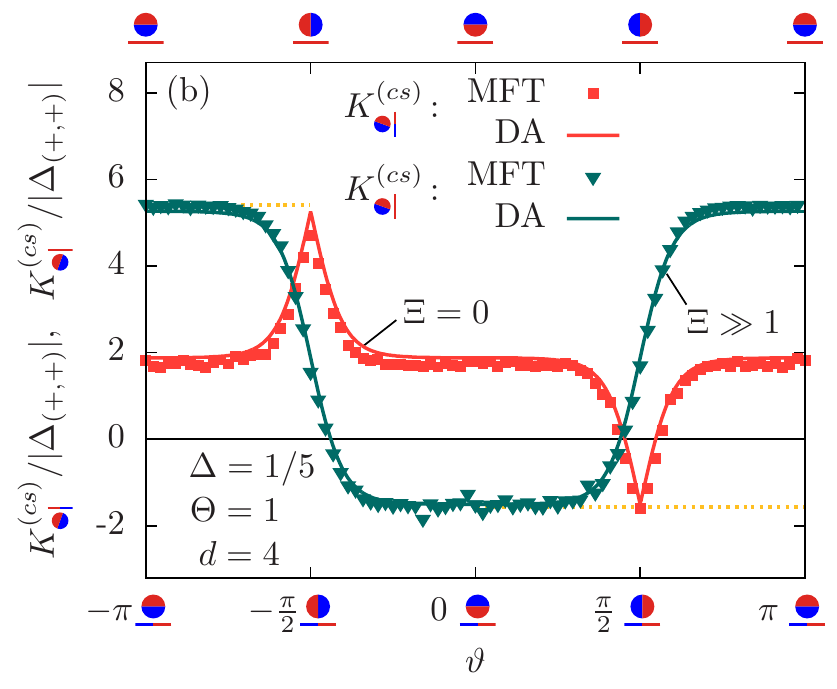}
  \caption{(a) Scaling function $K_{\symbJcS}^{(cs)}$ of the force between a Janus cylinder and a step on the substrate (in red), as a function of the particle orientation $\vartheta$ for $\Xi = 0$. The solid lines represent the results within DA, whereas the squares correspond to numerical MFT results for a separation $\Delta=D/R = 1$. The green lines and triangles represent the scaling function $K_{\symbJcH}^{(cs)}$ of the force which corresponds to the case of a homogeneous substrate, or equally, to the case of a step that is far away from the particle, i.e., $|\Xi|\gg 1$ [see Eq. \eqref{eq:DA_K_cyl_step}]. (b) The same, but for $\Delta=1/5$. Both in (a) and (b), the MFT values of the scaling functions $K^{(cs)}_{(+,+)}$ and $K^{(cs)}_{(+,-)}$ [Eq.~\eqref{eq:DA_K_cyl_hom_finite}] for the fully attractive ($<0$) and repulsive ($>0$) cases, respectively, of a homogeneous cylinder and homogeneous substrate are indicated by dotted golden lines. At the top of the panels, we indicate configurations with the Janus cylinder above a homogeneous substrate corresponding to certain points of the green curve for $\Xi\gg 1$. Similarly, at the bottom of the panels, configurations are shown with the Janus particle directly above the step corresponding to the red curve, i.e., $\Xi=0$.}
  \label{fig:cyl_da_mft_comparison}
\end{figure}
\fi

From these findings we conclude that the DA, although for $\Delta \gtrsim 1$ it deviates quantitatively from the MFT results in $d=4$, exhibits no basic flaws. In fact, studying the implication of the use of the DA in this section has revealed that the parameters $\Xi$ and $\vartheta$, associated with the positions of the chemical steps on the substrate and on the Janus cylinder, are related according to $\Xi(\vartheta)=\Delta^{-1/2}\cos\vartheta$. The modified scaling variable $\tilde\Xi(\vartheta)$ (Eq.~\eqref{eq:DAext_janus_step_rel}) improves quantitatively the agreement with the full MFT results.
We consider these properties as a justification to study below two Janus particles based on DA only.
\section{Two Janus cylinders}
\label{sec:janus_cyls}
Reassured by the result that DA can be used reliably for describing the force acting on a single Janus particle near a substrate, in this section we determine the force and the effective potential between two Janus cylinders within DA and without substrate. For reasons of simplicity, we assume the long axes of the two cylinders to be parallel to each other, i.e., the positions and rotations of the cylinders are confined to a plane. This amounts to consider effectively discs in a two-dimensional system but with interactions corresponding to an embedding solvent in $d=3$ or $d=4$. In view of the experimental interest in such Janus particles, in the following figures we depict the scaling function in $d=3$. This is accomplished by taking the wall-wall scaling functions $k_{(a,b)}$, which are needed as input for the DA, from Ref. \cite{Vasilyev:2009}, i.e., from numerical simulations in $d=3$.

\begin{figure}[t!]
  \centering
  \hspace{-2.75cm}(a)\hspace{3.5cm}(b)\\
  \includegraphics[width=6cm]{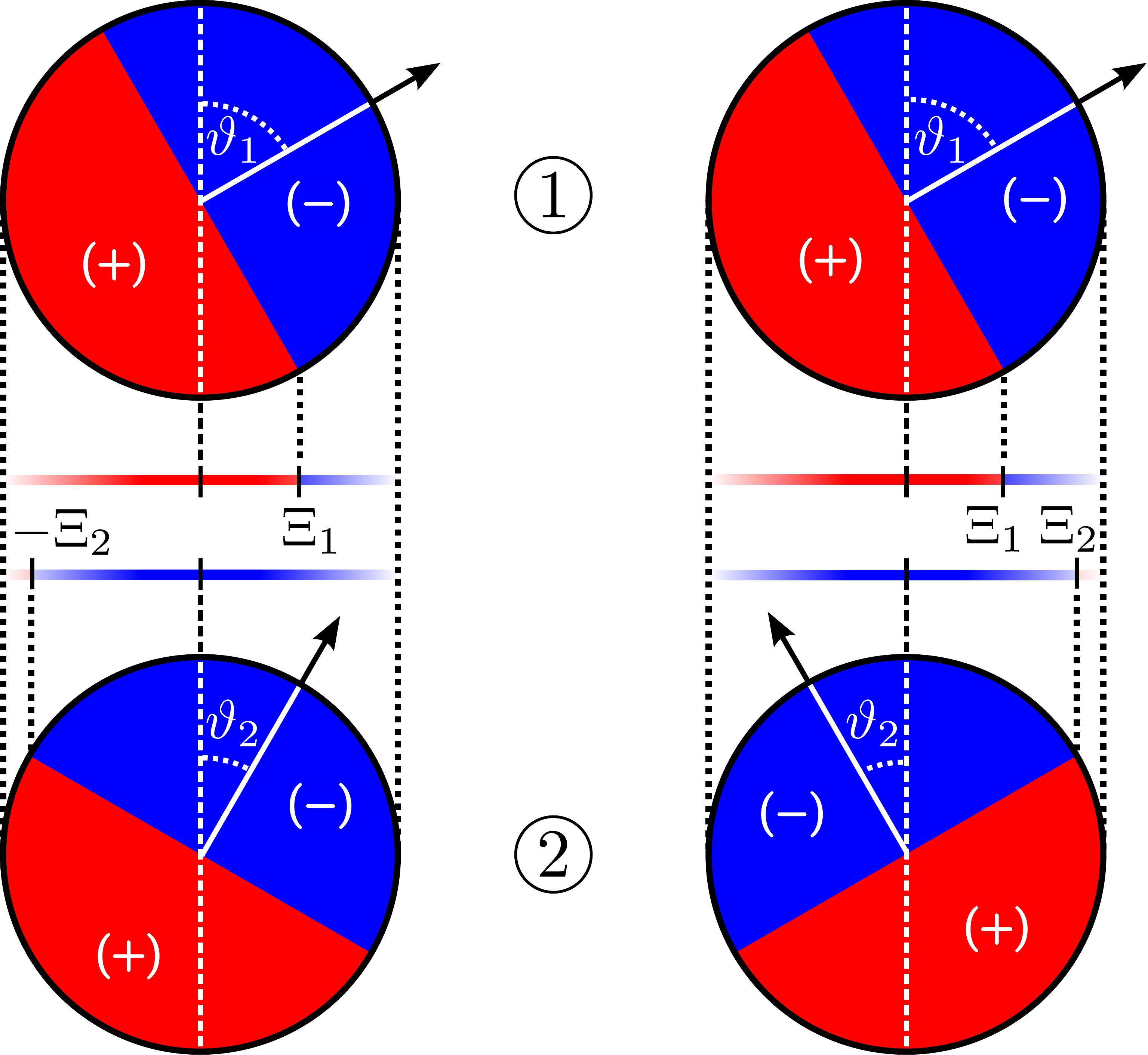}\ifTwocolumn\vspace{1em}\fi
  \caption{Sketch of the geometry for the Derjaguin approximation concerning the force between two Janus cylinders \numcirc{1} and \numcirc{2} for $\vartheta_2 > 0$ in (a) and $\vartheta_2 < 0$ in (b). The cylinder axes are supposed to extend out of the plane of view. The angles $\vartheta_1$ and $\vartheta_2$ of the orientation are relative to the axis connecting the centers of the two particles. All orientations can be mapped onto the principal domain $\vartheta_{1,2}\to\hat{\vartheta}_{1,2}\in[-\pi/2,\pi/2)$. The middle parts show the unrolled surfaces of the Janus cylinders opposing each other. The construction of the DA for two Janus cylinders is akin to the interaction between two structured substrates interacting \cite{Sprenger:2006, Parisen:2010}, considering, however, only that portion of the chemical structure which ranges from $-R$ to $+R$, i.e., from $-\Delta^{-1/2}$ to $+\Delta^{-1/2}$ in terms of the scaling variable, and using the appropriate local surface-to-surface distance. In its straightforward version, the DA projects the Janus equators to step positions at $\Xi_{1,2}\equiv\Xi(\vartheta_{1,2})=\Delta^{-1/2}\cos(\vartheta_{1,2})$. Additionally, depending on $\mathrm{sign}(\vartheta_1\,\vartheta_2)$, either the left or the right edge of the equatorial plane enters into the projection, leading to opposite step positions $\pm\Xi_{1,2}$.\ifTwocolumn\vspace{1em}\fi}
  \label{fig:sketch_cyl_da}
\end{figure}

According to renormalization group theory the singular part of the force between two Janus cylinders consists of prefactors which produce the unit of a force $(k_B T /R)$, the reduced length $(\mathcal L / R^{d-2})$ and a dimensionless scaling function $K^{(cc)}_{\symbJcJc}/\Delta^{d-1/2}$, as in the case of two homogeneous cylinders \cite{Law:2015}:
\begin{equation}
  F^{(cc)}_{\symbJcJc}(\vartheta_1, \vartheta_2, D, R, T) = k_B T\,\frac{\mathcal L}{R^{d-1}}\,\frac{K^{(cc)}_{\symbJcJc}(\vartheta_1, \vartheta_2, \Delta, \Theta)}{\Delta^{d-1/2}};
  \label{eq:cyl_force_scalform}
\end{equation}
$K^{(cc)}_{\symbJcJc}$ is the universal scaling function of the force between two Janus cylinders with equal radius $R$, $\vartheta_1$ and $\vartheta_2$ are the orientations of the two Janus particles, and $\Delta = D/R$ and $\Theta = \pm D/\xi_\pm(T)$ are dimensionless scaling variables for the surface-to-surface distance and the temperature, respectively.
This scaling form is of the same type as the one in Eq.~\eqref{eq:scalform_F_cyl_step} for a homogeneous cylindrical particle above a stepped substrate and to the one in Eq.~\eqref{eq:scalform_F_januscyl_hom} for a Janus cylinder above a homogeneous substrate; however, the corresponding scaling functions are distinct. Within DA, for certain configurations $K^{(cc)}_{\symbJcJc}$ and, e.g., $K^{(cs)}_{\symbJcS}$ can attain the same values as the corresponding scaling functions for homogeneous particles. Within the corresponding expressions (see Eq.~\eqref{eq:DA_K_Jcyl_step}), $K_{(+,\pm)}^{(cs)}$ for a  homogeneous cylindrical particle and a homogeneous substrate is stronger by a factor of $\sqrt{2}$ compared with $K_{(+,\pm)}^{(cc)}$ for two homogeneous cylinders.

\subsection{Derjaguin approximation}


Within DA, the force $F^{(cc)}_{\symbJcJc}$ between two Janus cylinders orientated top-to-bottom [$(\vartheta_1, \vartheta_2)=(0,0)$ and $(\pm\pi,\pm\pi)$], bottom-to-bottom [$(0,\pm\pi)$], or top-to-top [$(\pm\pi,0)$], is identical to the force between two homogeneous cylinders $F^{(cc)}_{(a, b)}$ \cite{Law:2015} with $(a,b)$ as the BC of the sides facing each other (compare Fig.~\ref{fig:sketch_cyl_da}). Upon construction this follows from the fact that for these configurations, the equatorial planes are orthogonal to the axis connecting the centers of the particles.
Analogously as in Eq.~\eqref{eq:DA_K_Jcyl_step}, we express the force between two Janus cylinders $F^{(cc)}_{\symbJcJc}$ relative to the force $F^{(cc)}_{(+, \pm)}$ between two homogeneous cylinders, yielding
\ifTwocolumn\begin{widetext}\fi
\begin{equation}
F^{(cc)}_{\symbJcJc}(\vartheta_1, \vartheta_2, D, R, T) = \begin{cases}
F^{(cc)}_{(\textcolor{red}{+},\textcolor{blue}{-})}(\Delta,\Theta)+\Delta F^{(cc)}_{\oslash\oslash}(\hat{\vartheta}_1, \hat{\vartheta}_2,\Delta,\Theta), & \text{for }\Xi(\vartheta_1)\,\Xi(\vartheta_2)> 0,\\
F^{(cc)}_{(\textcolor{red}{+},\textcolor{red}{+})}(\Delta,\Theta)-\Delta F^{(cc)}_{\oslash\oslash}(\hat{\vartheta}_1,\hat{\vartheta}_2,\Delta,\Theta), & \text{for }\Xi(\vartheta_1)\,\Xi(\vartheta_2) < 0
\end{cases}
\label{eq:cyl_force_separation}
\end{equation}
with $\Xi(\vartheta_i)=\Delta^{-1/2}\cos\vartheta_i$ and where, without loss of generality, we introduced the reduced angles $\hat{\vartheta}_{1,2}=\vartheta_{1,2}\mp\pi$ such that $\hat{\vartheta}_{1,2}\in[-\pi/2,\pi/2)$. Note that a shift of $\pm\pi$ amounts to reflecting the normals $\vec{n}_1$ and $\vec{n}_2$ at the corresponding equatorial plane of particle \numcirc{1} and \numcirc{2}, respectively. The subscript of  the excess scaling function $\Delta F^{(cc)}_{\oslash\oslash}(\hat{\vartheta}_1, \hat{\vartheta}_2,\Delta,\Theta)$ is not colored in order to emphasize that only the reduced angles enter.
As in the previous case, the form given by Eq.~\eqref{eq:cyl_force_separation} is manifestly invariant against that reflection while also exchanging the BCs of the particles.
For instance, for any configuration with $\vartheta_{1,2}\in[-\pi/2,\pi/2)$, i.e., $\Xi(\vartheta_1)\,\Xi(\vartheta_2) > 0$, one has
\begin{equation}
F^{(cc)}_{\symbJcJc}\left(\vartheta_1, \vartheta_2, D, R, T\right) = F^{(cc)}_{\includegraphics[height=1.2em, angle=90]{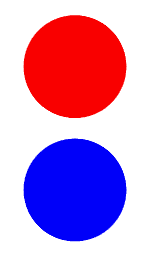}}(D, R, T) + \Delta F^{(cc)}_{\oslash\oslash}(\hat\vartheta_1=\vartheta_1, \hat\vartheta_2=\vartheta_2, D, R, T);
\end{equation}
if instead $\vartheta_2$ is reflected to $\vartheta'_2=\vartheta_2+\pi$, one has $\Xi(\vartheta'_2)=\Delta^{-1/2}\cos\vartheta'_2 < 0$, leading to
\begin{equation}
F^{(cc)}_{\includegraphics[height=1.2em, angle=90]{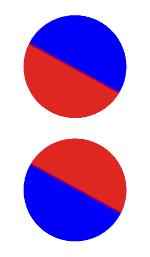}}\left(\vartheta_1, \vartheta'_2, D, R, T\right) = F^{(cc)}_{\includegraphics[height=1.2em, angle=90]{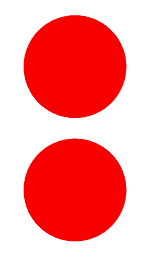}}(D, R, T) - \Delta F^{(cc)}_{\oslash\oslash}(\hat\vartheta_1=\vartheta_1, \hat{\vartheta_2}=\vartheta'_2-\pi=\vartheta_2, D, R, T),
\end{equation}
which corresponds to exchanging the BC.
\ifTwocolumn\end{widetext}\fi
The subscript of the forces $F^{(cc)}_{\includegraphics[height=1.2em, angle=90]{viz_cyls_a_b}} = F^{(cc)}_{(+,-)}$ and $F^{(cc)}_{\includegraphics[height=1.2em, angle=90]{viz_cyls_b_b}} = F^{(cc)}_{(+,+)}$ between homogeneous particles have been colored in order to visualize the BC.
We additionally enforce $\cos(\hat{\vartheta}_1)\leq\cos(\hat{\vartheta}_2)$. The expressions for $F^{(cc)}_{\symbJcJc}$ for all other configurations can be reduced to those in Eq.~\eqref{eq:cyl_force_separation} by exchanging the BC $(a,b)$, by appropriately setting $\hat{\vartheta}_i=\vartheta_i\mp \pi$ with $i=\{1,2\}$, and by switching the labeling $\hat{\vartheta}_1\leftrightarrow\hat{\vartheta}_2$.
In this sense, in the following we drop the hat of $\hat{\vartheta}$ in favor of a lighter notation.
Note that $\Xi(\vartheta_1)\,\Xi(\vartheta_2)=0$ is exempted from the cases considered in Eq.~\eqref{eq:cyl_force_separation}; in this limiting case the conditions should be read as abbreviations for $\lim\limits_{x\to\vartheta_1^+}\lim\limits_{y\to\vartheta_2^+}\Xi(x)\,\Xi(y)\gtrless 0$, i.e., the right-sided limit from above \footnote{The one-sided limit is required not because of any discontinuities in the scaling function, but for consistency with the discontinuous definition $\hat{\vartheta}_{1,2}\in[-\pi/2,\pi/2)$. The two-sided limit does exist and the scaling function is continuous, see Appendix \ref{sec:app_janus_cyls}.
Concerning the numerical evaluation, $\hat\vartheta_{1,2}$ can be equally defined as the remainder of the division of $x=\vartheta_{1,2}+\pi/2$ (dividend) by $y=\pi$ (divisor) minus $\pi/2$. The remainder is obtained from the floating point version of the modulo operation, provided in most programming languages. However, care needs to be taken as the floating point modulus is ambiguous for negative numbers. The right-sided limit is compatible only with the definition by ``floored division'' $x\ \mathrm{mod}\ y = x - y\big\lfloor\frac{x}{y}\big\rfloor$, where the integer quotient is given by the floor function $\big\lfloor\frac{x}{y}\big\rfloor$, ensuring that $0\leq x\ \mathrm{mod}\ \pi < \pi\ \forall x$; see D. Knuth, \textit{The Art of Computer Programming}, Vol. 1 (Addison-Wesley, 1997) p. 39}.

Dividing up the force as in Eq.~\eqref{eq:cyl_force_separation} leads, in conjunction with Eq.~\eqref{eq:cyl_force_scalform}, to an analogous separation of the scaling function of the critical Casimir force between two Janus cylinders. To this end, we introduce a new scaling function
\begin{equation}
  K^{(cc)}_{\symbJcJc}(\vartheta_1, \vartheta_2, \Delta, \Theta)=K_{(+,\mp)}^{(cc)}(\Delta, \Theta) \pm \Delta K_{\oslash\oslash}^{(cc)}(\vartheta_1, \vartheta_2, \Delta, \Theta),
  \label{eq:cyl_k_separation}
\end{equation}
where the signs adhere to Eq.~\eqref{eq:cyl_force_separation} and (compare Eq.~\eqref{eq:DA_K_cyl_hom_finite})
\begin{equation}
 K_{(+,\mp)}^{(cc)}(\Delta, \Theta) = \int_1^{1+\Delta^{-1}}\upd\alpha\,\frac{k_{(+,\mp)}(\alpha \Theta)}{\alpha^{d}\sqrt{\alpha-1}}
 \label{eq:cyl_K_hom}
\end{equation}
is the scaling function of the critical Casimir force between two homogeneous cylinders with $(+,\mp)$ BC \cite{Law:2015}. For homogeneous particles, the limit $\Delta\to 0$ , in which DA holds, can be carried out explicitly, so that in Eq.~\eqref{eq:cyl_K_hom} the upper limit of integration reaches infinity. On the other hand, $\Delta K_{\oslash\oslash}^{(cc)}$ depends on $\Delta$ via $\Xi(\vartheta_{1,2}) = \Delta^{1/2}\cos\vartheta_{1,2}$ within DA. 
In order for the separation in Eq. \eqref{eq:cyl_k_separation} to be consistent, both scaling functions $ K_{(+,\mp)}^{(cc)}$ and $\Delta K_{\oslash\oslash}^{(cc)}$ need to retain their dependence on $\Delta$. Nonetheless, the scaling functions within DA are expected to hold only for small but nonzero $\Delta$; keeping the dependence on $\Delta$ is not necessarily a refinement (see Sec.~\ref{sec:janus_vs_steps}).

The scaling function $\Delta K_{\oslash\oslash}^{(cc)}$ is constructed from the sum of surface elements as sketched in Fig.~\ref{fig:sketch_cyl_da}. This is similar to the case of two opposing structured substrates \cite{Sprenger:2006, Parisen:2010}, but with the appropriately varying distance between the surface elements. Thus, we introduce the chemical step-like (i.e., dependence on $\Xi$) force scaling function (compare Eqs.~\eqref{eq:DA_deltaK_cyl_step} and \eqref{eq:DA_deltaK_Jcyl_step})
\begin{equation}
	\Delta k^{(cc)}(\Xi,\Delta, \Theta) = \frac{1}{2} \int_{1+\Xi^2}^{1+\Delta^{-1}}\upd\alpha\,\frac{\Delta k(\alpha \Theta)}{\alpha^{d}\sqrt{\alpha-1}}
	\label{eq:cyl_deltaSmallK}
\end{equation}
with the scaling variable $\Xi$ determined by the projected lateral step position of the Janus equator. For simplicity, we use the DA projection $\Xi(\vartheta_{1,2})=\Delta^{-1/2}\cos(\vartheta_{1,2})$, instead of the improved relation discussed in Sec.~\ref{sec:janus_vs_steps}. The complete scaling function of the force is found to be given by
\begin{multeqline}
	\Delta K_{\oslash\oslash}^{(cc)}(\vartheta_1, \vartheta_2, \Delta, \Theta) = \Delta k^{(cc)}(|\Xi(\vartheta_1)|,\Delta,\Theta) \ifTwocolumn \\ \fi
	+ \mathrm{sign}(\vartheta_1\,\vartheta_2)\,\Delta k^{(cc)}(|\Xi(\vartheta_2)|,\Delta,\Theta).
\label{eq:cyl_deltaK}
\end{multeqline}
We point out the similarity between Eqs.~\eqref{eq:DA_deltaK_Jcyl_step} and \eqref{eq:cyl_deltaK}. However, in comparison, the sign-prefactor in Eq.~\eqref{eq:DA_deltaK_Jcyl_step} is superseded by the imposed restriction $|\cos\vartheta_1|\leq|\cos\vartheta_2|$. Moreover, the factor $-\mathrm{sign}(\vartheta_1)$ is replaced by $\mathrm{sign}(\vartheta_1\,\vartheta_2)$; a configuration $\vartheta_1>0$ and $\vartheta_2>0$ results in a projected step-step configuration with opposite signs for the step positions $\Xi(\vartheta_1)$ and $\Xi(\vartheta_2)$ (see Fig.~\ref{fig:sketch_cyl_da}(a)), thus, compared to  Eq.~\eqref{eq:DA_deltaK_Jcyl_step}, changing the sign of the term. This concise representation of $\Delta K_{\oslash\oslash}^{(cc)}$ in terms of the sign function is possible only for the reduced domain $\hat{\vartheta}_{1,2}\in[-\pi/2,\pi/2)$.
Special configurations of the two Janus cylinders, for which the scaling functions of the force require explicit considerations, are analyzed analytically in Appendix \ref{sec:app_janus_cyls}.

\begin{figure}[t!]
  \centering
  \includegraphics[width=0.47\textwidth]{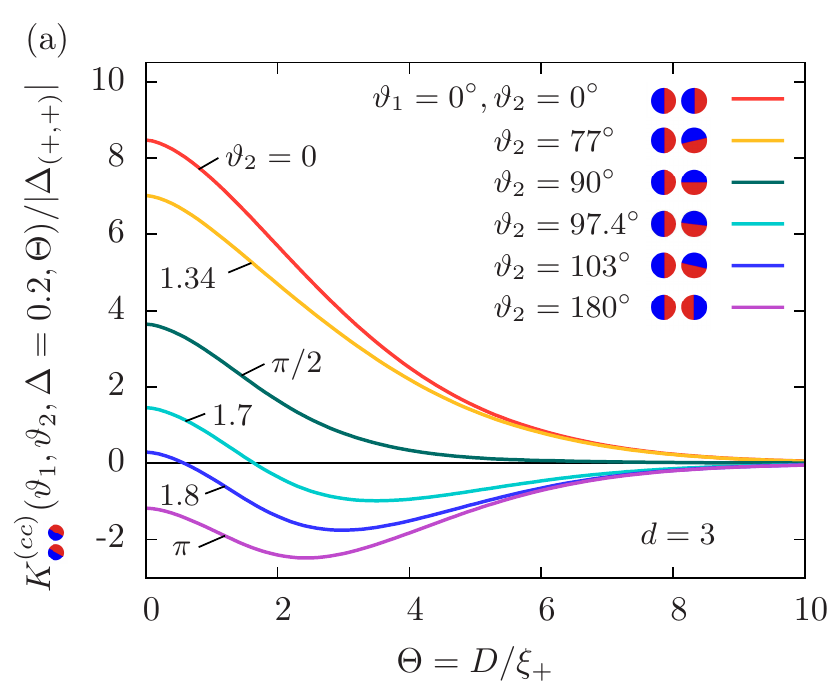}\hfill
  \includegraphics[width=0.47\textwidth]{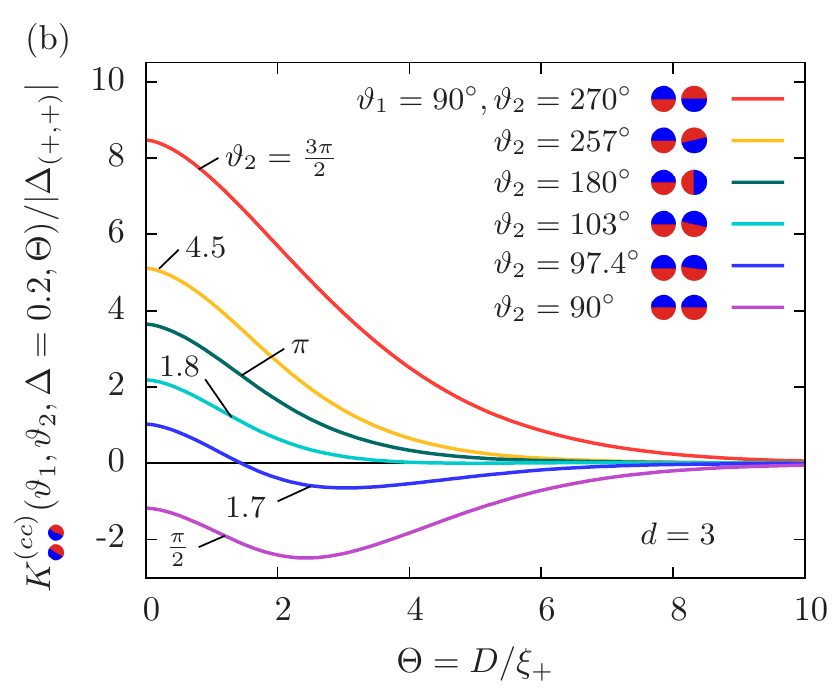}
  \caption{The scaling function $K_{\symbJcJc}^{(cc)}$ of the force between two Janus cylinders, within DA in $d=3$, as a function of scaled temperature $\Theta=D/\xi_+$ for various orientations. The scaling function is normalized by the absolute value of the universal critical Casimir amplitude $\Delta_{(+,+)}$. The wall-wall scaling function $k_{(a,b)}(L/\xi_\pm)$, which, inter alia, determines $\Delta_{(+,+)}$, is taken from MC results for the film geometry in Ref.~\cite{Vasilyev:2009}. (a) Configurations with $\vartheta_1 = 0$ for the orientation of particle \numcirc{1} for various orientation angles $\vartheta_2$ of particle \numcirc{2}, as visualized in the legend. (b) The case of $\vartheta_1 = \pi/2$ for various orientations $\vartheta_2$ of particle \numcirc{2}. The scaling function for the configuration $\vartheta_1=\vartheta_2=0$ matches the (repulsive) one of two homogeneous cylindrical particles with opposing BC, whereas the scaling function for the configuration $\vartheta_1=0, \vartheta_2=\pi$ equals the (attractive) one between two homogeneous cylinders with the same BC. The orientation angles belonging to the other curves have been chosen in order to visualize the particular sensitivity and insensitivity of the scaling function around the top-to-bottom, top-to-top, and bottom-to-bottom configurations, respectively. The angle $\vartheta_i$ is the one between the axis connecting the centers of the particles and the normal of the equatorial plane of particle $i$ (see Fig.~\ref{fig:sketch_cyl_da}).}
  \label{fig:cyl_K_da}
  \vspace{-0.5em}
\end{figure}

In Fig.~\ref{fig:cyl_K_da} we show the scaling function of the force between two Janus cylinders in $d=3$ as a function of the scaled temperature $\Theta=D/\xi_+ > 0$ (i.e., for $t>0$), for several orientations of the two Janus cylinders. The configuration $\vartheta_1=0$ and $\vartheta_2=0$ in Fig.~\ref{fig:cyl_K_da}(a) corresponds to the scaling function between two homogeneous cylinders with opposing BC which is repulsive for all temperatures $t>0$ (see Appendix \ref{sec:app_janus_cyls}). Variations of the orientations out of this configuration lead to only small changes of the force between the cylinders. Evan a significant rotation of $\vartheta_2=77^\circ$ results only in a small change in the scaling function. Around the perpendicular orientation $\vartheta_2=90^\circ$, the force is much more sensitive to small tilts. For $\vartheta_1=0$ and orientations of the second Janus cylinder close to $\vartheta_2=180^\circ$, the force is attractive for all temperatures $t>0$ (see again Appendix \ref{sec:app_janus_cyls}).
If, however, particle $1$ is rotated by $90^\circ$ relative to the axis connecting the centers, and $\vartheta_2=270^\circ$ (see Fig.~\ref{fig:cyl_K_da}(b)), the scaling function of the force is more sensitive to changes in the orientation. The configuration $\vartheta_1=90^\circ, \vartheta_2=180^\circ$ is geometrically equivalent to the one with $\vartheta_1=0^\circ, \vartheta_2=90^\circ$; in the former case the force is less sensitive to rotations of the second particle with $\vartheta_2=180^\circ$.


\subsection{Scaling function of the effective potential}
Concerning thermodynamic properties, the effective pair potential between particles is of even more direct importance than the force because it is experimentally more easily accessible.
The effective interaction potential $V_{\symbJcJc}^{(cc)}$ between two parallel Janus cylinders can be determined from the force $F$ according to
\ifTwocolumn
\begin{multline}
V_{\symbJcJc}^{(cc)}(\vartheta_1, \vartheta_2, D,R,T) = \int_D^\infty\upd z\,F_{\symbJcJc}^{(cc)}(\vartheta_1, \vartheta_2, z,R,T)\\
=k_B T \frac{\mathcal L}{R^{d-1}}\int_D^\infty\upd z\,\frac{K_{\symbJcJc}(\vartheta_1, \vartheta_2, z/R, z/\xi_\pm)}{(z/R)^{d-1/2}},
\end{multline}
\else
\begin{align}
V_{\symbJcJc}^{(cc)}(\vartheta_1, \vartheta_2, D,R,T) &= \int_D^\infty\upd z\,F_{\symbJcJc}^{(cc)}(\vartheta_1, \vartheta_2, z,R,T)\\
&=k_B T \frac{\mathcal L}{R^{d-1}}\int_D^\infty\upd z\,\frac{K_{\symbJcJc}(\vartheta_1, \vartheta_2, z/R, z/\xi_\pm)}{(z/R)^{d-1/2}}, \nonumber
\end{align}
\fi
with $z$ and $D$ as surface-to-surface distances. This can be cast into the scaling form
\begin{equation}
V_{\symbJcJc}^{(cc)}(\vartheta_1, \vartheta_2, D, R, T) = k_B T\, \frac{\mathcal{L}}{R^{d-2}}\, \frac{\Phi_{\symbJcJc}^{(cc)}(\Theta, \Delta, \vartheta_1, \vartheta_2)}{\Delta^{d-3/2}}
\end{equation}
where the scaling function $\Phi_{\symbJcJc}^{(cc)}$ of the potential follows the same partition as the force in Eq.~\eqref{eq:cyl_force_separation} so that
\ifTwocolumn\begin{widetext}\fi
\begin{equation}
\Phi_{\symbJcJc}^{(cc)}(\vartheta_1, \vartheta_2, \Delta, \Theta) =\begin{cases} \Phi_{(+,-)}^{(cc)}(\Delta, \Theta) + \Delta \Phi_{\oslash\oslash}^{(cc)}(\hat{\vartheta}_1, \hat{\vartheta}_2, \Delta, \Theta), & \text{for }\Xi(\vartheta_1)\,\Xi(\vartheta_2)> 0,\\
\Phi_{(+,+)}^{(cc)}(\Delta, \Theta) - \Delta \Phi_{\oslash\oslash}^{(cc)}(\hat{\vartheta}_1, \hat{\vartheta}_2, \Delta, \Theta), & \text{for }\Xi(\vartheta_1)\,\Xi(\vartheta_2)< 0\,
\end{cases}
\label{eq:cyl_theta_separation}
\end{equation}
with $\Xi(\vartheta_i)=\Delta^{-1/2}\cos\vartheta_i$ and where
\begin{align}
\Phi_{(+,\pm)}^{(cc)}(\Delta, \Theta) =& 2\int_1^\infty\upd\beta\,\sqrt{\beta - 1}\, \beta^{-d}\,k_{(+,\pm)}(\beta \Theta) \nonumber\\
&-2\int_{1+\Delta^{-1}}^\infty \upd \beta\,\left(\sqrt{\beta - 1} - \Delta^{-1/2}\right)\,\beta^{-d}\, k_{(+,\pm)}(\beta\Theta)
\end{align}
is the scaling function of the effective interaction potential between two homogeneous cylinders \cite{Law:2015}.
\ifTwocolumn\end{widetext}\fi
The Janus-induced excess scaling function $\Delta \Phi_{\oslash\oslash}^{(cc)}$ in the reduced domain of $\hat{\vartheta}_{1,2}$ (see the previous subsection) follows from integrating Eqs.~\eqref{eq:cyl_deltaSmallK} and \eqref{eq:cyl_deltaK}. This keeps the general structure of the force scaling function, leading to
\ifTwocolumn
\begin{multline}
	\Delta \Phi_{\oslash\oslash}^{(cc)}(\vartheta_1, \vartheta_2, \Delta, \Theta) = \Delta \Phi_s^{(cc)}(|\Xi_1|,\Delta, \Theta) \\
	+ \mathrm{sign}(\vartheta_1\,\vartheta_2)\,\Delta \Phi_s^{(cc)}(|\Xi_2|,\Delta, \Theta),
	\label{eq:cyl_deltaTheta}
\end{multline}
\else
\begin{equation}
	\Delta \Phi_{\oslash\oslash}^{(cc)}(\vartheta_1, \vartheta_2, \Delta, \Theta) = \Delta \Phi_s^{(cc)}(|\Xi_1|,\Delta, \Theta)
	+ \mathrm{sign}(\vartheta_1\,\vartheta_2)\,\Delta \Phi_s^{(cc)}(|\Xi_2|,\Delta, \Theta),
	\label{eq:cyl_deltaTheta}
\end{equation}
\fi
with $\Xi_i=\Xi(\vartheta_i)$ and with the chemical step-like scaling function of the potential (compare Eq.~\eqref{eq:cyl_deltaSmallK}):
\ifTwocolumn
\begin{multline}
	\Delta \Phi_s^{(cc)}(\Xi,\Delta,\Theta) =\hspace{-0.3em} \int_{1+\Xi^2}^{\infty}\hspace{-0.5em}\upd\beta\,\left(\sqrt{\beta - 1} - \Xi\right)\,\beta^{-d}\, \Delta k(\beta\Theta)\\
-\int_{1+\Delta^{-1}}^\infty \upd \beta\,\left(\sqrt{\beta - 1} - \Delta^{-1/2}\right)\,\beta^{-d}\, \Delta k(\beta\Theta).
	\label{eq:cyl_deltaSmallTheta}
\end{multline}
\else
\begin{align}
	\Delta \Phi_s^{(cc)}(\Xi,\Delta,\Theta) =& \int_{1+\Xi^2}^{\infty}\upd\beta\,\left(\sqrt{\beta - 1} - \Xi\right)\,\beta^{-d}\, \Delta k(\beta\Theta)\nonumber\\
&-\int_{1+\Delta^{-1}}^\infty \upd \beta\,\left(\sqrt{\beta - 1} - \Delta^{-1/2}\right)\,\beta^{-d}\, \Delta k(\beta\Theta).
	\label{eq:cyl_deltaSmallTheta}
\end{align}
\fi

\begin{figure}[t!]
  \centering
  \begin{minipage}{0.47\textwidth}
  	\includegraphics[width=\textwidth]{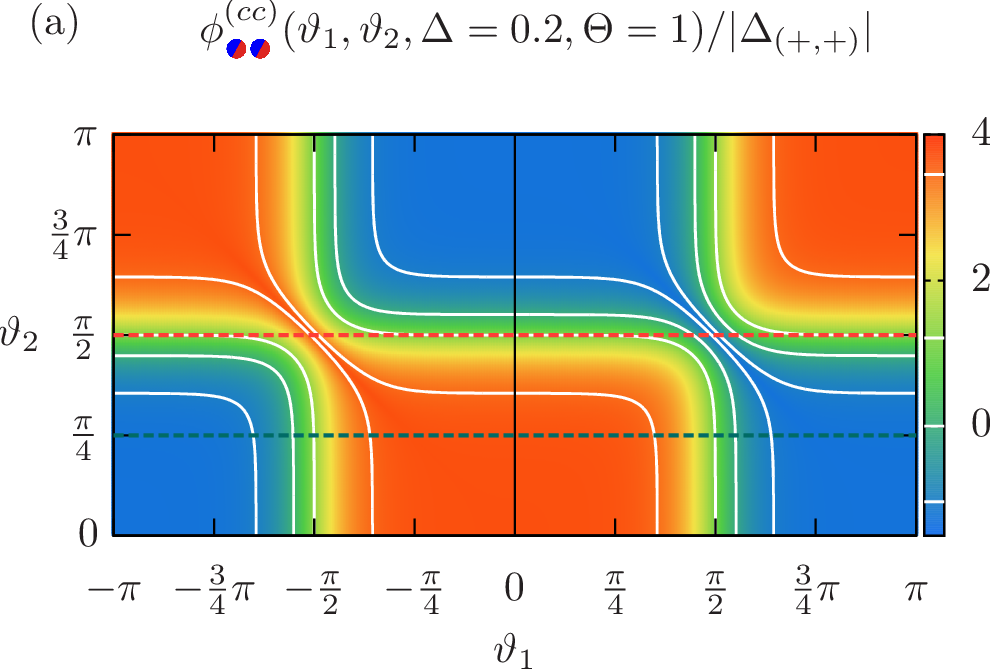}
  \end{minipage}
  \hfill
  \begin{minipage}{0.47\textwidth}
  	\includegraphics[width=\textwidth]{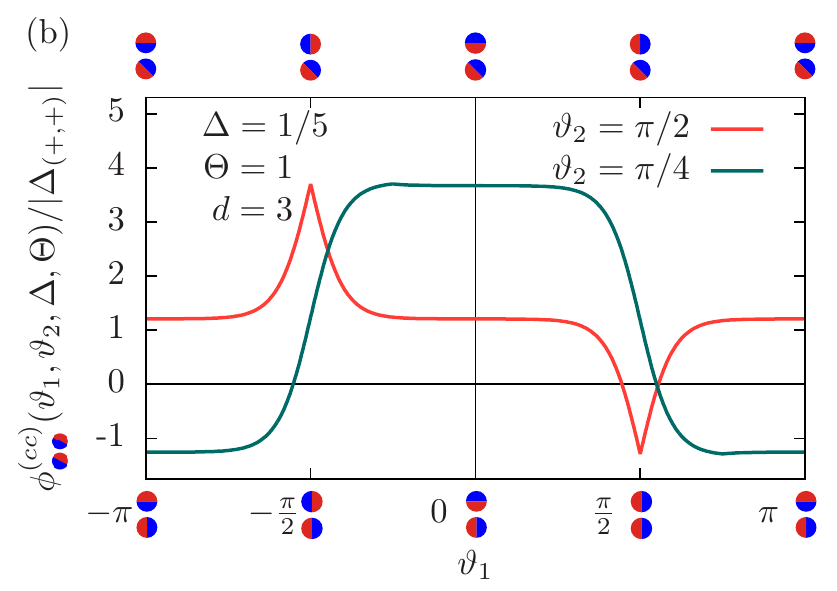}
  \end{minipage}
  \caption{The scaling function $\Phi_{\symbJcJc}^{(cc)}$ of the effective potential between two Janus cylinders in $d=3$ presented in (a) as a free energy landscape in terms of the orientations $\vartheta_1$ and $\vartheta_2$ for the fixed scaled temperature $\Theta=1$ and the fixed reduced distance $\Delta=1/5$. The value of the scaling function is color coded ranging from blue for a minimum in the energy to red for a maximum in the energy. The landscape exhibits broad and flat plateaus; four free energy isolines (white, each consisting of three disconnected pieces), indicating a low value, zero, the mean value between the minimum and the maximum energy, and a high value, are drawn as a guide to the eye. Two straight horizontal paths are shown as a dashed red ($\vartheta_2=\pi/2$) and as a dashed green line ($\vartheta_2=\pi/4$). The scaling function of the potential along these paths is shown in (b) as a function of $\vartheta_1$. The points of each curve correspond to the scaling function for a specific geometric configuration $(\vartheta_1, \vartheta_2)=(\pi/2,\pi/4)$. For the green curve selected configurations are indicated at the top of the panel, for the red curve examples are provided at the bottom. For each pair of particles the upper one is called \numcirc{1} and the lower one \numcirc{2}.}
  \label{fig:cyl_pot_janus}
\end{figure}

Since the dimensionless scaling function $\Phi_{\symbJcJc}^{(cc)}/\Delta^{d-3/2}$ of the potential describes an energy in units of $k_B T$ and per reduced length $\mathcal{L}/R^{d-1}$ of the cylinders, it is useful to present it as an energy landscape in terms of the orientation angles $\vartheta_1$ and $\vartheta_2$, for a fixed scaled temperature $\Theta$ and a fixed reduced distance $\Delta=D/R$ (see Fig.~\ref{fig:cyl_pot_janus}). This facilitates a thermodynamic interpretation. For example, it relates to the typical experimental setup in which the temperature is fixed and the formation of clusters is attributed to a minimum of the effective interaction potential. Invoking the critical Casimir effect alone, which leads to an irreversible aggregation, is insufficient to describe the distance dependence of effective interactions between colloids in suspension. Typically, in addition to the critical Casimir effect, there are repulsive electrostatic interactions due to counterions, which are short-ranged and prevent coagulation of the colloids. If, however, these additional interactions are basically isotropic (e.g., due to a more or less homogeneous charge distribution on the colloid surface), the critical Casimir potential considered here is the only orientation dependent interaction and thus fully responsible for any orientation dependent behavior during the clustering of Janus particles close to $T_c$.

The free energy landscape in Fig.~\ref{fig:cyl_pot_janus}(a) exhibits two broad minima (blue color) around $(\vartheta_1,\vartheta_2)=(\pm\pi,0)$ and $(\vartheta_1,\vartheta_2)=(0,\pi)$, which correspond to the two equally stable configurations of the blue sides facing each other, i.e., $(-,-)$ BC, and the red sides facing each other, i.e., $(+,+)$ BC, respectively (see Figs.~\ref{fig:sketch_cyl_da} and \ref{fig:cyl_K_da}). These two ground states are connected by a narrow valley along the line $\vartheta_2=\pi-\vartheta_1$, which corresponds to configurations which emerge from $(\vartheta_1=0,\vartheta_2=\pi)$ upon counter-rotating both particles. In Fig.~\ref{fig:cyl_pot_janus}(a) two selected paths are marked by dashed lines with the scaling function shown, as function of $\vartheta_1$ only, in Fig.~\ref{fig:cyl_pot_janus}(b). The resulting red and green curves are, unsurprisingly, reminiscent of the scaling function for the force acting between a Janus cylinder and a step (see Fig.~\ref{fig:cyl_da_mft_comparison}). This extends the equivalence of substrate steps and Janus particles discussed before. The wedge-like shape of the valley trough is an artifact of DA. Based on previous findings (Fig.~\ref{fig:cyl_da_mft_comparison}), we expect the actual potential landscape to be similar, but slightly broadened and smoother.
The broad plateaus correspond to repulsive configurations (colored in red) with opposing BC on those surface parts of the particles which face each other. The free energy landscape can be continued periodically with respect to $\vartheta_1$ and $\vartheta_2$, resulting in a checkerboard pattern. The occurrence of broad maxima and minima gives credit to the on-off ``bond-like'' character of the interaction used in the Kern-Frenkel model \cite{Kern:2003}. Interestingly however, for the critical Casimir potential less than $50\%$ of all configurations are actually attractive despite the Janus characteristic.
There are additional characteristic features of the critical Casimir potential with important thermodynamic consequences, such as the discrepancy in strength of attraction and repulsion, the discussion of which we leave to future studies.

\section{Janus spheres}
\label{sec:janus_spheres}
The effective interaction between parallel, cylindrical Janus particles is conveniently described by only two orientational degrees of freedem. While this constrained setup poses an additional experimental challenge, the behavior of spherical colloids can, instead, be studied straightforwardly. Therefore, in the following we determine the scaling function of the force and of the effective potential between two spherical Janus particles, without constraints on the orientation.

We consider a conventional sphere in $d=3$, for which the Janus characteristics are unambiguous. In $d=4$, we consider a three-dimensional sphere extended along an extra dimension with a length $L_4$, which is formally called a hyper-cylinder (rather than a hyper-sphere). This definition is distinct from the hyper-cylinder discussed before. In the context of spheres, $\mathcal{L}$ denotes $\mathcal{L}=1$ in $d=3$ and $\mathcal{L}=L_4$ in $d=4$.
\subsection{Scaling function of the force}
The force between two Janus spheres depends, in principle, on their orientation vectors $\vec{n}_1$ and $\vec{n}_2$ and the vector $\vec{r}_{12}$ connecting their centers \footnote{We only consider orientations of the Janus spheres in $d=3$ and disregard the possible, but contrived case of orientations in $d=4$ which would violate the invariance in the extra dimension.}. The force takes the scaling form
\begin{multeqline}
F_{\symbJcJc}^{(ss)}(\vec{n}_1, \vec{n}_2, \vec{r}_{12}, R, T) = \ifTwocolumn \\ \fi
 k_B T\, \frac{\mathcal{L}}{R^{d-2}}\,\frac{K_{\symbJcJc}^{(ss)}(\vec{n}_1, \vec{n}_2, \hat{\vec{r}}_{12}, \Delta, \Theta)}{\Delta^{d-1}},
\label{eq:scalform_F_spheres}
\end{multeqline}
where for the scaling function, the connecting vector $\vec{r}_{12} = (D + 2 R)\,\hat{\vec{r}}_{12}=R(\Delta+2)\hat{\vec{r}}_{12}$ is expressed in terms of the surface-to-surface distance $D$ along the direction $\hat{\vec{r}}_{12}=\vec{r}_{12}/|\vec{r}_{12}|$. Note that in the case of two spheres, at $T_c$ the force decays as $\Delta^{-(d-1)}$ with distance \cite{Gambassi:2009}, compared to $\Delta^{-(d-1/2)}$ for the force between two cylinders (see Eq.~\eqref{eq:cyl_force_scalform}).
As in the case of Janus cylinders, we decompose the scaling function $K_{\symbJcJc}^{(ss)}$ of the force into a part given by the scaling function $K_{(+,\pm)}^{(ss)}$ between two homogeneous spheres \cite{Gambassi:2009}
\begin{equation}
K_{(+,\pm)}^{(ss)}(\Delta,\Theta)=\pi\int_1^{1+\Delta^{-1}} \upd\alpha\,\alpha^{-d}\,k_{(+,\pm)}\left(\alpha\Theta\right),
\label{eq:sp_k_hom}
\end{equation}
and an excess scaling function $\Delta K_{\oslash\oslash}^{(s)}$:
\begin{multeqline}
K_{\symbJcJc}^{(ss)}(\vec{n}_1, \vec{n}_2, \hat{\vec{r}}_{12}, \Delta, \Theta) = \ifTwocolumn \\ \fi
 K_{(+,+)}^{(ss)}(\Delta, \Theta) - \Delta K_{\oslash\oslash}^{(ss)}(\vec{n}_1, \vec{n}_2, \hat{\vec{r}}_{12}, \Delta, \Theta).
\label{eq:sp_k_separation}
\end{multeqline}
In contrast to the preceding sections, here we do not bear out explicitly all possible cases within the scaling function $K_{\symbJcJc}^{(ss)}$. Instead we absorb them into $\Delta K_{\oslash\oslash}^{(ss)}$, because the underlying symmetries are less intuitive and transparent. This leaves one with the arbitrary choice of whether to relate $\Delta K_{\oslash\oslash}^{(ss)}$ to $K_{(+,+)}^{(ss)}(\Delta, \Theta)$  or $K_{(+,-)}^{(ss)}(\Delta, \Theta)$; we follow the definition in Eq.~\eqref{eq:sp_k_separation}. Note that it is not necessary to express $\Delta K_{\oslash\oslash}^{(ss)}$ in terms of reduced angles, because as a spherical coordinate $\vartheta_{1,2}\in[0,\pi]$ is a reduced angle by definition. Again, the uncolored subscript emphasizes invariance with respect to the shift $\vartheta_i\to\vartheta_i\pm\pi$.

\begin{figure}
\centering
\includegraphics[width=0.35\textwidth]{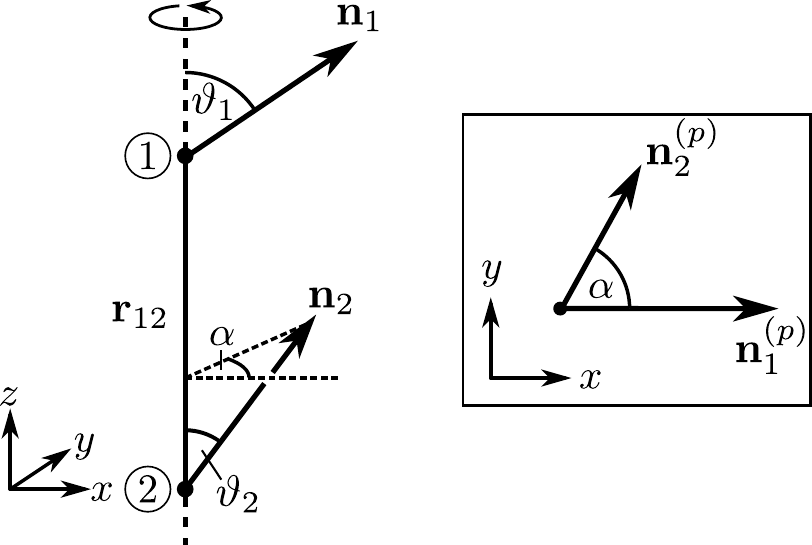}
\caption{Generic sketch of the orientations of the Janus spheres \numcirc{1} and \numcirc{2} defining the azimuthal angle $\alpha=\phi_2-\phi_1$, and the polar angles $\vartheta_1$ and $\vartheta_2$ of the relative coordinate system which has the z-axis aligned with the vector $\vec{r}_{12}$ and orientated such that $\phi_1=0$. Left: side view with a slight perspective in order to depict $\alpha$. Right: top view of the same configuration, with $\vec{n}_i^{(p)}$ the projection of $\vec{n}_i$ onto the xy plane. The DA considers pairs of surface elements projected along $\vec{r}_{12}$, thus effectively representing a top-down view. Rotating the frame of reference, so that $\phi_1\neq 0$ but $\alpha=\phi_2-\phi_1$ is kept constant, does not affect the interaction in that case.}
\label{fig:spherical_vectors}
\end{figure}

Determining completely the excess scaling function $\Delta K_{\oslash\oslash}^{(ss)}$ requires careful considerations of all possible orientations.
It turns out that within DA, the force necessarily depends only on the relative coordinates, because the interaction is expressed via the overlap of surface elements projected along the connecting vector $\vec{r}_{12}$. This is worked out in detail in Appendix \ref{sec:app_janus_spheres}, using spherical coordinates $\vec{n}_1=(\phi_1, \vartheta_1)$ and $\vec{n}_2=(\phi_2, \vartheta_2)$. Thus the interaction depends only on the polar angles $\vartheta_1$ and $\vartheta_2$, and the dependence on $\phi_1$ and $\phi_2$ reduces to one on the angle difference $\alpha=\phi_2-\phi_1$ (see Fig.~\ref{fig:spherical_vectors}).
For comparison, we briefly consider the pair potential between two point dipoles of strength $\mu$:
\begin{equation}
V_{dip} = -\frac{\mu^2}{r_{12}^3}\,\left[3\left(\vec{n}_1\cdot\hat{\vec{r}}_{12}\right)\left(\vec{n}_2\cdot\hat{\vec{r}}_{12}\right)-\vec{n}_1\cdot\vec{n}_2\right]
\end{equation}
Written similarly in the relative coordinate system connecting the two dipoles, these render $\vec{n}_i\cdot\vec{r}_{12}=\cos\vartheta_i$ and $\vec{n}_1\cdot\vec{n}_2 = \cos\vartheta_1\cos\vartheta_2 + \sin\vartheta_1\sin\vartheta_2\cos(\phi_1 - \phi_2)$. Thus, concerning the dependence on the orientations, the critical Casimir interaction between two Janus spheres exhibits the same level of complexity as the dipole-dipole interaction.

Here, we provide the excess scaling function $\Delta K_{\oslash\oslash}^{(ss)}$ as a function of $\vartheta_1$, $\vartheta_2$, and the relative coordinate $\alpha$ (see Appendix \ref{sec:app_janus_spheres}):
\ifTwocolumn\begin{widetext}\fi
\begin{align}
\Delta K_{\oslash\oslash}^{(ss)}&(\alpha,\vartheta_1,\vartheta_2, \Delta, \Theta) =
\pi\,H\left(\coscos\right)\int_1^{1+\Delta^{-1}r_s^2} \upd x\,x^{-d}\,\Delta k\left(x \Theta\right) \nonumber\\
&-\mathrm{sign}\left(\coscos\right)\Bigg[\int_{1+\Delta^{-1}\cos^2\vartheta_1}^{1+\Delta^{-1} r_s^2} \upd x\,\arccos\left(|\cot\vartheta_1|\sqrt{\frac{1}{\Delta(x-1)}-1}\right)\,x^{-d}\,\Delta k\left(x \Theta\right) \nonumber\\
&\hspace{5.5em}+c(\alpha,\vartheta_1,\vartheta_2)\int_{1+\Delta^{-1}\cos^2\vartheta_2}^{1+\Delta^{-1} r_s^2} \upd x\,\arccos\left(|\cot\vartheta_2|\sqrt{\frac{1}{\Delta(x-1)}-1}\right)\,x^{-d}\,\Delta k\left(x \Theta\right) \Bigg] \nonumber\\
&+\alpha\int_{1+\Delta^{-1} r_s^2}^{1+\Delta^{-1}} \upd x\,x^{-d}\,\Delta k\left(x \Theta\right).
\label{eq:DeltaK_sp_janus_DA}
\end{align}
\ifTwocolumn\end{widetext}\fi
As before one has $\Delta k = k_{(+,+)}-k_{(+,-)}<0$. The first term with the Heaviside step function $H\left(\coscos\right)$ as a prefactor effectively serves the same purpose as the case analysis within the scaling functions of previous geometries (see, e.g., Eq.~\eqref{eq:cyl_force_separation}, with $\Xi(\vartheta_1)\,\Xi(\vartheta_2)\propto \coscos$). 
Additionally, $\Delta K_{\oslash\oslash}^{(ss)}$ depends non-trivially on $\alpha,\vartheta_1$, and $\vartheta_2$, inter alia, via the dimensionless radius $r_s=R_s(\alpha,\vartheta_1,\vartheta_2)/R$ (see Eq.~\eqref{eq:app_ellipse_intersection_leqs}) of a particular ring of surface elements occurring within DA in the subdivision of the surfaces. The projection of the equatorial steps of both Janus spheres onto a common plane, normal to the axis connecting the colloids, results in two half-ellipses corresponding to each configuration. The surface ring with radius $R_s$ intersects the projections of the equatorial steps of both Janus spheres in a single point. Thus, the scaled radius $r_s$ is defined as $r_s=\sqrt{x^2+y^2}$, with the intersection point $(x,y)$ of the two ellipses determined by a particular solution of a system of two equations. For details, we refer to Appendix \ref{sec:app_janus_spheres}.

Certain configurations of the two Janus particle give rise to forces which consist of force contributions of the same strength, but of opposite signs. All these cases can be subsumed by Eq.~\eqref{eq:DeltaK_sp_janus_DA} via the common prefactor $\mathrm{sign}\left(\coscos\right)$ and via the sign picking function
\ifTwocolumn\begin{widetext}\fi
\begin{equation}
	c(\alpha, \vartheta_1, \vartheta_2) = \begin{cases}
	\mathrm{sign}(\cos\alpha), & \text{if } \coscos = 0, \\
	1, & \text{if } \alpha \leq \arccos\left(-(\tan\vartheta_2)\,(\cot\vartheta_1)\right)\leq\pi H\left(\coscos\right) \\
	& \quad \text{or } \pi H\left(\coscos\right)\leq \arccos\left(-(\tan\vartheta_2)\,(\cot\vartheta_1)\right)\leq\alpha,\\
	-1 & \text{otherwise,}
	\end{cases}
\label{eq:sp_signpick_func}
\end{equation}
which assumes without loss of generality that $0\leq\alpha\leq \pi$; otherwise $\alpha > \pi$ is replaced by $\alpha \to 2\pi-\alpha$.
\ifTwocolumn\end{widetext}\fi

\begin{figure}[t!]
  \centering
  \includegraphics[width=0.47\textwidth]{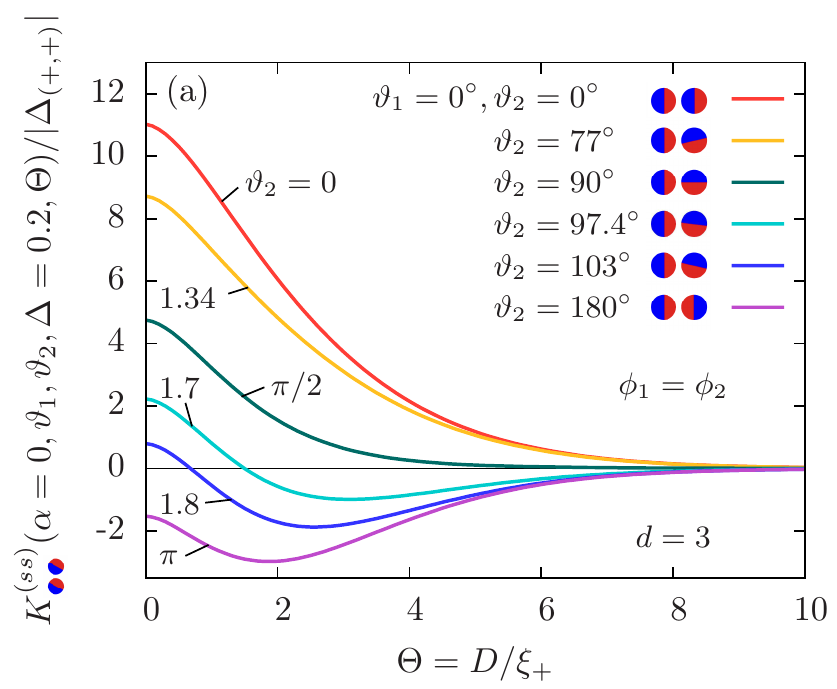}\hfill
  \includegraphics[width=0.47\textwidth]{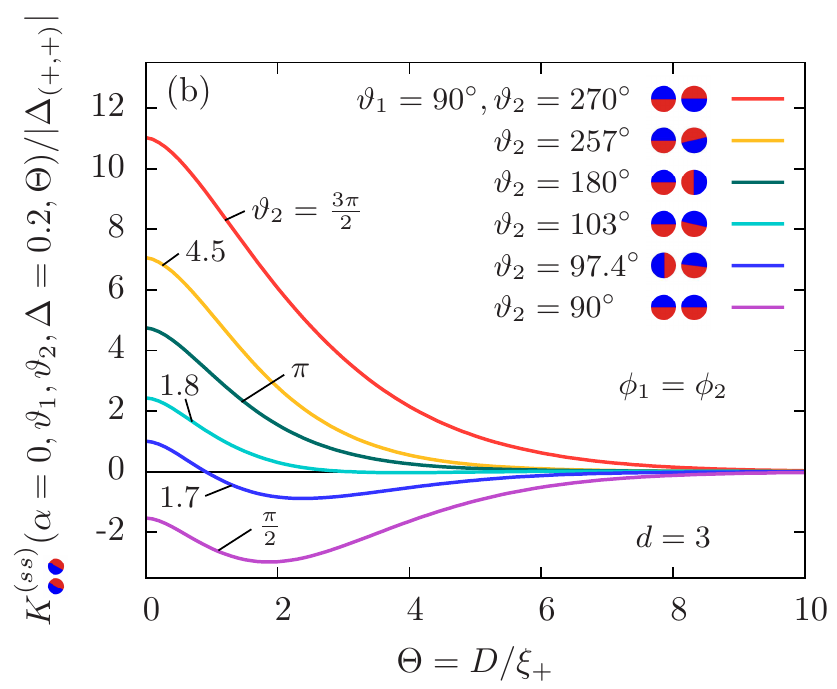}
  \caption{The normalized scaling function of the force $K_{\symbJcJc}^{(ss)}$ between two Janus spheres within DA in $d=3$, as a function of $\Theta=D/\xi_+$ for the same orientations as in Fig.~\ref{fig:cyl_K_da}. (a) Configurations with $\vartheta_1 = 0$ for the orientation of the left particle \numcirc{1} for various orientation angles $\vartheta_2$ of the right particle \numcirc{2}, as visualized in the legend. (b) The case of $\vartheta_1 = \pi/2$ for various orientations $\vartheta_2$ of the second particle. In order to facilitate a transparent comparison with Fig.~\ref{fig:cyl_K_da}, the azimuthal angle $\alpha$ is set to $\alpha=0$, i.e., $\phi_1=\phi_2$, which restricts the orientation vectors $\vec{n}_1$ and $\vec{n}_2$ to lie in a common plane, as it is the case in our analysis of parallel cylinders.}
  \label{fig:K_sp_da}
\end{figure}

The scaling function $K_{\symbJcJc}^{(ss)}(\alpha,\vartheta_1,\vartheta_2, \Delta, \Theta)$ of the force (given by Eqs.~\eqref{eq:sp_k_hom}--\eqref{eq:DeltaK_sp_janus_DA}) is shown in Fig.~\ref{fig:K_sp_da} for various configurations with $\alpha=0$, i.e., $\phi_1=\phi_2$, akin to Fig.~\ref{fig:cyl_K_da} for parallel cylindrical Janus particles. In accordance with Fig.~\ref{fig:spherical_vectors}, $\alpha=0$ implies that the two orientation vectors $\vec{n}_1$ and $\vec{n}_2$ lie in the same plane, as for parallel cylinders, so that the corresponding equatorial planes are rotated with respect to each other $(\vartheta_1\neq\vartheta_2)$, but not tilted (see Fig.~\ref{fig:sketch_cyl_da}). On first sight, the scaling functions of the force for Janus spheres and for Janus cylinders appear to be qualitatively very similar. 
Quantitatively, the force between spheres appears to be stronger than the force between parallel cylinders. However, one has to take into account that the force between two Janus cylinders is proportional to their length.
A fair comparison of the strengths of the forces requires to consider a cylinder length which is comparable with the size of the sphere, i.e., $L\approx2R$. In this case the force between two parallel cylinders is stronger. 
Additionally, the scaling function for Janus spheres decays slightly faster as function of $\Theta$. 
Generally, the scaling function of the force between two Janus spheres is slightly more sensitive to small rotations of one particle than the one for cylinders.

\subsection{Scaling function of the effective potential}
As in the case of the Janus cylinders, the effective potential between two Janus spheres of radius $R$ can be determined from the critical Casimir force in the relative coordinate system according to
\begin{multeqline}
V_{\symbJcJc}^{(ss)}(\vec{n}_1, \vec{n}_2, \vec{r}_{12}=(D+2R)\vec{e}_z, R, T) = \ifTwocolumn \\ \fi
\int_D^\infty\upd z\,F_{\symbJcJc}^{(ss)}(\vec{n}_1, \vec{n}_2, \vec{r}_{12}=(z+2R)\vec{e}_z, R, T).
\label{eq:sp_phi_def}
\end{multeqline}
After inserting Eq.~\eqref{eq:scalform_F_spheres}, this can be cast into the scaling form
\begin{multeqline}
V_{\symbJcJc}^{(ss)}(\vec{n}_1, \vec{n}_2, \vec{r}_{12}=(D+2 R)\vec{e}_z, R, T) = \ifTwocolumn \\ \fi
 k_B T\, \frac{\mathcal{L}}{R^{d-3}}\, \frac{\Phi_{\symbJcJc}^{(ss)}(\alpha, \vartheta_1, \vartheta_2, \Delta,\Theta)}{\Delta^{d-2}}.
\end{multeqline}
Following Eq.~\eqref{eq:sp_k_separation}, the scaling function $\Phi_{\symbJcJc}^{(ss)}$ of the potential is divided up into the two contributions
\begin{multeqline}
\Phi_{\symbJcJc}^{(ss)}(\alpha, \vartheta_1, \vartheta_2, \Delta,\Theta) = \ifTwocolumn \\ \fi
 \Phi_{(+,+)}^{(ss)}(\Delta, \Theta) - \Delta \Phi_{\oslash\oslash}^{(ss)}(\alpha, \vartheta_1, \vartheta_2, \Delta,\Theta),
\label{eq:sp_theta_separation}
\end{multeqline}
where
\ifTwocolumn
\begin{multline}
\Phi_{(+,\pm)}^{(ss)}(\Delta, \Theta) = \pi\int_1^\infty\upd x\,(x - 1) x^{-d}\,k_{(+,\pm)}(x\,\Theta)\\
-\pi\int_{1+\Delta^{-1}}^\infty \upd x\,(x - 1 - \Delta^{-1})\,x^{-d}\, k_{(+,\pm)}(x\,\Theta)
\label{eq:sp_theta_hom}
\end{multline}
\else
\begin{align}
\Phi_{(+,\pm)}^{(ss)}(\Delta, \Theta) = &\pi\int_1^\infty\upd x\,(x - 1) x^{-d}\,k_{(+,\pm)}(x\,\Theta)\nonumber\\
&-\pi\int_{1+\Delta^{-1}}^\infty \upd x\,(x - 1 - \Delta^{-1})\,x^{-d}\, k_{(+,\pm)}(x\,\Theta)
\label{eq:sp_theta_hom}
\end{align}
\fi
is the scaling function of the potential between two homogeneous spheres, and $\Delta \Phi_{\oslash\oslash}^{(ss)}$ is the Janus-induced excess scaling function. In view of the known expression for $\Phi_{(+,\pm)}^{(ss)}(\Delta\to 0,\Theta)$ \cite{Hanke:1998, Gambassi:2009}, we again retain the explicit dependence on $\Delta$ in the scaling function of the homogeneous case for reasons of consistency with the orientation dependent excess scaling function in Eq.~\eqref{eq:sp_theta_separation}. The previous caveats regarding the dependence on $\Delta$ within DA apply here, too.

Upon inserting the scaling function of the force (Eq.~\eqref{eq:DeltaK_sp_janus_DA}) into Eqs.~\eqref{eq:sp_phi_def}-\eqref{eq:sp_theta_hom}, the excess scaling function of the potential is given by (see Appendix \ref{sec:app_janus_spheres_pot})
\ifTwocolumn
\begin{multline}
\Delta \Phi_{\oslash\oslash}^{(ss)}(\alpha,\vartheta_1,\vartheta_2,\Delta,\Theta) =\\
\pi\,H\left(\coscos\right)\Delta u^{(ss)}(r_s^2,0,\Delta, \Theta)\\
-\mathrm{sign}\left(\coscos\right)\Big[\Delta v^{(ss)}(r_s^2,\vartheta_1, \Delta, \Theta)\\
+c(\alpha,\vartheta_1,\vartheta_2)\Delta v^{(ss)}(r_s^2,\vartheta_2, \Delta, \Theta)\Big]\\
+\alpha\,\Delta u^{(ss)}(1,r_s^2,\Delta, \Theta)
\label{eq:DeltaPot_sp_janus_DA}
\end{multline}
\else
\begin{align}
\Delta \Phi_{\oslash\oslash}^{(ss)}(\alpha,\vartheta_1,\vartheta_2,\Delta,\Theta) =&\pi\,H\left(\coscos\right)\Delta u^{(ss)}(r_s^2,0,\Delta, \Theta)\nonumber\\
&-\mathrm{sign}\left(\coscos\right)\Big[\Delta v^{(ss)}(r_s^2,\vartheta_1, \Delta, \Theta)\\
&\hspace{5.5em}+c(\alpha,\vartheta_1,\vartheta_2)\Delta v^{(ss)}(r_s^2,\vartheta_2, \Delta, \Theta)\Big]\nonumber\\
&+\alpha\,\Delta u^{(ss)}(1,r_s^2,\Delta, \Theta)\nonumber
\label{eq:DeltaPot_sp_janus_DA}
\end{align}
\fi
with $c(\alpha,\vartheta_1,\vartheta_2)$ defined by Eq.~\eqref{eq:sp_signpick_func} and where
\ifTwocolumn
\begin{multline}
\Delta u^{(ss)}(a,b,\Delta, \Theta)=\int_{1+b/\Delta}^\infty\upd y\,(y-1-b/\Delta)\,y^{-d}\Delta k(y\,\Theta)\\
-\int_{1+a/\Delta}^\infty\upd y\,(y-1-a/\Delta)\,y^{-d}\Delta k(y\,\Theta)
\end{multline}
\else
\begin{align}
\Delta u^{(ss)}(a,b,\Delta, \Theta)=&\int_{1+b/\Delta}^\infty\upd y\,(y-1-b/\Delta)\,y^{-d}\Delta k(y\,\Theta)\nonumber\\
&-\int_{1+a/\Delta}^\infty\upd y\,(y-1-a/\Delta)\,y^{-d}\Delta k(y\,\Theta)
\end{align}
\fi
and (see Eq.~\eqref{eq:app_ellipse_intersection_leqs} concerning $r_s$)
\ifTwocolumn
\begin{multline}
\Delta v^{(ss)}(r_s^2,\vartheta, \Delta, \Theta) =\\
 \Delta^{-1}\int_{1+\cos^2\vartheta/\Delta}^{1+r_s^2/\Delta}\upd y\,g\big(\Delta(y-1),\vartheta\big)\,y^{-d}\,\Delta k\left(y\,\Theta\right)\\
+\Delta^{-1}\int_{1+r_s^2/\Delta}^\infty\upd y\,g(r_s^2,\vartheta)\,y^{-d}\,\Delta k\left(y\,\Theta\right)
\end{multline}
\else
\begin{align}
\Delta v^{(ss)}(r_s^2,\vartheta, \Delta, \Theta) = &\Delta^{-1}\int_{1+\cos^2\vartheta/\Delta}^{1+r_s^2/\Delta}\upd y\,g\big(\Delta(y-1),\vartheta\big)\,y^{-d}\,\Delta k\left(y\,\Theta\right)\nonumber\\
&+\Delta^{-1}\int_{1+r_s^2/\Delta}^\infty\upd y\,g(r_s^2,\vartheta)\,y^{-d}\,\Delta k\left(y\,\Theta\right)
\end{align}
\fi
are excess scaling functions of Janus spheres (vaguely analogous to the chemical step-like scaling functions for Janus cylinders).
The integrand of the latter scaling function $\Delta v^{(ss)}$ contains a geometry specific expression
\ifTwocolumn
\begin{align}
g(u,\vartheta) = &\int_{\cos^2\vartheta}^u\upd w\,\arccos\left(|\cot\vartheta|\sqrt{\frac{1}{w}-1}\right)\\
=&\ u \arccos\left(|\cot\vartheta|\sqrt{\frac{1}{u}-1}\right) \nonumber\\
& - |\cos\vartheta|\arccos\left(|\csc\vartheta|\sqrt{1-u}\right),\enspace \cos^2\vartheta \leq u \nonumber.
\end{align}
\else
\begin{align}
g(u,\vartheta) = &\int_{\cos^2\vartheta}^u\upd w\,\arccos\left(|\cot\vartheta|\sqrt{\frac{1}{w}-1}\right)\\
=&\ u \arccos\left(|\cot\vartheta|\sqrt{\frac{1}{u}-1}\right) - |\cos\vartheta|\arccos\left(|\csc\vartheta|\sqrt{1-u}\right),\enspace \cos^2\vartheta \leq u \nonumber.
\end{align}
\fi

\begin{figure}[t!]
  \centering
  \begin{minipage}{0.47\textwidth}
  	\includegraphics[width=\textwidth]{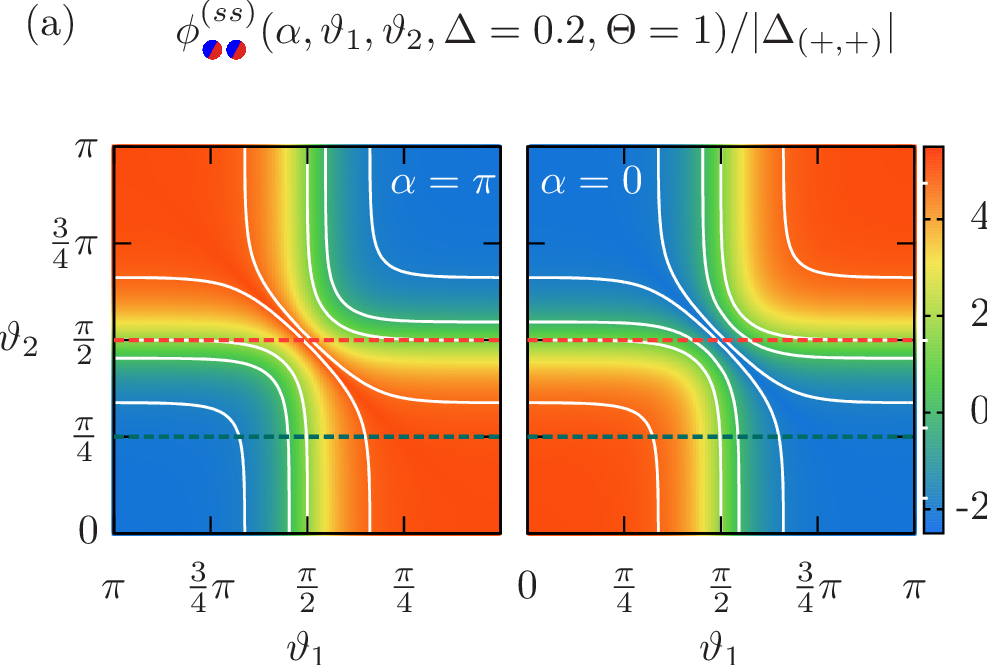}
  \end{minipage}
  \hfill
  \begin{minipage}{0.47\textwidth}
  	\includegraphics[width=\textwidth]{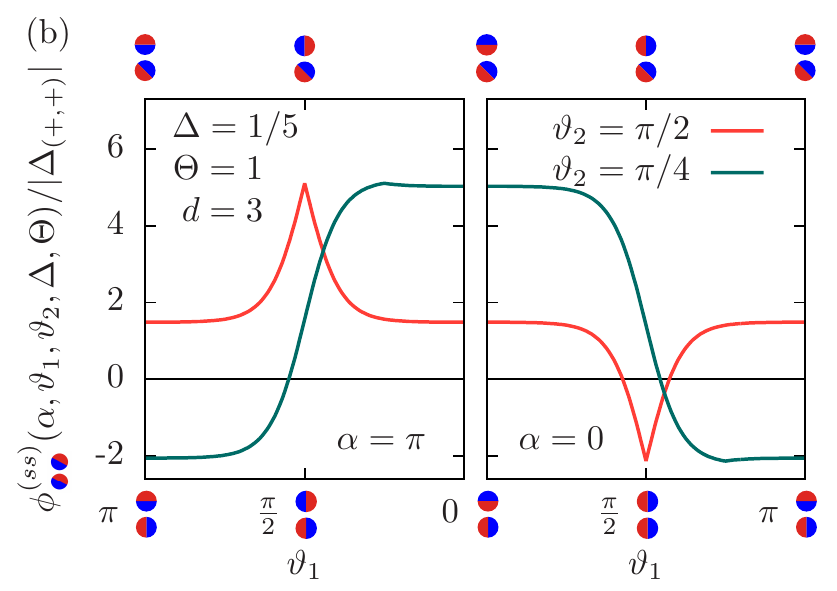}
  \end{minipage}
  \caption{The scaling function $\Phi_{\symbJcJc}^{(ss)}$ of the effective pair potential between two Janus spheres in $d=3$ for $\alpha=\pi$ and $\alpha=0$ presented (a) as a free energy landscape in terms of $\vartheta_1$ and $\vartheta_2$ for a fixed scaled temperature $\Theta=1$, and (b) as a function of $\vartheta_1$ along the two paths $\vartheta_2=\pi/2$ (red dashed line) and $\vartheta_2=\pi/4$ (green dashed line). At the top of the panel, the geometric configurations indicate those which correspond to points of the green curve; configurations corresponding to the red curve are indicated at the bottom.
The comparison with Fig.~\ref{fig:cyl_pot_janus} tells that the free energy landscapes for cylinders and spheres are qualitatively very similar. Note that for $\alpha=\pi$ in (a) and (b) the horizontal axes are inverted in order to emphasize the geometric correspondence of $\alpha=\pi$ and $\vartheta_1>0$ in spherical coordinates to $\vartheta_1<0$ in cylindrical coordinates. An increase of $\alpha$ affects the potential only within a limited angular range around $\vartheta_1=\vartheta_2=\pi/2$, changing the potential in that range from being attractive ($\alpha=0$) to being repulsive ($\alpha=\pi$). This means that upon increasing $\alpha$ the potential gradually develops a potential barrier (see the red curve in (b)).}
  \label{fig:sp_pot_janus}
\end{figure}

The free energy landscape of the scaling function $\Phi_{\symbJcJc}^{(ss)}$ of the pair potential between two Janus spheres can be presented in a single plot only as a function of two variables, but not for the full set $\alpha, \vartheta_1, \vartheta_2$ of three variables. Accordingly, in Fig.~\ref{fig:sp_pot_janus} we choose to show the scaling function of the pair potential between Janus spheres for the two values $\alpha = 0$ and $\alpha=\pi$.
For $\alpha=0$, in the range $\vartheta_1>0$ the scaling function of the potential is very similar to the one for cylinders shown in Fig.~\ref{fig:cyl_pot_janus}. On the other hand, for Janus spheres, the case of $\alpha=\pi$ in Fig.~\ref{fig:sp_pot_janus} is similar to the one of $\vartheta_1<0$ for Janus cylinders. Obviously, in spherical coordinates an orientation vector with $\alpha=\pi$ and $\vartheta_1\in[0, \pi]$ lies in the same plane as an orientation vector with $\alpha=0$, and can be mapped to a cylindrical angle $\vartheta_1\in[-\pi,0]$.
The scaling function of the pair potential between Janus spheres is also dominated by the attractive minima and the repulsive plateaus of interaction (Fig.~\ref{fig:sp_pot_janus}(a)). The variation of the relative azimuthal angle $\alpha$ affects the potential only locally around $\vartheta_1=\vartheta_2=\pi/2$. Upon increasing $\alpha$, the potential energy smoothly changes from having the potential minima connected by a valley to having the plateaus bridged.


With the scaling function of the potential at our disposal, inter alia we are able to elucidate a certain experimental aspect.
A general issue concerning experimental studies of colloidal aggregation consists of the influence of the unavoidable presence of a substrate. It can be used deliberately, e.g., for the gravity induced formation of a monolayer of homogeneous particles on the bottom wall of the sample. Experimentally, the particles can be prevented from sticking to the substrate by applying a surface treatment of the substrate such that it becomes repulsive at small distances between the particles and the wall. 
For Janus particles, the experimental situation can be more intricate. Typically, the interaction with the wall is biased towards favoring one side of the colloid over the other. If the attractive interaction with the wall dominates over the inter-particle interaction (or similarly, if the substrate is repulsive towards only one of the two sides of the Janus particle), a scenario can prevail according to which all Janus particles orientate with one and the same side towards the substrate.

Within this line of reasoning, let us suppose that the interaction with the substrate has been reduced substantially, but is still present, resulting in a small biased tilt of all Janus particles relative to the substrate normal. Depending on the setup, this tilt might be barely noticable, but would still affect the experimental determination of the effective pair potential between the particles.

\begin{figure}[t!]
  \centering
  \includegraphics[width=0.47\textwidth]{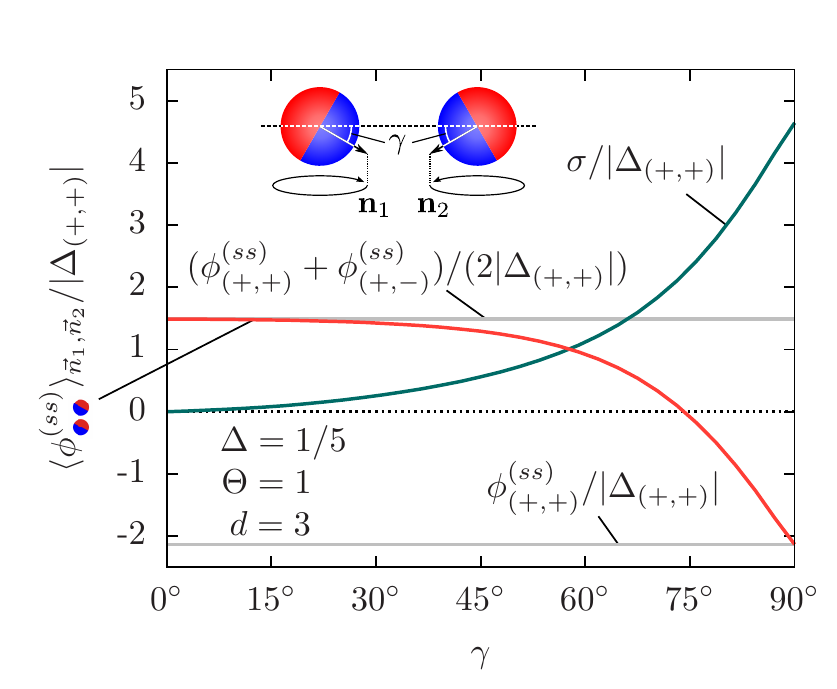}
  \caption{Angularly averaged and normalized scaling function $\Phi_{\symbJcJc}^{(ss)}$ of the effective pair potential for two Janus particles, which are considered to be at equal height above a substrate (not depicted), but far enough so that the influence of the substrate is weaker than that of the pair interaction between the particles. The orientations $\vec{n}_1$ and $\vec{n}_2$ are tilted by a common angle $\gamma$ towards the substrate and out of the plane which contains both particles centers and is parallel to the substrate. However, the influence of the substrate is taken to be isotropic in the remaining lateral directions. This is supposed to mimic a typical experimental setup. Thus, we consider the average $\langle\Phi_{\symbJcJc}^{(ss)}\rangle$ taken over $\vec{n}_1$ and $\vec{n}_2$ (see the main text), such that the tips of $\vec{n}_1$ and $\vec{n}_2$ form circles lying in a common plane parallel to the substrate surface (see the inset). The influence of the externally imposed tilt $\gamma$ on the effective pair potential is visualized by the dependence on $\gamma$ of the averaged scaling function $\langle\Phi_{\symbJcJc}^{(ss)}\rangle$ (red curve) and its standard deviation $\sigma$ with respect to the scaling function for $\gamma=0$ (green curve; see the main text). For $\gamma\to 0$ the average approaches the simple mean $(\Phi_{(+,+)}^{(ss)}+\Phi_{(+,-)}^{(ss)})/2$ of attraction and repulsion of homogeneous spheres (upper gray curve). For $\gamma=90^\circ$ the Janus equators are tilted such that they are parallel to the substrate and thus unaffected by rotations around the normal of the substrate, leading to an average $\Phi_{(+,+)}^{(ss)}$ (lower gray curve). All quantities are normalized by $|\Delta_{(+,+)}|$.}
  \label{fig:sp_scalfunc_tilted}
\end{figure}

In Fig.~\ref{fig:sp_scalfunc_tilted}, we show the scaling function of the effective potential between two Janus spheres, which are tilted by a common angle $\gamma$ relative to the axis connecting the centers of the two particles, due to the effects of a hypothetical substrate below the particles and parallel to the axis. Within this model, the horizontal components of the orientations $\vec{n}_1$ and $\vec{n}_2$ of the two Janus spheres are distributed isotropically in a plane parallel to the substrate; but the tilt $\gamma$ is fixed to a given value, corresponding to an equilibrium configuration of the Janus colloids relative to the substrate. Thus, the tips of $\vec{n}_1$ and $\vec{n}_2$ fluctuate on circles in a plane parallel to the substrate.
Note that for $\gamma>0$ a rotation of the whole configuration around the normal of the plane corresponds to a non-trivial trajectory in the three-dimensional space of the relative spherical coordinates $\alpha, \vartheta_1, \vartheta_2$, so that determining the average $\langle\Phi_{\oslash\oslash}^{(ss)}\rangle$ requires knowledge of the full scaling function of the potential. Due to problems associated with the multivalued nature of the transformation functions, we refrain from providing an explicit parametrization of the orientations $\vec{n}_1$ and $\vec{n}_2$ in terms of the new coordinates which would include $\gamma$.
Instead, for a fixed value of the tilt angle $\gamma$, we evaluate the scaling function $\Phi_{\symbJcJc}^{(ss)}$ numerically on a discretized set of $64\times 64$ orientations $\vec{n}_1$ and $\vec{n}_2$, each of them describing a circular path on the unit sphere. The set is expressed in terms of Cartesian coordinates and then transformed into spherical coordinates determining $\alpha, \vartheta_1$, and $\vartheta_2$ \footnote{Although the transformation $(x, y, z)\to(\alpha,\vartheta_1,\vartheta_2)$ also involves inverse trigonometric functions, which are multivalued within the principal domains $\alpha\in[0,2\pi)$ and $\vartheta_i\in[0,\pi]$ the only points which give rise to ambiguities are the ``north'' and ``south'' pole at $\vartheta_i=0$ and $\vartheta_i=\pi$, for which the value of $\alpha$ is completely arbitrary. However, for tilt angles $\gamma>0$ these poles do not lie on the circular paths of $\vec{n}_1$ and $\vec{n}_2$ and therefore are avoidable by this transformation.}.
The average $\langle\Phi_{\symbJcJc}^{(ss)}(\alpha, \vartheta_1, \vartheta_2)\rangle_{\vec{n}_1,\vec{n}_2}$ of the scaling function $\Phi_{\symbJcJc}^{(ss)}$ of the effective potential, i.e., the arithmetic mean of the data set, is plotted as a function of the tilt angle $\gamma$, together with the standard deviation $\sigma=\sqrt{\left\langle\left(\Phi_{\symbJcJc}^{(ss)}-\langle\Phi_{\symbJcJc}^{(ss)}\rangle_{\gamma=0}\right)^2\right\rangle}$ relative to the averaged scaling function for $\gamma=0$.
For $\gamma=0$, the average is taken such that both $\vec{n}_1$ and $\vec{n}_2$ describe a great circle on each sphere. They can be parameterized unambiguously by the relative coordinates $\alpha=0$, $0\leq\vartheta_{1,2}\leq \pi$, and $\alpha=\pi$, $0<\vartheta_{1,2} < \pi$ (i.e., both free energy landscapes shown in Fig.~\ref{fig:sp_pot_janus}(a) enter into the mean value), resulting within DA in the average $(\Phi_{(+,+)}^{(ss)}+\Phi_{(+,-)}^{(ss)})/2$ due to the symmetry of the potential.

The presence of a planar substrate effectively leads to a tilt $\gamma>0$. In the extreme case of a strongly dominant substrate force, a tilt of $\gamma=90^\circ$ towards the substrate rotates the two Janus equators into a configuration in which both of them are parallel to the substrate surface. In this case, the rotation around the normal of the substrate does not affect the pair interaction, so that always equal boundary conditions face each other. Accordingly, within DA, the average is simply given by $\Phi_{(+,+)}^{(ss)}$. Figure~\ref{fig:sp_scalfunc_tilted} tells that even intermediate tilt angles $\gamma$ do not alter the effective interaction drastically. Up to $\gamma\approx 30^\circ$ the mean value and the standard deviation remain rather constant and small, respectively. The deviations become significant only above $\gamma\approx 45^\circ$, which can be expected to be an experimentally detectable tilt. For smaller angles $\gamma$, ignoring the tilt entirely turns out to be a safe approximation.

The weak influence of small tilt angles on the appearance of the effective pair potential is associated with the flat plateaus in the energy landscape of the scaling function of the potential (see, e.g., Fig.~\ref{fig:sp_pot_janus}(a)). However, the proper average takes fully into account the trough- and ridge-like extrema occuring for orthogonal orientations (see, e.g., Fig.~\ref{fig:sp_pot_janus}(b)). This shows that the critical Casimir interaction is not only rather insensitive to small tilts for specific configurations, but even for a statistical ensemble of orientations. However, experimentally observed aggregation structures may be driven by additional effects not captured by the DA-based effective pair potential, such as the occurence of order parameter bridges between the particles  (see Ref.~\cite{Law:2015}). Thus, the aggregation of Janus particles into a complex spatial structure should still be analyzed carefully by taking into account the relevance of substrate induced tilting beyond the DA.
\section{Summary, Conclusions, and Outlook}
\label{sec:conclusions}
The aim of this study has been to determine theoretically the effective pair interaction between Janus particles immersed in a near-critical solvent and thus interacting via critical Casimir forces.
First, we have calculated the critical Casimir force acting on a single cylindrical Janus particle in the presence of a homogeneous substrate (Sec.~\ref{sec:janus_vs_steps} A) both by using the Derjaguin approximation (DA), and by applying mean field theory (MFT), which is valid in $d=4$ spatial dimensions. 

The DA implies a close relation between the critical Casimir forces for distinct geometries. Indeed, a comparison of DA with results from full MFT in $d=4$ reveals that, in the limit $\Delta=D/R\to 0$ of the ratio of the distance $D$ and radius $R$, the DA holds equally both for the force between a Janus cylinder and a substrate and for the force between a homogeneous cylinder and a substrate with a chemical step (see Figs.~\ref{fig:sketch_cyls} and \ref{fig:Kcyl_step_Theta}). However, as shown in Fig.~\ref{fig:Kcyl_step_to_janus}, the MFT scaling functions for the two geometries are distinct for nonzero $\Delta$. This caused us to address the question whether the relation between these two geometries has any merit beyond the limit $\Delta \to 0$ in which DA holds.

In the case of nonzero $\Delta$, for a homogeneous cylinder above a homogeneous substrate with opposing BC it is known that the MFT scaling function deviates quantitatively from the DA scaling function  \cite{Labbe:2014}. In Ref. \cite{Labbe:2014}, the deficiencies of the DA for $\Delta > 0$ have been traced back to an implicit assumption of the DA about the location and the shape of the interface which appears in this system for $t=(T-T_c)/T_c<T_c$, i.e., in the demixed phase. 
Following this argument, in Section \ref{sec:janus_vs_steps} B we have inspected the MFT order parameter (OP) profiles shown in Fig.~\ref{fig:profiles_janus_vs_step} for both aforementioned types of configurations and for two scaled temperatures $\Theta = D/\xi_+(t)=t^\nu D/\xi_0^+$ for $t > 0$, i.e., in the mixed phase. Thus, there is no interface present and the OP profile $\phi(\vec r, t)$ is mostly small. Still, the opposing BC of $\phi\to +\infty$ and  $\phi\to -\infty$ on the surface of the (Janus) particle and of the (stepped) substrate impose the occurrence of a line at which $\phi(\vec{r})$ crosses zero. DA makes implicit assumptions about the OP profile based on the one between a homogeneous particle and a homogeneous substrate with opposing BC and at the same temperature. We have found that the isoline $\phi(\vec{r})=0$ indeed follows closely the profile for homogeneous surfaces, however it smoothly bends towards the particle or the substrate, which is unaccounted for within DA. In Fig.~\ref{fig:profiles_janus_vs_step}, one can compare visually examples of the order parameter profiles for these two configurations, each of which gives rise to a force equal in strength to that in the other case for the same temperature $\Theta$. For the MFT results one can find systematically via interpolation for each scaled position $\Xi=X/\sqrt{R D}$ of the substrate step a certain rotational orientation $\vartheta$ of a Janus cylinder close to a homogeneous substrate which results in the same strength of the critical Casimir force, yielding numerically a discrete set $\{\Xi, \vartheta\}$. Visually, it appears that the forces are equal in strength whenever the bending and the extension of the isolines $\phi(\vec{r})=0$ are similar for both configurations.
An improved model has been introduced by applying the DA for a fictional, larger colloid, scaled by a parameter $p$, in order to incorporate the bending of the isoline into DA by fiat (see Fig.~\ref{fig:profile_sketch}). This translates to a relation $\tilde{\Xi}(\theta)$ in Eq.~\eqref{eq:DAext_janus_step_rel} for the rescaled step position $\tilde{\Xi}$, with $p$ as the only parameter. The model fits well to the discrete set $\{\Xi,\vartheta\}$ determined from the MFT results for $p\approx 1/4$, roughly independent of temperature. This corresponds to placing the fictional, rescaled colloid surface halfway between the physical surface and the isoline $\phi(\vec{r})=0$ as depicted in Fig.~\ref{fig:profile_sketch}.
The improvement achieved using the relation $\tilde{\Xi}(\theta)$ in Eq.~\eqref{eq:DAext_janus_step_rel} with $p=1/4$ is demonstrated in Fig.~\ref{fig:Kcyl_step_to_janus_mapped}, where the two scaling functions $K^{(cc)}_{\symbCS}$ and $K^{(cc)}_{\symbJcH}$ for a homogeneous cylinder close to a chemically stepped substrate and a Janus cylinder close to a homogeneous substrate, respectively, coincide even for $\Delta=1$, far away from the DA limit $\Delta\to0$. Thus, the correspondence between these two configurations holds also within MFT, albeit this is based on a relation $\tilde{\Xi}(\theta)$ which differs from the one obtained by using the original, unmodified DA.

The correspondence of Janus particles and chemical steps on a substrate is also relevant for Section \ref{sec:janus_vs_steps} C which discusses the scaling function of the force between a Janus cylinder and a substrate with a chemical step. The MFT scaling function in Fig.~\ref{fig:cyl_da_mft_comparison} for a Janus cylinder and a step at a lateral position $X=0$ is qualitatively similar to the dependence of the scaling function of the force between two patterned substrates on a lateral shift \cite{Sprenger:2006}. This configuration reveals a deficiency of DA: For an orthogonal orientation of the Janus particle, i.e., when the Janus equator faces the substrate at $\vartheta=\pm\pi/2$, within DA the scaling function of the force exhibits cusplike extrema of attraction or repulsion as a function of the particle orientation, whereas the MFT results are smooth. However, for $\Delta=1/5$, i.e., close to the DA limit $\Delta\to 0$, the agreement between DA and full MFT is surprisingly good even for this pathological case (see fig.~\ref{fig:cyl_da_mft_comparison}(b)).

In Sec. \ref{sec:janus_cyls} we have used DA in order to obtain the scaling function of the force between two parallel Janus cylinders (Eqs.~\eqref{eq:cyl_k_separation}--\eqref{eq:cyl_deltaK}). With a view on the experimental interest in Janus particles, in Fig.~\ref{fig:cyl_K_da} the results for the corresponding scaling function of the force are given in $d=3$ spatial dimensions. We find that the force between two Janus cylinders can be attractive and repulsive, depending on their orientations. The strongest attraction is found in the case of the two Janus cylinders facing each other with the same face, whereas the strongest repulsion occurs when they are orientated in line. The force is rather insensitive against tilts out of these two configurations.
Based on the scaling function of the force we have also determined the scaling function of the effective pair potential between two Janus cylinders (Eqs.~\eqref{eq:cyl_theta_separation}--\eqref{eq:cyl_deltaSmallTheta}). In Fig.~\ref{fig:cyl_pot_janus} we present it as an energy landscape in terms of the particle orientations $\vartheta_1$ and $\vartheta_2$. There are two shallow and stable minima in the potential energy, which are connected by a narrow trough representing counter-rotating orientations of the Janus particles. The large plateaus of repulsive orientational states corresponding to opposing BC yield a checkerboard landscape pattern.

Similarly, in Sec. \ref{sec:janus_spheres} we have derived the scaling function of the force between two Janus spheres in a relative coordinate system as a function of three spherical coordinates $\alpha=\phi_2-\phi_1$, $\vartheta_1$, and $\vartheta_2$ (Fig.~\ref{fig:spherical_vectors} and Eqs.~\eqref{eq:sp_k_separation}--\eqref{eq:DeltaK_sp_janus_DA}). The details of this derivation, accounting for all possible orientations, are provided in Appendix \ref{sec:app_janus_spheres}. The result is shown in Fig.~\ref{fig:K_sp_da}, which is rather similar to the case of two Janus cylinders. The scaling function of the force between two Janus spheres decays faster as function of $\Theta$ and, for the same tilt out of the attractive (repulsive) configuration of the particles facing each other (orientated in line), the force is less attractive (less repulsive) compared to the same configuration for two cylinders.
Also here, the force has been integrated in order to determine the scaling function of the effective pair potential between two Janus sphere. Since it is a function of three spherical coordinates, one cannot visualize, within a single plot, the full dependences of the potential. In Fig.~\ref{fig:sp_pot_janus}, we show the energy landscape for the two cases $\alpha = 0$ and $\alpha=\pi$. The free energy landscape is qualitatively similar to that in Fig.~\ref{fig:cyl_pot_janus} for two Janus cylinders. For spheres and $\alpha=\pi$, the two orientation vectors (and the axis connecting the centers of the particles) form the same plane as in the case of $\alpha = 0$, thus the two configurations $\alpha=0$ and $\alpha=\pi$ correspond to $\vartheta_1>0$ and $\vartheta_1<0$ for cylinders, respectively. For $0<\alpha<\pi$ the scaling function of the effective potential varies primarily only around orientations $\vartheta_i=\pi/2$ for the two particles $i=\{1,2\}$ (Fig.~\ref{fig:sp_pot_janus}). However, the pronounced plateau structure is largely unaffected by changes of $\alpha$.

We have used the scaling function of the effective potential in order to address the special experimental situation in which the particle positions and orientations are confined to a plane parallel to the planar surface of a substrate, however such that the substrate does not alter the pair interaction among the particles. Using the full pair interaction potential, we have analyzed how the effective influence of the substrate, incorporated as an externally imposed common tilt $\varphi$ of all Janus particles, changes the effective pair interaction among the Janus particles. The deviations turn out to be small for tilts $\varphi \lesssim30^\circ$ and still acceptable for $\varphi \lesssim 45^\circ$ (Fig.~\ref{fig:sp_scalfunc_tilted}). Under this condition, concerning the interaction among the particles the substrate induced interaction can be discarded.

Thus our findings are to a certain extent compatible with the on-off ``bond-like'' interaction underlying the popular Kern-Frenkel model \cite{Kern:2003}. However, so far here we have discussed only the orientational part of the interaction at fixed spatial distance between the particles. Whereas the Kern-Frenkel model is based on a short-ranged square-well potential, close to $T_c$ the critical Casimir interaction is long-ranged. Furthermore, the critical Casimir potential carries both attractive and repulsive contributions. Since the repulsion is stronger than the attraction, less than half of all configurations are actually attractive (see Fig.~\ref{fig:sp_pot_janus}), despite the overall Janus character.
All these aspects contribute to the thermodynamic properties of suspensions of Janus particles with a critical solvent via integrals of the effective potential over both orientations and the radial distance.

Future studies could focus more on identifying those features of the effective interactions which are unique to the critical Casimir effect. Upon approaching the critical point $(T\to T_c)$, the scaling function of the critical Casimir force increases non-monotonically (see Figs.~\ref{fig:cyl_K_da} and \ref{fig:K_sp_da}), and close to $T_c$ the repulsive contributions become much stronger than the attractive ones. Additionally, the range of interaction increases significantly near $T_c$ and diverges at $T_c$. Considering an actual suspension of Janus particles, the critical Casimir interaction competes with other effects such as electrostatic repulsion and van-der-Waals attraction. However, only the critical Casimir part is singular as a function of $t=(T-T_c)/T_c$ \cite{Dantchev:2007}. The tuneable range of the critical Casimir interaction has profound effects on the aggregation behavior of chemically homogeneous colloids \cite{Mohry:2012, Hobrecht:2015, Edison:2015, Nguyen:2016}. It is expected that this holds also, maybe a fortiori, for Janus particles, deserving both experimental and theoretical investigations. On the other hand, for simulations such as molecular dynamics, in practice, it would be beneficial to employ a computationally more efficient model of the pair interaction than the one provided here. However, reducing the complexity of the pair interaction between Janus particles while keeping its distinguishing features intact poses a significant challenge which requires further work.
\section{Acknowledgments}
M.L.L. would like to thank Ludger Harnau and Matthias Tr\"{o}ndle for helpful discussions.
\appendix
\section{Special cases for the force between two Janus cylinders}\label{sec:app_janus_cyls}
(i)  
In the limit of both $\vartheta_1\to 0$ and $\vartheta_2\to 0$, one has $\Xi_1=\Xi_2=\Delta^{-1/2}$, so that in Eq.~\eqref{eq:cyl_deltaK} both terms involving $\Delta k^{(cc)}$ have the same values of their arguments and according to Eq.~\eqref{eq:cyl_deltaSmallK} one finds $\Delta k^{(cc)}(\Delta^{-1/2},\Delta,\Theta) = 0$, so that
\begin{equation}
\Delta K_{\oslash\oslash}^{(cc)}(\vartheta_1=0,\vartheta_2=0,\Delta,\Theta) = 0.
\end{equation}
This correctly resolves the issue that $\mathrm{sign}(\vartheta_1\,\vartheta_2)$ in Eq.~\eqref{eq:cyl_deltaK} depends on the direction of the limit, i.e., whether $\vartheta_{1,2}\to 0^+$ or $\vartheta_{1,2}\to 0^-$, but the resulting force should not. With this, the scaling function of the force is given by (see Eqs.~\eqref{eq:cyl_force_separation} and \eqref{eq:cyl_k_separation})
\begin{equation}
K_{\symbJcJc}^{(cc)}(\vartheta_1=0, \vartheta_2=0, \Delta, \Theta)=K_{(+,-)}^{(cc)}(\Delta, \Theta)>0,
\end{equation} 
as expected.

If instead $\vartheta_1=0$ and $\vartheta_2=\pi$, one has $\Xi_1=\Delta^{-1/2}=-\Xi_2$. Similarly, this leads to $\Delta k^{(cc)}(\pm\Delta^{-1/2},\Delta,\Theta) = 0$ (Eq.~\eqref{eq:cyl_deltaSmallK}) and $\Delta K_{\oslash\oslash}^{(cc)}(\vartheta_1=0,\vartheta_2=\pi,\Delta,\Theta) = 0$. According to Eq.~\eqref{eq:cyl_force_separation}, the scaling function in Eq.~\eqref{eq:cyl_k_separation} has to be evaluated for the second case of $\Xi(\vartheta_1)\Xi(\vartheta_2)<0$, which reduces to
$K_{\oslash\oslash}^{(cc)}(\vartheta_1=0, \vartheta_2=\pi, \Delta, \Theta)=K_{(+,+)}^{(cc)}(\Delta, \Theta)<0$.

(ii)
The case $\vartheta_1=\vartheta_2=\pi/2$ corresponds to $\Xi_{1,2}=0$ and Eq. \eqref{eq:cyl_deltaK} reduces to (see Eq.~\eqref{eq:cyl_deltaSmallK})
\begin{multeqline}
 \Delta K_{\oslash\oslash}^{(cc)}(\vartheta_1=\pi/2, \vartheta_2=\pi/2, \Delta, \Theta)
\ifTwocolumn \\ \fi 
=\int_{1}^{1+\Delta^{-1}}\upd\alpha\,\frac{\Delta k(\alpha \Theta)}{\alpha^{d}\sqrt{\alpha-1}}.
 \label{eq:app_DeltaK_pi2pi2}
\end{multeqline}
In this case, a careful analysis of $\Xi(\vartheta_1)\,\Xi(\vartheta_2)\gtrless 0$, which appears in Eq.~\eqref{eq:cyl_force_separation}, is required. If both $\vartheta_1\to\pi/2$ and $\vartheta_2\to\pi/2$ approach the limit from the same direction (both from above, or both from below), one has $\Xi(\vartheta_1)\,\Xi(\vartheta_2)\to 0^+$ and the total scaling function in Eq.~\eqref{eq:cyl_k_separation} between such Janus cylinders reduces to (compare Eq.~\eqref{eq:app_DeltaK_pi2pi2}, where $\Delta k=k_{(+,+)}-k_{(+,-)}$, with Eq.~\eqref{eq:cyl_K_hom})
\ifTwocolumn
\begin{multline}
  K_{\symbJcJc}^{(cc)}(\vartheta_1\to\pi/2, \vartheta_2\to\pi/2, \Delta, \Theta) \\
 = K_{(+,-)}^{(cc)}(\Delta, \Theta) + \Delta K_{\oslash\oslash}^{(cc)}(\vartheta_1\to\pi/2, \vartheta_2\to\pi/2, \Delta, \Theta)\\
 = K_{(+,+)}^{(cc)}(\Delta, \Theta)<0.
  \label{eq:app_K_pi2pi2}
\end{multline}
\else
\begin{align}
  K_{\symbJcJc}^{(cc)}(\vartheta_1\to\pi/2, \vartheta_2\to\pi/2, \Delta, \Theta)&=K_{(+,-)}^{(cc)}(\Delta, \Theta) + \Delta K_{\oslash\oslash}^{(cc)}(\vartheta_1\to\pi/2, \vartheta_2\to\pi/2, \Delta, \Theta) \nonumber\\
  &= K_{(+,+)}^{(cc)}(\Delta, \Theta)<0.
  \label{eq:app_K_pi2pi2}
\end{align}
\fi
If, however, $\vartheta_1\to\pi/2^+$ and $\vartheta_2\to\pi/2^-$, or vice versa, one has $\Xi(\vartheta_1)\Xi(\vartheta_2)\to 0^-$; thus the second case in Eq.~\eqref{eq:cyl_force_separation} maps $\vartheta_1=\pi/2+\epsilon$ (where $\epsilon\to 0^+$) to $\hat{\vartheta}_1=-\pi/2+\epsilon$. The excess scaling function of the force $\Delta K_{\oslash\oslash}^{(cc)}(\Theta, \Delta, \vartheta_1=-\pi/2, \vartheta_2=\pi/2)$ in Eq.~\eqref{eq:cyl_deltaK} vanishes because $\mathrm{sign}(\vartheta_1\,\vartheta_2)=-1$ and one finds $K_{\symbJcJc}^{(cc)}(\vartheta_1\to\pi/2^+, \vartheta_2\to\pi/2^-, \Delta, \Theta) = K_{(+,+)}^{(cc)}(\Delta, \Theta)$ as in Eq.~\eqref{eq:app_K_pi2pi2}.
Therefore, the two sided limit exists and the force in this configuration is attractive for all temperatures $t>0$.

(iii)
Similarly, for $\vartheta_1\to-\pi/2^+$ and $\vartheta_2\to\pi/2^-$, one also has $\Xi(\vartheta_1)\,\Xi(\vartheta_2)\to 0^+$ and $\Delta K_{\oslash\oslash}^{(cc)}(-\pi/2, \pi/2, \Delta, \Theta)=0$. Thus, the scaling function between the Janus cylinders in this case is
\begin{equation}
  K_{\symbJcJc}^{(cc)}(\vartheta_1=-\pi/2, \vartheta_2=\pi/2, \Delta, \Theta)=K_{(+,-)}^{(cc)}(\Delta, \Theta)>0,
\end{equation}
i.e., the force in this configuration is repulsive for all temperatures $t>0$.

(iv)
The last limit we discuss is $\vartheta_1=\pm\pi/2$ and $\vartheta_2=0$, which implies $\Xi_{1}=0$ and $\Xi_{2}=\Delta^{-1/2}$. Note that in this case, the numbering of the particles \numcirc{1} and \numcirc{2} is not interchangeable due to the restriction $|\cos\vartheta_1|<|\cos\vartheta_2|$. 
Since $\Delta k^{(cc)}(\Xi_{2}=\Delta^{-1/2},\Delta,\Theta) = 0$, only the first term in Eq.~\eqref{eq:cyl_deltaK} contributes, thus with $\Delta k=k_{(+,+)}-k_{(+,-)}$ resulting in
\ifTwocolumn
\begin{align}
 \Delta K_{\oslash\oslash}^{(cc)}&(\vartheta_1=\pm\pi/2, \vartheta_2=0, \Delta, \Theta) \nonumber\\
&=\frac{1}{2} \int_{1}^{1+\Delta^{-1}}\upd\alpha\,\frac{\Delta k(\alpha \Theta)}{\alpha^{d}\sqrt{\alpha-1}}\\
&= \frac{1}{2}\left(K_{(+,+)}^{(cc)}(\Delta, \Theta) - K_{(+,-)}^{(cc)}(\Delta, \Theta)\right). \nonumber
\end{align}
\else
\begin{align}
 \Delta K_{\oslash\oslash}^{(cc)}(\vartheta_1=\pm\pi/2, \vartheta_2=0, \Delta, \Theta) &= \frac{1}{2} \int_{1}^{1+\Delta^{-1}}\upd\alpha\,\frac{\Delta k(\alpha \Theta)}{\alpha^{d}\sqrt{\alpha-1}}\\
&= \frac{1}{2}\left(K_{(+,+)}^{(cc)}(\Delta, \Theta) - K_{(+,-)}^{(cc)}(\Delta, \Theta)\right). \nonumber
\end{align}
\fi
Regardless of the sign of $\vartheta_1=\pm\pi/2$ as well as whether $\Xi(\vartheta_1)\Xi(\vartheta_2)\to 0^\pm$, for these orientations due to Eqs.~\eqref{eq:cyl_force_separation} and \eqref{eq:cyl_k_separation} the scaling function of the force between two Janus cylinders reduces to the mean value of the attractive and the repulsive force between homogeneous cylinders:
\begin{multeqline}
  K_{\symbJcJc}^{(cc)}(\vartheta_1=\pm\pi/2, \vartheta_2=0, \Delta, \Theta)
 \ifTwocolumn \\ \fi 
 =\frac{1}{2}\left(K_{(+,+)}^{(cc)}(\Delta, \Theta) + K_{(+,-)}^{(cc)}(\Delta, \Theta)\right).
\end{multeqline}
\section{Derjaguin approximation for the force between two Janus spheres}
\label{sec:app_janus_spheres}
Concerning the geometry of two homogeneous, i.e., isotropic spheres, the Derjaguin approximation consists of subdividing their surfaces into infinitesimal thin rings of area $2\pi\rho\,\upd\rho$, parameterized by their radius $\rho$ \cite{Gambassi:2009}. This has been used successfully in several studies, such as Refs. \cite{Hanke:1998, Gambassi:2009, Troendle:2010}, generally in conjunction with the so-called ``parabolic distance approximation'' for the local distance $L(\rho)$ between surface elements of the two colloids:
\begin{equation}
L(\rho) = D+2 R - 2\sqrt{R^2-\rho^2}\approx D\left(1+\frac{\rho^2}{R D}\right).
\end{equation}
Building on that, for Janus spheres the corresponding step in BC has to be incorporated additionally, depending on the particle orientations.
Within DA, the overlap of pairs of surface elements on both spheres is determined after the projection along the vector $\vec{r}_{12}$ connecting the centers of the two spheres. We choose to express this geometry in terms of a local coordinate system, the $z$ axis of which passes through the centers of the two colloids, so that $\vec{r}_{12} = (D + 2 R)\,\hat{\vec{r}}_{12}$ with $\hat{\vec{r}}_{12} = (0,0,1)$ (see Fig.~\ref{fig:spherical_vectors}).
The orientations of the colloids can be represented by orientation vectors $\vec{n}_1$ and $\vec{n}_2$, which can be chosen to point either into the direction of the $(+)$ (red) or the $(-)$ (blue) side. As far as the figures in the main text are concerned, the orientation vector is chosen to point towards the $(-)$ (blue) cap. However, regarding the general approach in the present appendix, we shall use the more abstract notions of ``north'' and ``south'', which are supposed to underscore the arbitrariness of this choice.

Without loss of generality, we define the coordinate system such that the orientation of the first particle has an azimuthal angle $\phi_1=0$ and a polar angle $\vartheta_1$; the orientation $(\alpha, \vartheta_2)$ of the second particle is taken relative to the ``prime meridian'' of the first (i.e., $\alpha=\phi_2 -\phi_1$). Rotations of the coordinate system while keeping $(\alpha,\vartheta_1,\vartheta_2)$ fixed do not change the interaction between the particles. Still, there remains a choice in the numbering of the particles. We implement this such that $|\cos\vartheta_1| < |\cos\vartheta_2|$, as it shortens the notation below; otherwise one can exchange the labels $(1)$ and $(2)$ and rotate the frame of reference around the $y$ axis by $180^\circ$ (see Fig.~\ref{fig:spherical_vectors}).

\begin{figure}[t!]
  \centering
  \includegraphics[height=7.35cm]{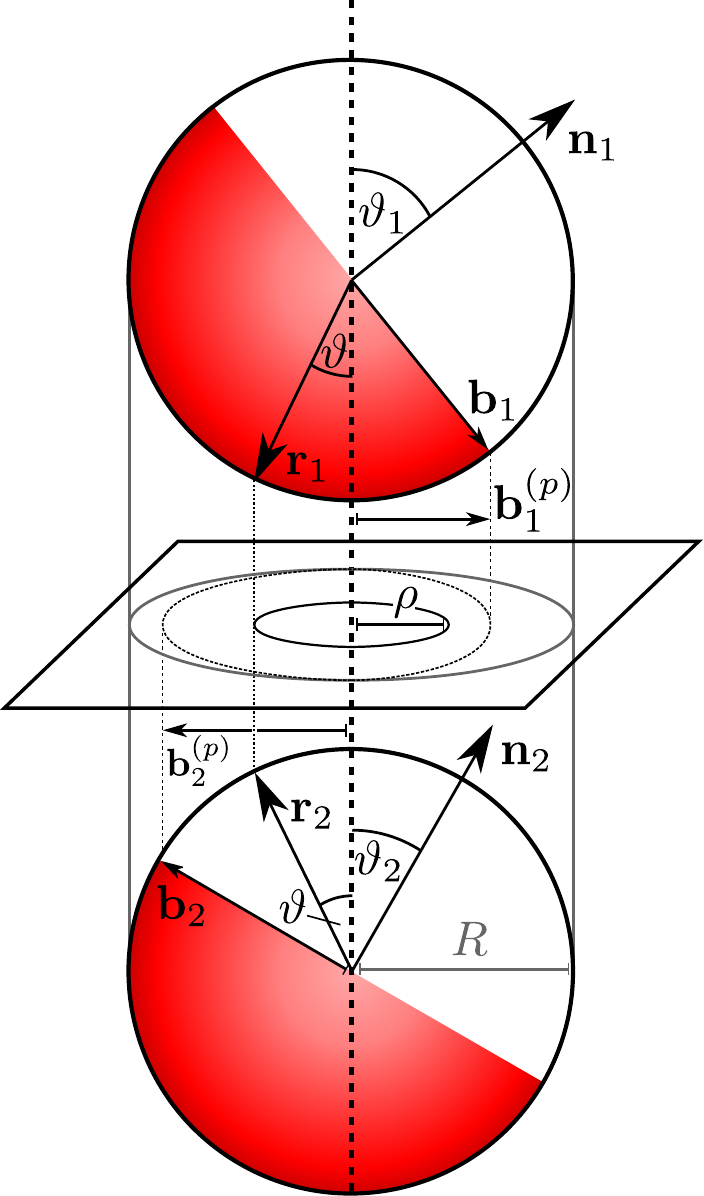}
  \caption{Two Janus spheres within the Derjaguin approximation. The two orientations of the two particles are given by the direction vectors $\vec{n}_1$ and $\vec{n}_2$ which are normals of the respective equatorial planes. In the relative coordinate system, given by the axis through the centers of both particles, the orientations can be represented by the two polar angles $\vartheta_1$ and $\vartheta_2$ and the relative azimuthal angle $\alpha$; for simplicity, here we depict the case $\alpha=0$ in which the two equatorial planes are rotated with respect to each other but not tilted (see Fig.~\ref{fig:spherical_vectors} for a reduced schematic drawing with $\alpha\neq 0$). A pair of surface elements at $\vec{r}_1(\phi,\vartheta)$ and $\vec{r}_2(\phi,\vartheta)$ on the two Janus spheres, such that they are mirror images of each other, i.e., $(\vec{r}_1)_z=-(\vec{r}_2)_z$, share the same ``northern'' BC if $\vec{r}_1\cdot\vec{n}_1 > 0$ and $\vec{r}_2\cdot\vec{n}_2 > 0$. Likewise, two surface elements share the same ``southern'' BC if $\vec{r}_1\cdot\vec{n}_1 < 0$ and $\vec{r}_2\cdot\vec{n}_2 < 0$; otherwise for the selected pair of surface elements the BC on the two Janus spheres differ. Surface elements at $\vec{r}_1$ and $\vec{r}_2$ with equal distance between them (dotted line parallel to the axis through the centers of both particles and connecting the tips of $\vec{r}_1$ and $\vec{r}_2$) form a ring with radius $\rho=R\cos\vartheta$ (here, the inner black circle) which is shown in the midplane between the particles. The equatorial steps of the Janus spheres \numcirc{1} and \numcirc{2} form half-ellipses when projected onto the same midplane. The vectors $\vec{b}_1$ and $\vec{b}_2$ lie in the equatorial plane of the corresponding particles and thus are orthogonal to $\vec{n}_1$ and $\vec{n}_2$, respectively. Their direction is chosen to point to that point on each equator which is closest in sight of the opposite particle. The projections $\vec{b}_1^{(p)}$ and $\vec{b}_2^{(p)}$ of the vectors $\vec{b}_1$ and $\vec{b}_2$, respectively, onto the midplane render the semi-minor axes of the half-ellipses.}
  \label{fig:sp_da_projection}
\end{figure}

The orientations $\vec{n}_{1,2}$ and two mirror points $\vec{r}_1$ and $\vec{r}_2$ on the surface of colloid $1$ and $2$, respectively, are parameterized within the relative coordinate system by
\begin{align}
\vec{n}_1 &= \begin{pmatrix}\sin\vartheta_1 \\ 0 \\ \cos\vartheta_1\end{pmatrix},&\quad
\vec{n}_2 &= \begin{pmatrix}\cos\alpha\sin\vartheta_2 \\ \sin\alpha\sin\vartheta_2 \\ \cos\vartheta_2\end{pmatrix},\label{eq:app_sp_n_vecs}\\
\vec{r}_1 &= R\begin{pmatrix}\cos\phi\sin\vartheta \\ \sin\phi\sin\vartheta \\ -\cos\vartheta\end{pmatrix},&\quad
\vec{r}_2 &= R\begin{pmatrix}\cos\phi \sin\vartheta\\ \sin\phi \sin\vartheta\\ \cos\vartheta\end{pmatrix},
\end{align}
where $\vartheta_1$ is the polar angle of the first particle, $(\alpha, \vartheta_2)$ are the azimuthal and polar angle of the second particle, and $(\phi, \vartheta)$ are the spherical coordinates of the vectors $\vec{r}_{1}$ and $\vec{r}_{2}$ of a pair of surface elements, where $\vec{r}_1$ and $\vec{r}_2$ are mirror images of each other with respect to the midplane orthogonal to $\hat{\vec{r}}_{12}=\vec{e}_z$, such that $(\vec{r}_1)_z=-(\vec{r}_2)_z$ (see Fig. \ref{fig:sp_da_projection}). After the projection into the midplane by using the orthogonal projection matrix
\begingroup
\begin{equation}
\mathbf{P}_{z}=\begin{pmatrix}
1 & 0 & 0 \\ 
0 & 1 & 0 \\ 
0 & 0 & 0
\end{pmatrix}, 
\end{equation}
\endgroup
surface elements with equal distance from their mirror element on the other particle form a ring with polar coordinates $(\rho=R\sin\vartheta, \phi)$ and a fixed value of $\vartheta$.

The force between the Janus spheres, as constructed within DA, depends on the combination of BC for a pair of surface elements. A selected pair of of surface elements will share the ``northern'' BC if $\vec{r}_1\cdot\vec{n}_1 > 0\ \text{and}\ \vec{r}_2\cdot\vec{n}_2 > 0$. Likewise, they will both have the ``southern'' BC if $\vec{r}_1\cdot\vec{n}_1 < 0\ \text{and}\ \vec{r}_2\cdot\vec{n}_2 < 0,$
otherwise the surface elements have different BCs.

In our parameterization and with $f_1(\phi) \mathrel{\mathop:}= \vec{r_1}\cdot\vec{n_1}=-\cos\vartheta\cos\vartheta_1+\cos\phi\sin\vartheta\sin\vartheta_1$ and $f_2(\phi) \mathrel{\mathop:}= \vec{r_2}\cdot\vec{n_2}=\cos\vartheta\cos\vartheta_2+\cos(\alpha - \phi)\sin\vartheta\sin\vartheta_2$, the two conditions above read
\begin{subequations}
\label{eq:sphere_da_BC_cond}
\begin{equation}
\text{same BC (``north'')}\Leftrightarrow\ f_1(\phi) > 0\ \land\ f_2(\phi) > 0\hphantom{.} \text{\hspace{0.2em} or}\vspace{-0.6em}
\end{equation}
\begin{equation}
\text{same BC  (``south'')}\Leftrightarrow\ f_1(\phi)  < 0\ \land\ f_2(\phi) < 0.\hphantom{\text{\hspace{0.2em} or}}
\end{equation}
\end{subequations}
There are two more conditions representing opposing BC, with opposite signs of $f_1(\phi)\gtrless 0$ and $f_2(\phi)\lessgtr 0$. For any value of $\phi$, one and only one of these four conditions is fulfilled.
Thus, these four conditions hold in four intervals. Determining the zeroes of $f_1$ and $f_2$ as functions of $\phi$ renders four possible values, separating the intervals (note that four points naturally enclose three closed intervals, and one more interval due to the periodicity in $\phi$):
\begin{subequations}
\begin{numcases}{f_1(\phi)=0\ifTwocolumn\Rightarrow\hspace{-0.25em}\else\quad\Rightarrow\quad\fi}
\ifTwocolumn\hspace{-0.25em}\fi\phi_1 = \arccos(\cot\vartheta\cot\vartheta_1),\\
\ifTwocolumn\hspace{-0.25em}\fi\phi_2 =  - \arccos(\cot\vartheta\cot\vartheta_1)\ifTwocolumn\hspace{0.3em}\else\quad\fi(+ 2\pi);\hphantom{-\alpha;}
\end{numcases}
\begin{numcases}{f_2(\phi)=0\ifTwocolumn\Rightarrow\hspace{-0.25em}\else\quad\Rightarrow\quad\fi}
\ifTwocolumn\hspace{-0.25em}\fi\phi_3 = \alpha - \arccos(-\cot\vartheta\cot\vartheta_2) \ifTwocolumn\hphantom{\hspace{0.3em}(+ 2\pi),}\\
\hspace{12em}(+ 2\pi),\nonumber \else \quad (+ 2\pi), \fi\\
\ifTwocolumn\hspace{-0.25em}\fi\phi_4 = \alpha + \arccos(-\cot\vartheta\cot\vartheta_2).
\end{numcases}
\label{eq:sphere_phi_roots}
\end{subequations}

Strictly speaking, Eq. \eqref{eq:sphere_da_BC_cond} has an infinite number of solutions, because any solution shifted by $\pm 2\pi$ is also a solution. With $(+2\pi)$ we indicate that $\phi_2$ and $\phi_3$ may need to be shifted such that all four given solutions are the relevant ones within the principal interval $[0,2\pi]$.

\begin{figure}[t!]
  \centering
  \begin{minipage}[t]{3.25cm}\raggedright (a)\\\vspace{-0.5em}\includegraphics[width=3.25cm]{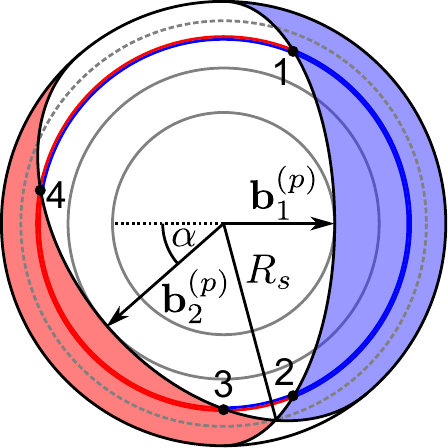}\end{minipage}\hspace{0.75cm}
  \begin{minipage}[t]{3.25cm}\raggedright (b)\\\vspace{-0.5em}\includegraphics[width=3.25cm]{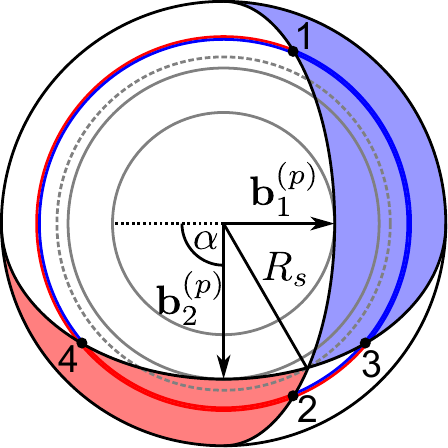}\end{minipage}
  \vspace{2em}\\
  \begin{minipage}[t]{3.25cm}\raggedright (c)\\\vspace{-0.5em}\includegraphics[width=3.25cm]{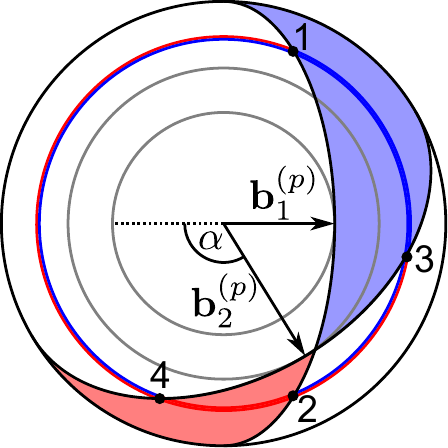}\end{minipage}\hspace{0.75cm}
  \begin{minipage}[t]{3.25cm}\raggedright (d)\\\vspace{-0.5em}\includegraphics[width=3.25cm]{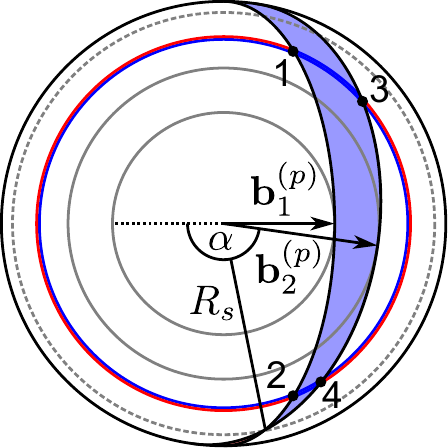}\end{minipage}
  \caption{A top-down plan view, as rendered by the projection matrix $\mathbf{P}_{z}$, of the geometry of two Janus spheres, which is the same as in Fig.~\ref{fig:sp_da_projection}, but highlights the significance of the angles $\phi_1$ to $\phi_4$ of the DA procedure given in Eq.~\eqref{eq:sphere_phi_roots}. The two half-elliptical curves running through $1$ and $2$ and through $3$ and $4$, respectively, represent the projection of the Janus equators onto the midplane. Their semi-minor axes are given by the projections $\vec{b}_1^{(p)}$ and $\vec{b}_2^{(p)}$ of the vectors $\vec{b}_1$ and $\vec{b}_2$, respectively, shown in Fig.~\ref{fig:sp_da_projection}, which enclose the angle $\pi-\alpha$. The full gray circles have radii $|\vec{b}_i^{(p)}|$. Here, the parameters of the particle orientations $\vec{n}_{1,2}\ (\perp\vec{b}_{1,2})$ are $\vartheta_1=\pi/3$ and $\vartheta_2 = \pi/4$, and $\alpha$ is varied from (a) $\alpha=0.7$, (b) $\alpha=\pi/2$, and (c) $\alpha=2.1$ to (d) $\alpha=3.0$.
In this projection, two surface elements forming a pair at $\vec{r}_1$ and $\vec{r}_2$ lie on top of each other, rendering a single point within the circular area.
The projected area, indicated in blue, corresponds to those pairs of surface elements which share the ``northern'' BC. Likewise, the projected area within which both surface elements feature the ``southern'' BC is indicated in red. The white areas correspond to pairs of surface elements with opposite BC. 
As a function of $\phi$ and for a fixed value of $\vartheta$, in projection the pairs of surface elements form a ring of radius $\rho=R\sin(\vartheta)$ (see Fig.~\ref{fig:sp_da_projection}). We depict the case $\vartheta=1$ so that $\rho=0.84\,R$ (color-coded ring). The points $1$ to $4$ mark the intersections of the color-coded ring with the projected equatorial steps of the BC, which are given by the polar coordinates $(\rho, \phi_1)$ through $(\rho, \phi_4)$. Both the thick red and the thick blue arcs of this ring represent equal BC on both particle surfaces, whereas those arcs being half blue and half red correspond to opposite BC. Additional explanations, such as the meaning of $R_s$, are given in the main text.}
    \label{fig:sphere_proj}
\end{figure}

Figure~\ref{fig:sphere_proj} puts the meaning of these four values of $\phi$ given by Eq. \eqref{eq:sphere_phi_roots} into proper perspectives. 
Figure~\ref{fig:sphere_proj} shows a schematic (top-down) plan view of the geometry shown in Fig.~\ref{fig:sp_da_projection} which is rendered by the projection matrix $\mathbf{P}_{z}$ for four different values of $\alpha$ and with additional details, visualizing how the projected surface elements entering the DA are partitioned by Eq.~\eqref{eq:sphere_phi_roots} (compare also Fig.~\ref{fig:spherical_vectors}). 
The spherical colloids are drawn with non-occluding outlines and the equatorial step is indicated only partially.
The projection of the equatorial steps between the ``north'' and the ``south'' Janus BC on each sphere results in two ellipses.
This follows from noting that the two equators can be parameterized as circles $\mathbf{p}_i=(\cos\phi_i, \sin\phi_i, 0)$, tilted by a rotation matrix
\begin{equation}
\mathbf{R}_i=\begin{pmatrix}
1 & 0 & 0 \\ 
0 & \cos\vartheta_i & -\sin\vartheta_i \\ 
0 & \sin\vartheta_i & \cos\vartheta_i
\end{pmatrix}.
\end{equation}
One finds that $\mathbf{P}_{z}\cdot\mathbf{R}_i\cdot\mathbf{p}_i=(\cos\phi_i, \cos\vartheta_i\sin\phi_i, 0)$ fulfills the ellipse equation $\frac{x^2}{a^2}+\frac{y^2}{b^2}=1$ for $a=1$ and $b=|\cos\vartheta_i|$.
Of the two elliptical projections, we draw only that half facing the other colloid, resulting in two half-elliptical curves, which are intersecting for $0<\alpha<\pi$ (i.e., they do not intersect for $\alpha=0$ and $\alpha=\pi$).
The semi-minor axes of the half-ellipses are indicated by the projections $\vec{b}_1^{(p)}$ and $\vec{b}_2^{(p)}$ of the vectors $\vec{b}_1$ and $\vec{b}_2$, respectively, which have a projected length of $R \cos\vartheta_{1,2}$ and form the angle $\pi-\alpha$ between them.
The projected Janus steps divide the circular area of radius $R$ into four regions (blue, white, red, white); a selected ring of fixed radius $\rho = R\sin\vartheta$ (corresponding to the color colored circle in Fig.~\ref{fig:sphere_proj}) is divided into four arcs by points with the polar coordinates $(\rho,\phi_1)$ to $(\rho,\phi_4)$. In the case of small $\alpha$ as shown in Fig.~\ref{fig:sphere_proj}(a), the numbering of the values $\phi_1$ to $\phi_4$ given in Eq.~\eqref{eq:sphere_phi_roots} corresponds to a clockwise counting of the intersections of the ring with the projected Janus steps (i.e., the half-ellipses). However, the order of their occurrence changes upon increasing $\alpha$ towards $\pi$ (see Figs.~\ref{fig:sphere_proj}(a)--(d)).

Within DA, the force due to each ring of surface elements of equal distance between them is proportional to its arc length and to the force between parallel walls corresponding to the respective combination of the BC. In Fig.~\ref{fig:sphere_proj}(a), the blue curve, representing a common ``northern'' BC, has an arc length of $(\phi_2 - \phi_1)\rho$, whereas the red arc represents a common ``southern'' BC with an arc length of $(\phi_4 - \phi_3)\rho$. In this case, using the relation $\arccos(-x)=\pi-\arccos(x)$, the total arc length of equal BC amounts to $\left[2\pi - 2\arccos\left(\cot\vartheta\cot\vartheta_1\right) - 2\arccos\left(\cot\vartheta\cot\vartheta_2\right)\right]\rho$.

The number and the order of the intersections between a ring of equidistant surface element pairs and the projected Janus equators depends on the radius of the ring. For $\rho < R\cos\vartheta_1$ (the inner gray circle in Fig.~\ref{fig:sphere_proj} indicates $\rho=R\cos\vartheta_1$), the ring does not cross the projected steps in BC at all. For $R\cos\vartheta_1<\rho<R \cos\vartheta_2$, there are two points of intersection (we recall that the labels \numcirc{1} and \numcirc{2} are chosen such that $|\cos\vartheta_1|<|\cos\vartheta_2|$). Starting from $\rho=R \cos\vartheta_2$ (indicated by the outer gray circle), for $\rho>R\cos\vartheta_2$ there are four points of intersection. However, at a specific radius $\rho=R_s$ (gray dashed line in Fig.~\ref{fig:sphere_proj}), the two half-ellipses intersect and the order of the values $\phi_1\ldots\phi_4$ changes (e.g., compare the order of the intersections in Figs.~\ref{fig:sphere_proj}(a) and (b)). 

The dimensionless radius $r_s = R_s/R = \sqrt{x^2+y^2}$ is determined by the intersection point $(x, y)$ of the two semi-ellipses, which is found from a solution of the general problem of the intersection between two co-centric ellipses: the first ellipse $(x/a_1)^2+(y/b_1)^2=1$ and the second ellipse $(x/a_2)^2+(y/b_2)^2=1$ rotated by an angle $\alpha$. Within their parametric representations the intersections follow from
\begin{equation}
	\begin{pmatrix}
		x\\
		y
	\end{pmatrix}
	=
	\begin{pmatrix}
		a_1 \cos t_1\\
		b_1 \sin t_1
	\end{pmatrix}
	\stackrel{!}{=}
	\begin{pmatrix}
		a_2 \cos t_2 \cos\alpha - b_2\sin t_2\sin\alpha\\
		a_2 \cos t_2 \sin\alpha + b_2\sin t_2\cos\alpha
	\end{pmatrix}.
	\label{eq:app_ellipse_intersection_leqs}
\end{equation}
Equation~\eqref{eq:app_ellipse_intersection_leqs} is a system of two equations for the two unknowns $t_1$ and $t_2$, which become functions of $a_1, b_1, a_2, b_{2}$, and $\alpha$.
For the present situation, and with $x$ and $y$ giving rise to a dimensionless factor $\sqrt{x^2+y^2}$ of the radius $R$, the problem reduces to the special case in which the semi-major axes are $a_1=a_2=1$ (i.e., the semi-major axes are touching the circle of radius $R$) and the semi-minor axes are the projected lengths $b_1=\big|\vec{b}_1^{(p)}\big|/R=\lvert\cos(\vartheta_1)\rvert$ and $b_2=\big|\vec{b}_2^{(p)}\big|/R=\lvert\cos(\vartheta_2)\rvert$. While in principle this system of equations can be solved analytically, it is not guaranteed that all solutions are real, because in degenerate cases (e.g., for $\alpha=0$ or $\alpha=\pi$ and $b_1=b_2$,  or $b_1=b_2=1$, or $b_1=b_2=0$) the number of physically acceptable solutions can be less than four. In the non-degenerate cases, out of these four general solutions of the intersection of two ellipses, only one gives the intersection of two half-ellipses. 
We have followed a pragmatic approach by solving Eq. \eqref{eq:app_ellipse_intersection_leqs} numerically within an \textit{a priori} chosen interval of $t_2$ in order to preselect the appropriate solution for the half-ellipses \footnote{This also allows us to use optimized numerical root finding algorithms operating within an interval in which the function changes sign. We have chosen Brent's root finding method, \url{http://mathworld.wolfram.com/BrentsMethod.html}, which is implemented in the SciPy library \url{http://docs.scipy.org/doc/scipy-0.16.0/reference/generated/scipy.optimize.brentq.html}.}.
We note that our definition enforces the relation $|\cos\vartheta_1|<|\cos\vartheta_2|$, so that the dimensionless radius $r_s$ corresponding to the point of intersection between the two half-ellipses is bounded by $|\cos\vartheta_2|\leq r_s \leq 1$, because any point on the second ellipse has a radial distance from its center, the value of which lies between the semi-minor axis $b_2=\cos\vartheta_2$ and the semi-major axis $a_2=1$, and so does the point of intersection.

Using this procedure, we have constructed the force between two Janus spheres within DA by integrating the force between the rings of surface elements of radius $\rho$, with attractive and repulsive contributions proportional to the respective four arc lengths determined by $\phi_1\ldots\phi_4$ in Eq.~\eqref{eq:sphere_phi_roots}, and using the numerically determined radius $R_s=R_s(\alpha, \vartheta_1, \vartheta_2)$ for each configuration, which governs the occurrence of the attractive and repulsive force contributions (depending on $\rho\lessgtr R_s$) by interchanging the order of $\phi_1\ldots\phi_4$. 
A thorough investigation of all geometric configurations reveals that the excess force takes the following form
\ifTwocolumn\begin{widetext}\fi
\begin{align}
\Delta F_{\oslash\oslash}^{(ss)}&(\vec{n}_1, \vec{n}_2, \vec{r}_{12}=(D+2R)\vec{e}_z, R, T) = \frac{k_B T}{D^d} \Bigg[\int_0^{R_s} \upd\rho\,\rho\frac{2\pi\,H\left(\coscos\right)}{(L(\rho)/D)^d}\,\Delta k\left(\frac{L(\rho)}{\xi_\pm}\right) \nonumber\\
&-\mathrm{sign}\left(\coscos\right)\int_{R\cos\vartheta_1}^{R_s} \upd\rho\,\rho\frac{2\arccos\left((\mathrm{sign}(\cos\vartheta_1))(\cot\vartheta)(\cot\vartheta_1)\right)}{(L(\rho)/D)^d}\,\Delta k\left(\frac{L(\rho)}{\xi_\pm}\right) \nonumber\\
&-c(\alpha,\vartheta_1,\vartheta_2)\mathrm{sign}\left(\coscos\right)\nonumber\\
&\quad\quad\quad\times\int_{R\cos\vartheta_2}^{R_s} \upd\rho\,\rho\frac{2 \arccos\left((\mathrm{sign}(\cos\vartheta_2))(\cot\vartheta) (\cot\vartheta_2)\right)}{(L(\rho)/D)^d}\,\Delta k\left(\frac{L(\rho)}{\xi_\pm}\right) \nonumber\\
&+\int_{R_s}^{R} \upd\rho\,\rho\frac{2\alpha}{(L(\rho)/D)^d}\,\Delta k\left(\frac{L(\rho)}{\xi_\pm}\right) \Bigg],
\label{eq:app_DeltaF_sp_janus_DA}
\end{align}
\ifTwocolumn\end{widetext}\fi
with $\rho = R\sin\vartheta$ and $\cot\vartheta = \frac{\cos\vartheta}{\sin\vartheta}=\frac{\sqrt{1-\sin^2\vartheta}}{\sin\vartheta} = \frac{R}{\rho}\sqrt{1-\frac{\rho^2}{R^2}}$, for $\vec{r}_{12}=(D+2 R)\vec{e}_z$ and $\vec{n}_{1,2}$ in relative coordinates (see Eq.~\eqref{eq:app_sp_n_vecs}).
The occurrence of various expressions in Eq.~\eqref{eq:app_DeltaF_sp_janus_DA} can be rationalized as follows: The combined arclength of equal BC is generally of the form $\pm\phi_4\mp\phi_3\pm\phi_2\mp\phi_1$ (i.e., different combinations of the signs). According to Eq.~\eqref{eq:sphere_phi_roots}, additional shifts of $2\pi$ might be required to ensure $\phi_i\in[0,2\pi)$. In fact the term $2\pi$ occurs only for rings of surface elements with radii $\rho < R_s$, provided $\coscos\geq 0$, which is expressed by the limits of integration of the first term in Eq.~\eqref{eq:app_DeltaF_sp_janus_DA} (see below also the note regarding the second and third term).
Similarly, the azimuthal angle $\alpha$ contributes in total as $2\alpha$ to the arclength if $\rho > R_s$, but it does not contribute if $\rho < R_s$, leading to the fourth and last term in Eq.~\eqref{eq:app_DeltaF_sp_janus_DA}.
The second and third term reproduce the functional dependence of the arclength on $\vartheta(\rho)$ and $\vartheta_{1,2}$. The changes of sign of the argument in the $\arccos$ functions in Eq.~\eqref{eq:sphere_phi_roots} generalize to $\mathrm{sign}(\cos\vartheta_{1,2})$ in Eq.~\eqref{eq:app_DeltaF_sp_janus_DA} due to the relation $2\arccos(-x)=2\pi-2\arccos(x)$. Note that the shift of $2\pi$ re-enters the first term; in Eq.~\eqref{eq:app_DeltaF_sp_janus_DA} the first term reflects the notation in the second and third term.
Analogously to the geometry of two Janus cylinders, we find a dependence of the sign of the second and third term on the sign of $\coscos$.
Furthermore, the sign picking function $c(\alpha,\vartheta_1,\vartheta_2)$ is given by
\ifTwocolumn\begin{widetext}\fi
\begin{equation}
	c(\alpha, \vartheta_1, \vartheta_2) = \begin{cases}
	\mathrm{sign}(\cos\alpha), & \text{if } \coscos = 0, \\
	1, & \text{if } \alpha \leq \arccos\left(-(\tan\vartheta_2)\,(\cot\vartheta_1)\right)\leq\pi H\left(\coscos\right)\\
	& \quad \text{or } \pi H\left(\coscos\right)\leq \arccos\left(-(\tan\vartheta_2)\,(\cot\vartheta_1)\right)\leq\alpha,\\
	-1 & \text{otherwise,}
	\end{cases}
\end{equation}
with the restriction that $\alpha$ is replaced by $2\pi-\alpha$ if $\alpha > \pi$.

Finally, the scaling function of the excess force is found from Eq.~\eqref{eq:app_DeltaF_sp_janus_DA} by using the distance function $L(\rho)$ within the ``parabolic distance approximation'' $L(\rho)=D\left(1+\frac{\rho^2}{R D}\right)$ and by applying the substitution $\rho\to x=1+\frac{\rho^2}{R D}$ with $\upd x =\frac{2\rho}{R D}\,\upd\rho$, which leads to $L(x)=D\,x$, $\cot\vartheta = \sqrt{\frac{1}{\Delta(x - 1)}-1}$, and
\begin{align}
\Delta K_{\oslash\oslash}^{(ss)}&(\alpha,\vartheta_1,\vartheta_2, \Delta, \Theta) =
\pi\,H\left(\coscos\right)\int_1^{1+\Delta^{-1}r_s^2} \upd x\,x^{-d}\,\Delta k\left(x \vartheta\right) \nonumber\\
&-\mathrm{sign}\left(\coscos\right)\Bigg[\int_{1+\Delta^{-1}\cos^2\vartheta_1}^{1+\Delta^{-1} r_s^2} \upd x\,\arccos\left(|\cot\vartheta_1|\sqrt{\frac{1}{\Delta(x-1)}-1}\right)\,x^{-d}\,\Delta k\left(x \Theta\right) \nonumber\\
&\hspace{5.5em}+c(\alpha,\vartheta_1,\vartheta_2)\int_{1+\Delta^{-1}\cos^2\vartheta_2}^{1+\Delta^{-1} r_s^2} \upd x\,\arccos\left(|\cot\vartheta_2|\sqrt{\frac{1}{\Delta(x-1)}-1}\right)\,x^{-d}\,\Delta k\left(x \Theta\right) \Bigg] \nonumber\\
&+\alpha\int_{1+\Delta^{-1} r_s^2}^{1+\Delta^{-1}} \upd x\,x^{-d}\,\Delta k\left(x \Theta\right).
\label{eq:app_DeltaK_sp_janus_DA}
\end{align}
with the abbreviation $r_s = R_s/R$, and the replacement of $(\mathrm{sign}(\cos\vartheta_{1,2}))\cot\vartheta_{1,2}=|\cot\vartheta_{1,2}|$, which holds in the domain of definition of the polar angles, i.e., for $\vartheta_{1,2}\in [0,\pi]$.

\section{Scaling function of the effective potential for two Janus spheres}
\label{sec:app_janus_spheres_pot}
The effective potential can be determined from the force in the relative coordinate system according to
\begin{align}
V_{\symbJcJc}^{(ss)}(\vec{n}_1, \vec{n}_2, \vec{r}_{12}&=(D+2R)\vec{e}_z, R, T) = \int_D^\infty\upd z\,F_{\symbJcJc}^{(ss)}(\vec{n}_1, \vec{n}_2, \vec{r}_{12}=(z+2R)\vec{e}_z, R, T)\\
&=k_B T \frac{\mathcal{L}}{R^{d-2}}\int_D^\infty\upd z\,\frac{K_{\symbJcJc}^{(ss)}(\alpha, \vartheta_1, \vartheta_2, z/R,z/\xi_\pm)}{(z/R)^{d-1}}. \nonumber
\end{align}
Substitution of $z=D\,\tilde z$ with $\upd z = D\,\upd \tilde z$ yields
\begin{equation}
V_{\symbJcJc}^{(ss)}(\vec{n}_1, \vec{n}_2, \vec{r}_{12}=(D+2 R)\vec{e}_z, R, T)=k_B T \frac{\mathcal{L}}{R^{d-3}}\,\Delta^{-(d-2)}\int_1^\infty\upd \tilde z\,\frac{K_{\symbJcJc}^{(ss)}(\alpha, \vartheta_1, \vartheta_2, \tilde z \Delta,\tilde z \Theta)}{\tilde z^{d-1}}.
\label{eq:app_phi_sp_from_force}
\end{equation}
This can be cast into the scaling form
\begin{equation}
V_{\symbJcJc}^{(ss)}(\vec{n}_1, \vec{n}_2, \vec{r}_{12}=(D+2 R)\vec{e}_z, R, T) = k_B T\, \frac{\mathcal{L}}{R^{d-3}}\, \frac{\Phi_{\symbJcJc}^{(ss)}(\alpha, \vartheta_1, \vartheta_2, \Delta,\Theta)}{\Delta^{d-2}},
\end{equation}
with the scaling function $\Phi_{\oslash\oslash}^{(ss)}$ of the effective potential,
\begin{equation}
\Phi_{\symbJcJc}^{(ss)}(\alpha, \vartheta_1, \vartheta_2, \Delta,\Theta) = \Phi_{(+,+)}^{(ss)}(\Delta, \Theta) - \Delta \Phi_{\oslash\oslash}^{(ss)}(\alpha, \vartheta_1, \vartheta_2, \Delta,\Theta),
\end{equation}
where
\begin{align}
\Phi_{(+,\pm)}^{(ss)}(\Delta, \Theta) = &\pi\int_1^\infty\upd x\,(x - 1) x^{-d}\,k_{(+,\pm)}(x\,\Theta)\nonumber\\
&-\pi\int_{1+\Delta^{-1}}^\infty \upd x\,(x - 1 - \Delta^{-1})\,x^{-d}\, k_{(+,\pm)}(x\,\Theta)
\end{align}
is the scaling function of the potential between two homogeneous spheres \cite{Gambassi:2009}, with an explicit dependence on $\Delta$ retained (in spite of the underlying DA limit $\Delta\to 0$) for consistency with the dependence on $\Delta$ of the orientation dependent term $\Delta \Phi_{\oslash\oslash}^{(ss)}$.

In order to obtain the excess scaling function $\Delta\Phi_{\oslash\oslash}^{(ss)}$ one has to integrate $\Delta K_{\oslash\oslash}^{(ss)}$ from Eq. \eqref{eq:app_DeltaK_sp_janus_DA} in accordance with Eq.~\eqref{eq:app_phi_sp_from_force}. The integral of $\Delta K_{\oslash\oslash}^{(ss)}$ features two generic types of integrals (here, omitting the tilde of the integration variable):
\begin{equation}
I_1\equiv\int_1^\infty\upd z\,\frac{1}{z^{d-1}}\int_{1+b/(z\,\Delta)}^{1+a/(z\,\Delta)}\upd x\,x^{-d}\,\Delta k\left(x\, z\,\Theta\right)
\end{equation}
with the first and last contribution to this integral [compare Eqs.~\eqref{eq:app_DeltaK_sp_janus_DA} and \eqref{eq:app_phi_sp_from_force}] being described by $a=r_s^2,\ b=0$ and $a=1,\ b=r_s^2$, respectively, and
\begin{equation}
I_2\equiv\int_1^\infty\upd z\,\frac{1}{z^{d-1}}\int_{1+\cos^2\vartheta_{1,2}/(z\,\Delta)}^{1+r_s^2/(z\,\Delta)} \upd x\,\arccos\left(|\cot\vartheta_{1,2}|\sqrt{\frac{1}{(z\,\Delta)(x-1)}-1}\right)\,x^{-d}\,\Delta k\left(x \,z\,\Theta\right).
\end{equation}
We represent integral $I_1$ by the function
\begin{equation}
I_1\equiv\Delta u^{(ss)}(a,b,\Delta, \Theta)=\int_1^\infty\upd z\,\frac{1}{z^{d-1}}\Bigg[\int_{1+b/(z\,\Delta)}^\infty\upd x\,x^{-d}\,\Delta k\left(x\,z\, \Theta\right)-\int_{1+a/(z\,\Delta)}^\infty\upd x\,x^{-d}\,\Delta k\left(x\,z\,\Theta\right)\Bigg].
\end{equation}
With the substitution $x\to w=z\,\Delta\,(x - 1)$ so that $\upd w=z\,\Delta\,\upd x$ one has
\begin{multline}
\Delta u^{(ss)}(a,b,\Delta,\Theta)=\Delta^{-1}\int_1^\infty\upd z\,\frac{1}{z^d}\Bigg[\int_{b}^\infty\upd w\,\left(1+\frac{w}{z\,\Delta}\right)^{-d}\,\Delta k\left(z\left(1+\frac{w}{z\,\Delta}\right)\Theta\right)\\
-\int_a^\infty\upd w\,\left(1+\frac{w}{z\,\Delta}\right)^{-d}\,\Delta k\left(z\left(1+\frac{w}{z\,\Delta}\right)\Theta\right)\Bigg]
\end{multline}
and with the substitution $z\to y=z+w/\Delta$ with $\upd y = \upd z$ one finds
\begin{equation}
\Delta u^{(ss)}(a,b,\Delta,\Theta)=\Delta^{-1}\Bigg[\int_b^\infty\upd w\,\int_{1+w/\Delta}^\infty\upd y\,y^{-d}\,\Delta k\left(y\,\Theta\right)-\int_a^\infty\upd w\,\int_{1+w/\Delta}^\infty\upd y\,y^{-d}\,\Delta k\left(y\,\Theta\right)\Bigg].
\end{equation}
After switching the order of the integrations according to
\begin{equation}
\int_b^\infty\upd w\,\int_{1+w/\Delta}^{\infty}\upd y = \int_{1+b/\Delta}^\infty\upd y\,\int_b^{\Delta(y-1)}\upd w
\end{equation}
the integration over $w$ can be carried out, resulting in
\begin{align}
\Delta u^{(ss)}(a,b,\Delta, \Theta)=&\int_{1+b/\Delta}^\infty\upd y\,(y-1-b/\Delta)\,y^{-d}\Delta k(y\,\Theta)\nonumber\\
&-\int_{1+a/\Delta}^\infty\upd y\,(y-1-a/\Delta)\,y^{-d}\Delta k(y\,\Theta).
\end{align}
Integral $I_2$ is represented by the function
\begin{align}
I_2\equiv\Delta v^{(ss)}(r_s,\vartheta, \Delta, \Theta) =& \nonumber\\
\int_1^\infty\upd z\,\frac{1}{z^{d-1}}\Bigg[
&\int_{1+\cos^2\vartheta/(z\,\Delta)}^\infty \upd x\,\arccos\left(|\cot\vartheta|\sqrt{\frac{1}{(z\,\Delta)(x-1)}-1}\right)\,x^{-d}\,\Delta k\left(x\,z\,\Theta\right)\nonumber\\
&-\int_{1+r_s^2/(z\,\Delta)}^\infty \upd x\,\arccos\left(|\cot\vartheta|\sqrt{\frac{1}{(z\,\Delta)(x-1)}-1}\right)\,x^{-d}\,\Delta k\left(x\,z\,\Theta\right)\Bigg].
\end{align}
As before, we first use the substitution $x\to w=z\,\Delta\,(x - 1)$ with $\upd w=z\,\Delta\,\upd x$, followed by the substitution $z \to y=z+w/\Delta$ with $\upd y = \upd z$. This renders
\begin{align}
\Delta v^{(ss)}(r_s,\vartheta, \Delta, \Theta) =\Delta^{-1}\Bigg[&\int_{\cos^2\vartheta}^\infty\upd w\,\int_{1+w/\Delta}^\infty \upd y\,\arccos\left(|\cot\vartheta|\sqrt{\frac{1}{w}-1}\right)\,y^{-d}\,\Delta k\left(y\,\Theta\right)\nonumber\\
&-\int_{r_s^2}^\infty\upd w\,\int_{1+w/\Delta}^\infty \upd y\,\arccos\left(|\cot\vartheta|\sqrt{\frac{1}{w}-1}\right)\,y^{-d}\,\Delta k\left(y\,\Theta\right)\Bigg].
\end{align}
We recall that the semi-minor axes of the two half-ellipses are given by $b_{1,2}=|\cos\vartheta_{1,2}|$ and that $r_s$ denotes the distance of the intersection point between the half-ellipses from the symmetry axis of the two particles. Obviously, the intersection point cannot be closer to the common origin than any semi-minor axis, so that $|\cos\vartheta_1|\leq r_s$ and $|\cos\vartheta_2|\leq r_s$. Based on Eq.~\eqref{eq:app_DeltaK_sp_janus_DA}, we need to evaluate $\Delta v^{(ss)}(r_s,\vartheta, \Delta, \Theta)$ for $\vartheta=\vartheta_1$ and $\vartheta=\vartheta_2$. For that reason, we consider only the case $|\cos\vartheta|\leq r_s$ and reorder the integrals:
\begin{align}
&\int_{\cos^2\vartheta}^\infty\upd w\,\int_{1+w/\Delta}^\infty\upd y\ - \int_{r_s^2}^\infty\upd w\,\int_{1+w/\Delta}^{\infty}\upd y\nonumber\\
=&\int_{1+\cos^2\vartheta/\Delta}^\infty\upd y\,\int_{\cos^2\vartheta}^{\Delta(y-1)}\upd w - \int_{1+r_s^2/\Delta}^\infty\upd y\,\underbrace{\int_{r_s^2}^{\Delta(y-1)}\upd w}_{\int_{\cos^2\vartheta}^{\Delta(y-1)}\upd w\ - \int_{\cos^2\vartheta}^{r_s^2}\upd w}\\
=&\int_{1+\cos^2\vartheta/\Delta}^{1+r_s^2/\Delta}\upd y\,\int_{\cos^2\vartheta}^{\Delta(y-1)}\upd w\ + \int_{1+r_s^2/\Delta}^\infty\upd y\,\int_{\cos^2\vartheta}^{r_s^2}\upd w \nonumber
\end{align}
so that finally
\begin{align}
\Delta v^{(ss)}(r_s,\vartheta, \Delta, \Theta) = \Delta^{-1}&\int_{1+\cos^2\vartheta/\Delta}^{1+r_s^2/\Delta}\upd y\,g\big(\Delta(y-1),\vartheta\big)\,y^{-d}\,\Delta k\left(y\,\Theta\right)\nonumber\\
&+\Delta^{-1}\int_{1+r_s^2/\Delta}^\infty\upd y\,g(r_s^2,\vartheta)\,y^{-d}\,\Delta k\left(y\,\Theta\right),\enspace |\cos\vartheta|\leq r_s,
\end{align}
and
\begin{align}
g(u,\vartheta) &=\int_{\cos^2\vartheta}^u\upd w\,\arccos\left(|\cot\vartheta|\sqrt{\frac{1}{w}-1}\right)\nonumber\\
&=\left[w\arccos\left(|\cot\vartheta|\sqrt{\frac{1}{w}-1}\right) +|\cos\vartheta|\arcsin\left(|\csc\vartheta|\sqrt{1-w}\right)\right]_{\cos^2\vartheta}^u \nonumber\\
&= u \arccos\left(|\cot\vartheta|\sqrt{\frac{1}{u}-1}\right) - |\cos\vartheta|\arccos\left(|\csc\vartheta|\sqrt{1-u}\right),\enspace \cos^2\vartheta \leq u.
\label{eq:app_sp_gdef}
\end{align}
Note that $g(u=\cos^2\vartheta,\vartheta)=0$. Concerning the derivation of Eq.~\eqref{eq:app_sp_gdef} we leave out the detailed case analysis for the sign of $\cot\vartheta$, which in the end, can be subsumed by taking the absolute values as stated in Eq.~\eqref{eq:app_sp_gdef}.
Putting the results together, the excess scaling function of the potential is given by
\begin{align}
\Delta \Phi_{\oslash\oslash}^{(ss)}(\alpha,\vartheta_1,\vartheta_2,\Delta,\Theta) =\,&\pi\,H\left(\coscos\right)\Delta u^{(ss)}(r_s^2,0,\Delta, \Theta)\\
&-\mathrm{sign}\left(\coscos\right)\Big[\Delta v^{(ss)}(r_s^2,\vartheta_1, \Delta, \Theta)+c(\alpha,\vartheta_1,\vartheta_2)\Delta v^{(ss)}(r_s^2,\vartheta_2, \Delta, \Theta)\Big]\nonumber\\
&+\alpha\,\Delta u^{(ss)}(1,r_s^2,\Delta, \Theta)\nonumber.
\label{eq:app_DeltaPot_sp_janus_DA}
\end{align}
\ifTwocolumn\end{widetext}\fi

\bibliography{bibliography}

\end{document}